\documentclass[preprint,12pt]{elsarticle}

\usepackage{amssymb}
\usepackage{amsmath}

\usepackage{lineno}

\usepackage{mlmodern} 

\journal{Journal of Computational Physics}

\usepackage[dvipsnames]{xcolor}

\newcommand{\notaJGM}[1]{\textcolor{Blue}{#1}}

\newcommand{\hide}[1]{}

\usepackage{hyperref}
\hypersetup{
	colorlinks=true,
	linkcolor=blue,
	filecolor=magenta,      
	urlcolor=cyan,
	pdftitle={N-fit},
	pdfpagemode=FullScreen,
}

\usepackage{graphicx, wrapfig, subcaption, setspace, booktabs}
\usepackage{enumitem}
\usepackage{tabularx}
\usepackage{arydshln}

\usepackage[nottoc]{tocbibind}

\usepackage{subcaption}
\usepackage{adjustbox}

\usepackage{stackengine}

\usepackage{bm}

\begin{document}

\begin{frontmatter}



\title{Deep Learning Framework for Enhanced Neutrino Reconstruction of Single-line Events in the ANTARES Telescope}







            

\author[IPHC,UHA]{A.~Albert}
\author[IFIC]{S.~Alves}
\author[UPC]{M.~Andr\'e}
\author[UPV]{M.~Ardid}
\author[UPV]{S.~Ardid}
\author[CPPM]{J.-J.~Aubert}
\author[APC]{J.~Aublin}
\author[APC]{B.~Baret}
\author[LAM]{S.~Basa}
\author[APC]{Y.~Becherini}
% ...

\address[IPHC]{\scriptsize{Universit\'e de Strasbourg, CNRS,  IPHC UMR 7178, F-67000 Strasbourg, France}}
\address[UHA]{\scriptsize Universit\'e de Haute Alsace, F-68100 Mulhouse, France}
\address[IFIC]{\scriptsize{IFIC - Instituto de F\'isica Corpuscular (CSIC - Universitat de Val\`encia) c/ Catedr\'atico Jos\'e Beltr\'an, 2 E-46980 Paterna, Valencia, Spain}}
\address[UPC]{\scriptsize{Technical University of Catalonia, Laboratory of Applied Bioacoustics, Rambla Exposici\'o, 08800 Vilanova i la Geltr\'u, Barcelona, Spain}}
\address[UPV]{\scriptsize{Institut d'Investigaci\'o per a la Gesti\'o Integrada de les Zones Costaneres (IGIC) - Universitat Polit\`ecnica de Val\`encia. C/  Paranimf 1, 46730 Gandia, Spain}}
\address[CPPM]{\scriptsize{Aix Marseille Univ, CNRS/IN2P3, CPPM, Marseille, France}}
\address[APC]{\scriptsize{Universit\'e Paris Cit\'e, CNRS, Astroparticule et Cosmologie, F-75013 Paris, France}}
\address[LAM]{\scriptsize{Aix Marseille Univ, CNRS, CNES, LAM, Marseille, France }}
% ...

\begin{abstract}

We present the $N$-fit algorithm designed to improve the reconstruction of neutrino events detected by a single line of the ANTARES underwater telescope, usually associated with low energy neutrino events ($\sim$ 100 GeV). $N$-Fit is a neural network model that relies on deep learning and combines several advanced techniques in machine learning --deep convolutional layers, mixture density output layers, and transfer learning. This framework divides the reconstruction process into two dedicated branches for each neutrino event topology --tracks and showers-- composed of sub-models for spatial estimation --direction and position-- and energy inference, which later on are combined for event classification. Regarding the direction of single-line events, the $N$-Fit algorithm significantly refines the estimation of the zenithal angle, and delivers reliable azimuthal angle predictions that were previously unattainable with traditional $\chi^2$-fit methods. Improving on energy estimation of single-line events is a tall order; $N$-Fit benefits from transfer learning to efficiently integrate key characteristics, such as the estimation of the closest distance from the event to the detector. $N$-Fit also takes advantage from transfer learning in event topology classification by freezing convolutional layers of the pretrained branches. Tests on Monte Carlo simulations and data demonstrate a significant reduction in mean and median absolute errors across all reconstructed parameters. The improvements achieved by $N$-Fit highlight its potential for advancing multimessenger astrophysics and enhancing our ability to probe fundamental physics beyond the Standard Model using single-line events from ANTARES data.

\end{abstract}



\begin{keyword}
Deep Neural Network \sep Neutrino Event Reconstruction \sep Submarine Telescope \sep Convolutional Neural Networks \sep Mixture Density Networks \sep Transfer Learning \sep Dimensionality Reduction \sep Event Classification \sep Multimessenger Astrophysics



\end{keyword}

\end{frontmatter}


{
\hypersetup{linkcolor=black}
\tableofcontents
}

\newpage

\section{Introduction}
\label{sec:Intro}

Neutrino telescopes, such as the ANTARES deep-sea detector \cite{ANTARES}, are pivotal in advancing our understanding of high-energy astrophysical neutrinos, offering insights into cosmic sources and fundamental physics beyond the Standard Model. Accurate reconstruction of neutrino-induced events is essential for extracting meaningful physical information from the collected data. Hence, the inherent complexity of neutrino interactions, coupled with challenges posed by high-dimensional, limited information and correlated detector signals, can benefit from the development of advanced reconstruction techniques that surpass traditional methods. In recent years, machine learning (ML), particularly deep learning (DL), has emerged as a transformative tool in particle physics and astrophysics. The capacity of these techniques to process large volumes of data and extract intricate patterns has proven invaluable in neutrino experiments. For instance, the NOvA experiment implemented convolutional neural networks (CNNs) for event classification, achieving substantial improvements in neutrino identification accuracy \cite{Psihas}. Similarly, the IceCube and KM3NeT neutrino telescopes applied deep learning methods to enhance the reconstruction of neutrino direction and energy, resulting in improved angular resolution and energy precision \cite{IC_DL, Km3_DL2, Km3_DL}. 

The integration of ML into neutrino physics is part of a broader trend where Artificial Intelligence (AI) accelerates scientific discovery across various domains. Notably, AI models based on artificial neural networks leveraging deep learning have been applied to physical systems reaching remarkable accuracy, enabling advancements in fields such as weather forecasting and climate modelling \cite{Jones2023}. In fundamental physics, ML has been instrumental in detecting rare events, such as observations of the Higgs boson using Boosted Decision Trees (BDT) \cite{Higgs}, and assessing potential cosmic phenomena, thereby expanding our understanding of the Universe \cite{Duev2019}. Conventional reconstruction algorithms used by the ANTARES experiment primarily rely on traditional techniques, such as $\chi^2$ regression \cite{BBfit, Tantra} and maximum likelihood approaches \cite{AAfit, AAfit2}. While effective for certain event types, these methods exhibit limitations in capturing the full complexity of event signatures, particularly in single-line (SL) events, i.e., events reconstructed with data from only one detection line (see \autoref{sec:antares}). In such cases, reconstructing the azimuthal angle has remained a significant challenge, limiting the precision of directional and energy estimations. To address these issues, we introduce a novel deep learning–based reconstruction model, $N$-Fit, designed to enhance the accuracy and robustness of neutrino event reconstruction for this type of events.

$N$-Fit integrates multiple advanced neural network architectures, including deep convolutional networks \cite{CNN}, mixture density networks \cite{MDN}, and transfer learning strategies \cite{Transfer}, to effectively process detector data. This modular framework decomposes the reconstruction task into dedicated bran\-ches for each neutrino event topology --tracks and showers-- with their own networks for directional estimation and energy inference. Both branches are ultimately combined for event classification through transfer learning, ensuring a comprehensive characterization of neutrino interactions. The results of our study, using supervised training through Monte Carlo (MC) simulations and validated by both simulations and real data, demonstrate a significant improvement in reconstruction accuracy compared to conventional methods. These advancements underscore the potential of deep learning methodologies in multimessenger astrophysics, offering new opportunities to explore the high-energy neutrino sky and investigate phenomena beyond the Standard Model.

Remarkably, $N$-fit surpasses the SL direction reconstruction over $\chi^2$-fit in two main aspects; first, by significantly improving zenithal angle estimates; and second, by reporting azimuthal angle estimates, which were impossible to disentangle in $\chi^2$-fit when all the hits are coplanar. In addition, $N$-fit benefitted from knowledge in pre-trained direction reconstruction through \emph{direct} and \emph{indirect} transfer learning (TL), leveraging improvements beyond direction reconstruction. In the next sections, we first describe the ANTARES detector and the characteristics of the dataset used in the study (\ref{sec:antares}), then fully describe $N$-fit components in detail (\ref{sec:approach}) and its application in ANTARES (\ref{sec:application}). After that, we compare its performance with respect to the standard reference for low-energy SL events in ANTARES (\ref{sec:results}), and illustrate the use of $N$-fit in physics analyses (\ref{sec:physics}).







\section{ANTARES dataset}
\label{sec:antares}


ANTARES, located 40 km offshore from Toulon at 2475 m depth, was the first undersea neutrino telescope \cite{ANTARES}. It was completed on May 29, 2008, becoming the largest neutrino telescope in the sea until February 2022, when it was decommissioned and KM3NeT took over. Results obtained by the ANTARES detector are summarized in \cite{legacy}. ANTARES was made of 12 flexible vertical lines anchored to the seabed. Each line had 25 storeys with a vertical distance of $14.5$ m. In each storey there were 3 Optical Modules (OMs) --consisting of a glass sphere containing a large photomultiplier (PMT)-- separated $120^\circ$ in the horizontal plane and looking downwards at $45^\circ$. The storey also contained a titanium cylinder for the electronics and other devices needed for the operation of the telescope, such as positioning and timing calibration systems and connections.

The data acquisition in ANTARES followed an \textit{all-data-to-shore} philosophy. This means that every PMT signal surpassing a predefined voltage threshold is digitized by an Analogue Ring Sampler (ARS) chip offshore and sent to the shore station where they are filtered, recording an event if certain triggers were fulfilled. A \textit{run} was defined as a time period in which the telescope took data with a given configuration, typically covering a few hours. The information for each event included the run number, the position of OMs, their direction, the environmental conditions and the information of the PMTs in terms of hits, i.e., discrete signals that were triggered when the amplitude of the PMT reached the predefined threshold of $0.3$ photoelectrons. The hit information was basically the time, voltage amplitude, and the OM identifier.

\newcommand{\numuCC}{%
  \makebox[0pt][l]{\raisebox{1.25ex}[0pt]{$\tiny{\scalebox{.4}{\textbf{(~~~~)}}}$}}\bar{\nu}_\mu^{\mathrm{CC}}
}

\newcommand{\numuNC}{%
  \makebox[0pt][l]{\raisebox{1.25ex}[0pt]{$\tiny{\scalebox{.4}{\textbf{(~~~~)}}}$}}\bar{\nu}_\mu^{\mathrm{NC}}
}

\newcommand{\nueCC}{%
  \makebox[0pt][l]{\raisebox{1.25ex}[0pt]{$\tiny{\scalebox{.4}{\textbf{(~~~~)}}}$}}\bar{\nu}_e^{\mathrm{CC}}
}

\newcommand{\nueNC}{%
  \makebox[0pt][l]{\raisebox{1.25ex}[0pt]{$\tiny{\scalebox{.4}{\textbf{(~~~~)}}}$}}\bar{\nu}_e^{\mathrm{NC}}
}

The information recorded in the runs is used to reconstruct the event parameters, such as the neutrino direction or energy. An event can be classified as a \textit{track} or as a \textit{shower} event, depending on the signature left in the detector. Track-like events originate mainly from charged current (CC) interactions of muon-(anti)neutrinos ($\numuCC$) where a muon is produced, which crosses the detector and typically leaves a straight line signature. Shower-like events originate from neutral current (NC) interactions ($\nueNC$ and $\numuNC$) and charged currents of electron-(anti)neutrinos ($\nueCC$). They can be seen as localized point sources of light in the detector. Every event is reconstructed as a track and as a shower, and then selected as one or the other for the physics analyses according to reconstruction quality criteria.

Tracks are the most used events in current physics analyses. There are two main methods developed by the ANTARES Collaboration to reconstruct track events. The $\lambda$-fit method \cite{AAfit, AAfit2} is based on maximum likelihood estimates and it is particularly suitable for mid- and high-energy events ($\gtrsim 150$ GeV), whereas the $\chi^2$-fit method is most efficient for lower energies ($\lesssim150$ GeV). The $\chi^2$-fit method relies on $\chi^2$ analyses \cite{BBfit}, and it allows discrimination between neutrino events that are detected by a SL vs. detection in multiple lines (multi-line). SL events can occur due to several factors, such as the energy of the neutrino, the distance of its interaction from the detector, or its high verticality. SL events for the current study were selected according to $\chi^2$-fit, i.e., those events reconstructed by $\chi^2$-fit with only htis from a single line. Shower-like events were reconstructed by a different $\chi^2$ method presented in \cite{Tantra}.

The ANTARES Collaboration developed Monte Carlo simulations of the detector performed \textit{run-by-run}, which are particularly useful to develop and test reconstruction algorithms \cite{MCsim}. These simulations take into account the telescope configuration, and the set of parameters ruling the run condition. Thus, they are well suited for a supervised learning approach. Given the huge amount of data available from the ANTARES Collaboration, we selected a subset ($\sim$10\%) of runs, to keep a good balance between training time and performance. We used charged and neutral current interactions of electron and muon neutrinos for the development of the method. This data was subdivided in training, validation and test datasets in a $3:1:1$ ratio. After the development phase, we also used other events types, such as simulations of atmospheric muons and random noise background events, for control analyses (see \autoref{subsec:robust} and \ref{app:robust}).

\begin{figure}[htbp]
	\centering
	\includegraphics[scale=0.4]{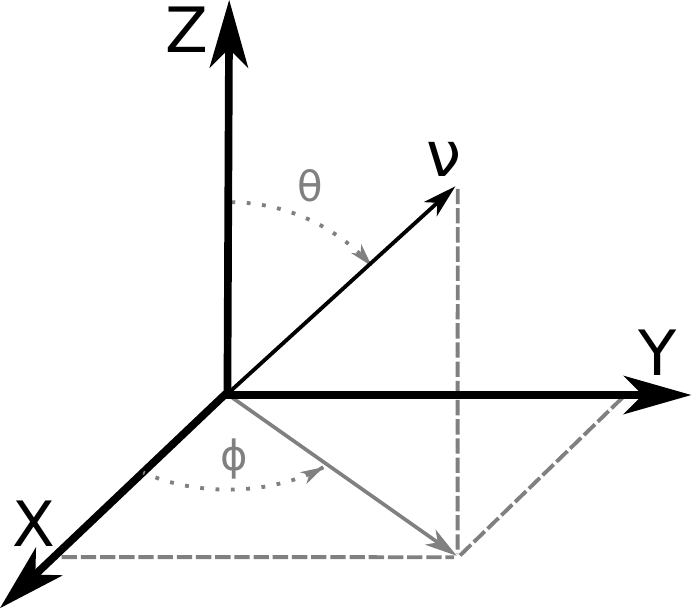}
	\caption{\label{fig:reco}Schematic representation of the angles $\theta$ and $\phi$ that define the direction of a neutrino, $\nu$, in the ANTARES coordinate system.}
\end{figure}

Before proceeding with the $N$-fit development, we must first clarify the coordinates we use in ANTARES. We describe the spherical coordinates $(\theta,\,\phi)$ as the zenithal and azimuthal angles, which do not strictly represent the same magnitudes as (Zenith, Azimuth), even though they are strongly correlated. The coordinates (Zenith, Azimuth) represent the direction where the neutrino comes from in the sky, while ($\theta$, $\phi$) represent the direction to where neutrino moves to (see \autoref{fig:reco}). The transformation between both coordinates is the following:

\begin{equation}
	\mathrm{Zenith} = 180^\circ-\theta \quad \quad \quad
	\mathrm{Azimuth} = \phi + 180^\circ \,.
\end{equation}

Note that the range of the Azimuth is $(0^\circ, \, 360^\circ)$, while the range $(-180^\circ,$ $180^\circ)$ is used for $\phi$. Both of these coordinates are in the Universal Transverse Mercator (UTM) system. In this system, the $X$ coordinate points to the UTM East, while the $Y$ coordinate points to the UTM North, which means that the Azimuth runs anti-clockwise.

\section{Components of \textit{N}-fit}
\label{sec:approach}

The key feature for the successful performance of the $N$-fit algorithm is the integration of several Deep Learning tools. In this section, we unravel the algorithm: first, we introduce the very basics of artificial neural networks, on which the algorithm is based; then, we describe deep convolutional networks (DCNs) and mixture density networks (MDNs), which play an essential role in $N$-fit's capabilities. Lastly, we describe how we utilized different aspects of transfer learning (TL) to enhance challenging reconstruction analyses, such as energy estimation and event classification.

\subsection{Basics of Artificial Neural Networks}
\label{subsubsec:ANNs}

\begin{figure}[htbp]
	\centering
	\includegraphics[width=.95\textwidth]{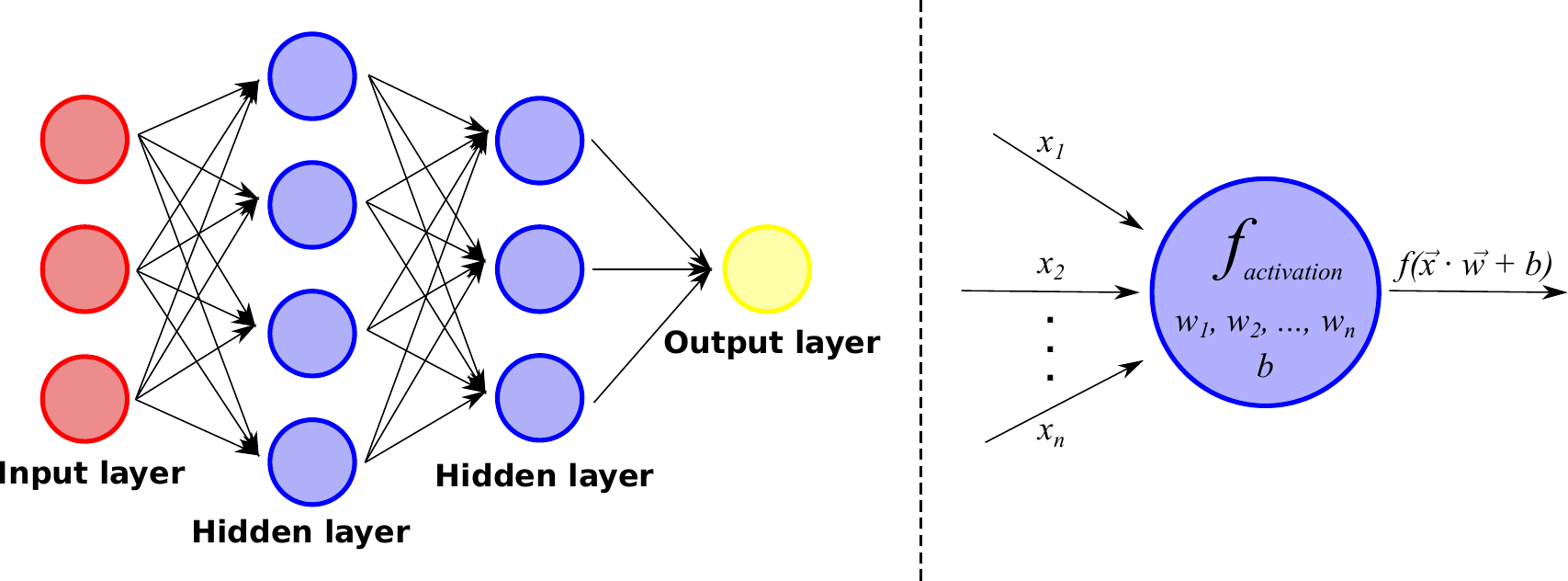}
	\caption{\label{fig:feed-forw}Schematic diagram of a simple feed-forward neural network. Circles represent neurons, and arrows represent synaptic connections. In general, more than a single output can be considered.}
\end{figure}

Neurons are the core units of neural network models. The basic neuron model in artificial neural networks is the perceptron \cite{perceptron}, which is inspired by the non-linear transduction of synaptic input summation in biological neurons towards action potential firing. Mathematically, the output ($y$) is described as the result of a nonlinear \textit{activation} function of a weighted linear combination of synaptic inputs: 

\begin{equation}
\label{eq:perceptron}
y = f\left( \sum_i w_i \cdot x_i + b \right) =  f\left( \vec{w}\cdot\vec{x} +b \right) \,,
\end{equation}
where $\vec{x}$, $\vec{w}$, and $b$ represent inputs, weights, and the neuron bias, respectively. Even though this represents a strong oversimplification of the nonlinear dynamics of biological neurons in the brain, such a computation is capable of mapping arbitrary input-output functions efficiently if multiple perceptrons are present in layers, creating a \textit{feed-forward} network (\autoref{fig:feed-forw}). In $N$-fit, the Rectified Linear Unit \cite{relu} --defined as $\mathrm{ReLU}(x) = \max\{0,x\}$-- is used as the activation function of neurons in the hidden layers.

Each layer in the network model processes a level of internal representation that transmits information from one layer to the next. One of the main characteristics of deep vs. shallow learning is precisely its power of abstraction, which is achieved by significantly increasing the number of layers. Data fixes the number of neurons in the input, whereas the number of outputs is concomitant to the question that the model is designed to address. In contrast, the number of hidden layers and the number of neurons in them can vary. These are examples of hyper-parameters that need to be explored while testing the network performance. Similarly, the learning process is also sensitive to the network initialization \cite{ini_weights}. We initialized weights and biases randomly following \cite{He} for all the feed-forward layers in the $N$-fit algorithm.

During training in supervised machine learning, the loss function computes the error committed between the network output and the desired output in data batches (of 64 events in $N$-fit). Weights and biases are updated following error backpropagation, which uses gradient descent to propagate the output loss backwards through the network from the output layer, through the hidden layers, to the input layer. While there are several optimization algorithms based on error backpropagation \cite{SGD}, $N$-fit uses the Adam algorithm \cite{Adam} because it dynamically self-regulates the learning rate, which controls the pace at which learnable parameters in the network adapt. The learning rate was initialized at $0.001$ in every $N$-fit training process.

Typically, the learning procedure iterates over multiple epochs until a predetermined number of cycles is completed, where an epoch constitutes one complete pass of the entire training dataset through the neural network. However, an optimization strategy called Early Stopping enables premature termination of training before reaching this preset epoch threshold in order to diminish the risk of over-fitting \cite{EarlyStop}. This method monitors the loss of the validation dataset after each epoch and halts the process when two conditions are met: the validation loss reaches a minimum value, and a specified number of subsequent epochs (termed the \textit{patience} parameter) elapse without further reduction in validation loss. Upon stopping, the algorithm automatically restores the model parameters corresponding to the optimal validation performance observed during training. We set the maximum number of epochs to 150 and the patience to 10 epochs.

Another technique employed in order to avoid overfitting is called Drop\-Out \cite{Dropout}. It consists of randomly switching off a preset percentage of random neurons within a layer so that the active neurons vary in each batch. In this way, DropOut prevents the neurons from becoming too specialized. In the $N$-fit algorithm, 20\% of the neurons from the convolutional block outputs were randomly dropped (i.e., their activation was set to 0) before feeding the first feed-forward layer.

\subsection{Deep Convolutional Networks}
\label{subsubsec:DCN}

Convolutional neural networks (CNNs) are a class of neural network models designed for processing structured grid data, such as images and image-like tensors \cite{CNN}. Inspired by the human visual system, they are built to recognize patterns and spatial hierarchies efficiently. Unlike traditional fully connected neural networks such as feed-forward ones, which treat every input as independent, CNNs leverage local connectivity, weight sharing, and hierarchical feature extraction to reduce the computational complexity while enhancing learning efficiency. At the heart of a CNN are convolutional layers, which are composed of small, trainable filters that slide over the input image to extract meaningful patterns.  These filters evolve over time through training, adapting to detect increasingly sophisticated features in deeper layers. Initially, filters may identify simple edges or gradients, but as they propagate through deeper layers, they start recognizing textures and shapes. The weight updates are governed by backpropagation and gradient descent, ensuring that the filters become more attuned to distinguishing relevant features from noise. After convolutional operations, the activation function introduces non-linearity to the network. In $N$-fit convolutional layers, ReLU is used, as well as in the fully connected layers. Without the activation function, CNNs would merely perform linear transformations, limiting their ability to capture complex representations.

Pooling layers follow the convolutional layers to refine the feature maps by reducing their spatial dimensions. These layers do not have trainable weights but instead apply fixed operations such as max pooling or average pooling. We used max pooling, the most common method, that retains the most prominent feature in a given region, preserving the strongest activations while discarding less relevant information. This downsampling process helps mitigate overfitting by reducing the number of parameters, thereby improving generalization. By compressing the feature maps, pooling layers also allow subsequent convolutional layers to focus on more abstract features without excessive computational overhead.

Deep Convolutional Networks (DCNs), meaning Deep CNNs, are the main tools used in the development of $N$-fit. DCNs extend traditional CNNs by stacking multiple convolutional, activation, and pooling layers to extract increasingly abstract and hierarchical features from input data. We processed ANTARES data generating image-like tensors that serve as inputs (see \autoref{subsec:data}). Weights initialization followed \cite{Glorot} for every convolutional layer in $N$-fit. 

While basic CNNs can capture low-level patterns such as edges and textures, DCNs leverage deeper architectures to detect complex structures, object parts and entire objects with high accuracy. As the depth increases, feature maps transition from simple to highly abstract representations, enabling superior generalization and performance in tasks like image recognition and object detection. Training DCNs introduces challenges such as vanishing gradients and computational inefficiency. They are addressed through techniques like batch normalization, which we considered in $N$-fit. By deepening the network, DCNs significantly enhance the ability to learn intricate patterns, making them the backbone of state-of-the-art deep learning applications. Lastly, the extracted features from the convolutional layers are fed into fully connected layers, such as those in the feed-forward network explained earlier. At this stage, the network shifts from learning spatial hierarchies to making final predictions. 

\subsection{Mixture Density Networks}
\label{subsubsec:MDN}

Mixture Density Networks (MDNs) extend traditional regression neural networks by predicting complex, multimodal probability distributions instead of deterministic single-point estimates \cite{MDN}, making them ideal for estimating uncertainty in data. A measure of uncertainty is essential for selecting the most reliable estimates in posterior physics analyses.

The outputs of a MDN model represent the necessary parameters to create the multimodal probability density function (PDF) from Gaussian kernels. Thus, instead of the typical output $\vec{y}_{p}$ for a single multidimensional prediction, the outputs of the network consist of means $\vec{\mu}_i$, variances $\sigma_i$ and mixture weights $\alpha_i$ (assuming that the outputs are statistically independent):
\begin{equation}
PDF(\vec{y}_p) = \sum_i^n \alpha_i \cdot N(\vec{\mu}_i, \sigma_i; \,\vec{y}_p), \quad \quad \text{such that} \quad \sum_i^n \alpha_i = 1 \,,
\end{equation}
where $N(\vec{\mu}, \sigma;\, \vec{y}_p)$ is a multi-dimensional isotropic Gaussian function. The number of Gaussian distributions ($n$) must be hand fixed and predicted variances must always be positive, using the appropriate activation function --$N$-fit uses $\text{ELU}+1$ (ELU being the Exponential Linear Unit function). Similarly, the mixture weights use a softmax function to ensure they add to one. The network model learns to predict the parameters of these Gaussians --means, variances, and mixture weights-- using a specialized loss function based on maximum likelihood estimation (MLE), where the network is optimized to maximize the probability of the observed data under the predicted mixture model (\ref{eq:logloss}):
\begin{equation}
\label{eq:logloss}
\mathcal{L} = -\ln \left\{ \sum_i^n  \frac{\alpha_i}{(2\pi)^{c/2}\sigma_i^c}\exp\left( -\frac{||\vec{y}_t-\vec{\mu}_i||^2}{2\sigma_i^2} \right) \right\} \,,
\end{equation}
where $c$ is the dimension of the output vector $\vec{y}_t$. In the development of $N$-fit, we only needed to parametrise each of our reconstructed parameters, following each a single Gaussian mode. With $c=1$, the loss function is reduced to: 
\begin{equation}
\label{eq:loglossfinal}
\mathcal{L} = \ln(\sqrt{2\pi}\sigma) + \left( \frac{(y_t-\mu)^2}{2\sigma^2} \right) \,.
\end{equation}

This is different from traditional loss functions like mean-squared error (MSE), as it allows the model to learn distributions that can accommodate multiple correct answers, such as in inverse problems, rather than forcing a single prediction. However, training MDNs can be challenging due to numerical instability, and difficulties in optimizing multiple parameters simultaneously. These effects can be mitigated using regularization techniques, careful initialization strategies, and robust optimization methods like Adam's. In addition to careful initialization and Adam's optimization, $N$-fit avoided numerical instability issues by adding a small contribution to $\sigma$'s activation function: $\text{ELU}+1+\epsilon$. The value $\epsilon = 10^{-5}$ was selected after a quick exploration, in which a balance was achieved between being small enough not to alter the results, but large enough to avoid instability.

By integrating neural networks with probabilistic mixture models, MDNs bridge the gap between deep learning and statistical modelling.

\subsection{Transfer Learning}
\label{subsubsec:TL}

Transfer learning (TL) refers to the technique of leveraging knowledge gained from previously trained neural network models to improve performance on related tasks \cite{Transfer}. In the context of $N$-fit, we adopt two distinct paradigms of TL: \emph{direct transfer learning}, based on model reuse through layer freezing, and \emph{indirect transfer learning}, relying on knowledge distillation via dimensionality reduction.

In direct TL, convolutional blocks from pretrained models are reused as fixed feature extractors in subsequent models. Specifically, networks trained for spatial reconstruction are repurposed for classification in $N$-fit between track and shower neutrino events by freezing their convolutional layers. These frozen components are connected in parallel to a feed-forward neural network (FFN), effectively enriching the classifier input space with spatially encoded representations without requiring retraining of the feature extractors. This strategy benefits from the spatial specialization of each pretrained model and mitigates overfitting by reducing the number of trainable parameters for classification.

Indirect TL in $N$-fit is achieved through knowledge distillation \cite{moslemi2024} from pretrained networks by extracting internal activations and transforming them into compact, informative representations using dimensionality reduction techniques. Specifically, this is accomplished using Principal Component Analysis (PCA), a linear dimensionality reduction technique that identifies orthogonal directions (principal components) along which data variability is maximized \cite{PCA}.

PCA is applied to neuron activations from the hidden layers of the networks trained for spatial reconstruction in $N$-fit, which are assumed to encode latent features relevant for energy estimation. These activations form a high-dimensional feature space, potentially redundant or noisy. With the PCA, we can reduce this space by projecting the original activations onto a lower-dimensional subspace defined by the principal components. Components are ordered by their explained variance, and only those contributing significantly to the total variance are retained. This selection ensures that the most informative and least correlated features are preserved.

The resulting low-dimensional vectors serve as inputs to a downstream FFN tasked with energy regression. This setup effectively transfers high-level abstract knowledge learned during direction and distance reconstruction, enabling the energy model to exploit interdependencies among event features not explicitly provided by the original inputs. More generally, this PCA-based approach enables a form of model-agnostic knowledge transfer, where the distilled representations serve as a bridge between pretrained networks and the target model.

Specific details on how Transfer Learning is applied to every reconstruction task are left to \autoref{subsec:workflow}.

\section{Application to the ANTARES dataset}
\label{sec:application}

\hide{
\notaJGM{Esto lo he cogido de la introducción que había antes de la sección 3, aunque creo que ha quedado muy justa. Podrías directamente borrarla o poner algo con más sentido.}

Here, we present how we preprocess the ANTARES input data to the algorithm. Then, we explain specific details in the development of the $N$-fit algorithm related to results.
}

\subsection{Dataset preprocessing}
\label{subsec:data}

The first step to apply $N$-fit to single-line (SL) neutrino events was to preprocess the available data as image-like tensors. In this view, the ANTARES telescope is akin to a camera collecting 3D images at each time step, having as many pixels as OMs. For SL events, we created colored 2D images, with time in one dimension (X-axis) and the storey of the line in the other (Y-axis). The time stamp of each event was relative to the reference hit of the $\chi^2$-fit strategy, which is the one that occurred first in time. Next, all hits recorded by the OMs on the line covering a time window of $[-200$, $600]$ ns relative to the reference hit are selected. Cleaning noisy or background hits is left to the networks themselves. Colors resulted from combining the information provided by the three evenly-distributed OMs in each storey. In each pixel of the 2D image, we inserted RGB channels to be informative of the event reconstruction in the horizontal (XY) plane, helping to estimate the azimuthal angle ($\phi$). More specifically, we weighted the angle to which each OM points in XY ($\alpha$) by its recorded voltage amplitudes ($A$). Mathematically, the transformation is represented as: 
\begin{equation}
\label{eq:RGB}
(R;G;B) = \left\{
\begin{aligned}
A\cdot(1-\tfrac{\alpha}{2\pi/3};\tfrac{\alpha}{2\pi/3};0) ~~~~~~
\quad\text{if}~~&\alpha \in \left[0, \tfrac{2\pi}{3}\right]
\\
A\cdot(0;2-\tfrac{\alpha}{2\pi/3};\tfrac{\alpha}{2\pi/3}-1)
\quad\text{if}~~&\alpha \in \left[\tfrac{2\pi}{3}, \tfrac{4\pi}{3}\right] \,.
\\
A\cdot(\tfrac{\alpha}{2\pi/3}-2;0;3-\tfrac{\alpha}{2\pi/3})
\quad\text{if}~~&\alpha \in \left[\tfrac{4\pi}{3}, \tfrac{6\pi}{3}\right]
\end{aligned}
\right.
\end{equation}

\begin{figure}[htbp]
	\centering
	\includegraphics[width=.5\textwidth]{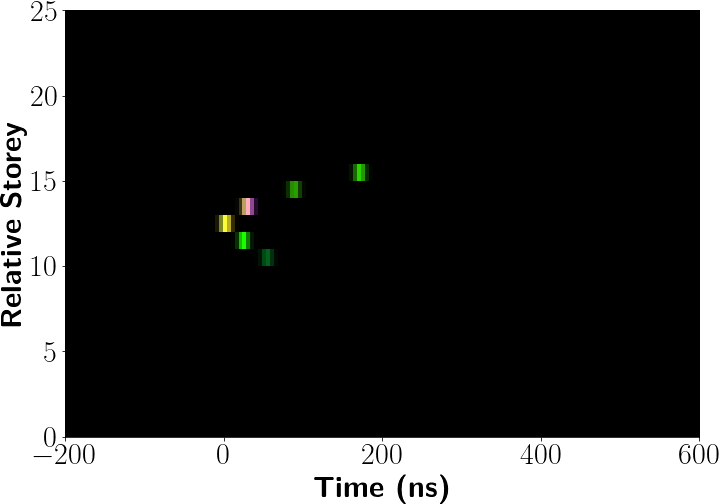}
	\caption{\label{fig:RGB}Example of a normalized RGB image from a track-like event ($\numuCC$).}
\end{figure}

From the discrete hits, we built a continuous signal representing the hit amplitudes over time by applying a Gaussian kernel smoothing with $\sigma = 5$ ns, and then discretized this signal at regular intervals of 5 ns for each individual color: $\{R(t);G(t);B(t)\}$. Finally, we re-centred the image based on the floor of the reference hit (a representative example is shown in \autoref{fig:RGB}).

The collected images of SL events were separated into track-like ($\numuCC$) and shower-like ($\numuNC$, $\nueCC$, $\nueNC$) MC simulations datasets, in order to train specialized models for each type of event. Before feeding images to the network, a Z-score normalization was separately applied to each set. For this, we computed the mean and standard deviation of the respective train sub-sets and applied the transformation to all sub-sets (including real data) to control for distribution shifts.

%
%
%
%
%
%
%
%
%

\subsection{$N$-fit modular organization}
\label{subsec:workflow}

The $N$-fit reconstruction strategy is subdivided into two branches specialized in each type of event: tracks and showers. Each branch covers the different aspects that fully describe neutrino events. The track-branch reconstructs the neutrino direction, the closest point of the secondary muon track to the detector line and the muon energy. The shower-branch also reconstructs the direction of the neutrino, the vertex point of the neutrino interaction and its energy. The two branches converge through TL in the classifier that divides the events in tracks or showers. The next subsections explain the details of these network models.

\subsubsection{Direction reconstruction}
\label{subsubsec:direction}

Several architectures of increasing complexity were analyzed during the optimization of direction reconstruction models for track events. The key steps which were followed to measure and improve the model performance are summarized below:

\begin{enumerate}[label=\textit{\roman*})]
    \item Baseline Control Model: Feed-forward network of four hidden layers implemented to fully reconstruct the Cartesian components of the direction unit vector: $(X;Y;Z)$.
    
    \item Components Separation: Next, the model was divided to reconstruct $Z$ and $\{X,Y\}$ separately. The reconstructions were sequential: first $Z$, then $\{X,Y\}$. We regularized the value of the components $\{X,Y\}$ to penalize deviations of $(X;Y;Z)$ from unit vectors.
    
    \item MDNs and $\theta$ Representation: Point predictions were replaced by inferring probability distributions (MDN). The change did not affect the performance but provided event uncertainty estimation, which is of major relevance as a quality metric in posterior physics analyses. Also, the network output $Z$ was replaced by the angle $\theta$, gaining a best estimation of the angle uncertainty, since no error propagation was needed to infer it. 
    
    \item Convolutional Layers: A significant improvement was observed by ad\-ding two convolutional layers before the dense network.
    
    \item Image Centring: Aligning events to a centred reference frame further enhanced performance, despite the architecture remained unaltered.
    
    \item Final Adjustments: The number of convolutional layers was optimized (see \autoref{tab:comparison}). From there on, additional layers provided only marginal changes in performance.

\end{enumerate}

\begin{figure}[htbp]
	\centering
	\includegraphics[scale=0.73]{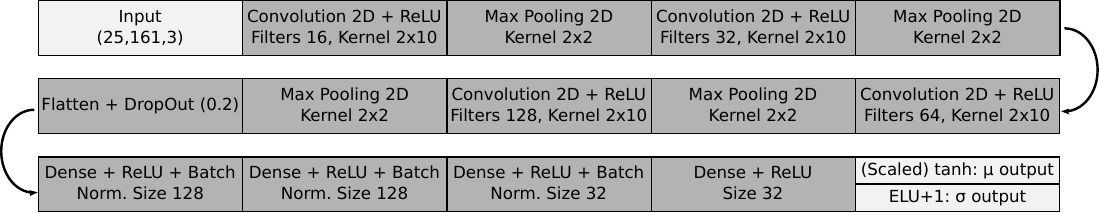}
	\caption{\label{fig:network}Details of the direction neural network architecture. Note that for the $\theta$ prediction, we scaled the hyperbolic tangent activation function for $\mu_\theta$, so that its values lay in $[0, \pi]$ radians. No scaling was necessary for the prediction of $\phi$ since its estimation was derived from the Cartesian $\{X,Y\}$ components of the unit vector.}
\end{figure}

The progression of the Mean Absolute Error (MAE) through these steps is presented in \autoref{tab:comparison}. Further analyses and results can be found in \autoref{sec:results}. The final architecture is illustrated in \autoref{fig:network}.

This optimized architecture for the track branch was then applied in the shower branch without further optimization.

\begin{table}[htbp]
    \centering
    \renewcommand{\arraystretch}{1.2} 
    \begin{tabular}{|c|cc:cccc|}
    \hline
    MAE & \textit{i} & \textit{ii} & \textit{iii } & \textit{iv} & \textit{v} & \textit{vi} \\ \hline
    $\theta$ & $11.2^\circ$ & $10.5^\circ$ & $10.5^\circ$ & $9.6^\circ$ & $8.4^\circ$ & $7.4^\circ$  \\ 
    $\phi$ & $50.1^\circ$ & $49.5^\circ$ & $49.5^\circ$ & $46.2^\circ$ & $44.1^\circ$ & $41.4^\circ$ \\ \hline
    \end{tabular}
    \caption{\label{tab:comparison}Evolution of the MAE of the test data set along the key steps which were followed in the development and optimization of the $N$-fit algorithm for the direction reconstruction of tracks. The estimation of $\sigma$ was introduced in step (\textit{iii}).}
\end{table}

The loss function applied to the angle $\theta$ becomes:

\begin{equation}
\label{eq:loss_theta}
\mathcal{L} = \ln (\sqrt{2\pi}\sigma_{\theta}) + \frac{1}{2}\frac{(\theta_{t}-\mu_{\theta} )^2}{\sigma_{\theta}^2} \,,
\end{equation}
where $\theta_t$ represents the true value of the angle $\theta$ for each neutrino event in the simulation.

A second network predicts the angle $\phi$ through its Cartesian coordinate values ($\mu_X$, $\mu_Y$) and their uncertainties ($\sigma_X$, $\sigma_Y$). The loss function was transformed consequently as:
\begin{equation}
\label{eq:loss_phi}
\begin{split}
\mathcal{L} = &\ln (\sqrt{2\pi}\sigma_{X}) + \frac{1}{2}\frac{(X_{t}-\mu_{X} )^2}{\sigma_{X}^2}+\\
+ &\ln (\sqrt{2\pi}\sigma_{Y}) + \frac{1}{2}\frac{(Y_{t}-\mu_{Y} )^2}{\sigma_{Y}^2}+\\
+ &(1-[\mu_{X}^2+\mu_{Y}^2+\cos^2(\mu_{\theta})])^2 \,.
\end{split}
\end{equation}

The last term in equation (\ref{eq:loss_phi}) regularizes the output of the network, penalizing deviations of $(X;Y;Z)$ from unit vectors. 
To infer the uncertainty of $\phi$, we performed a quadratic error propagation:
\begin{equation}
\label{eq:error_prop}
\sigma_{\phi}^2 = \left(\frac{\partial \phi}{\partial \mu_X}\cdot \sigma_X\right)^2 + \left(\frac{\partial \phi}{\partial \mu_Y}\cdot \sigma_Y\right)^2 \Rightarrow
\sigma_{\phi} = \frac{\sqrt{\sigma_X^2\cdot \mu_Y^2+\sigma_Y^2\cdot \mu_X^2}}{\mu_X^2+\mu_Y^2 } \,.
\end{equation}

\begin{figure}[htbp]
	\centering
	\includegraphics[width=.48\textwidth]{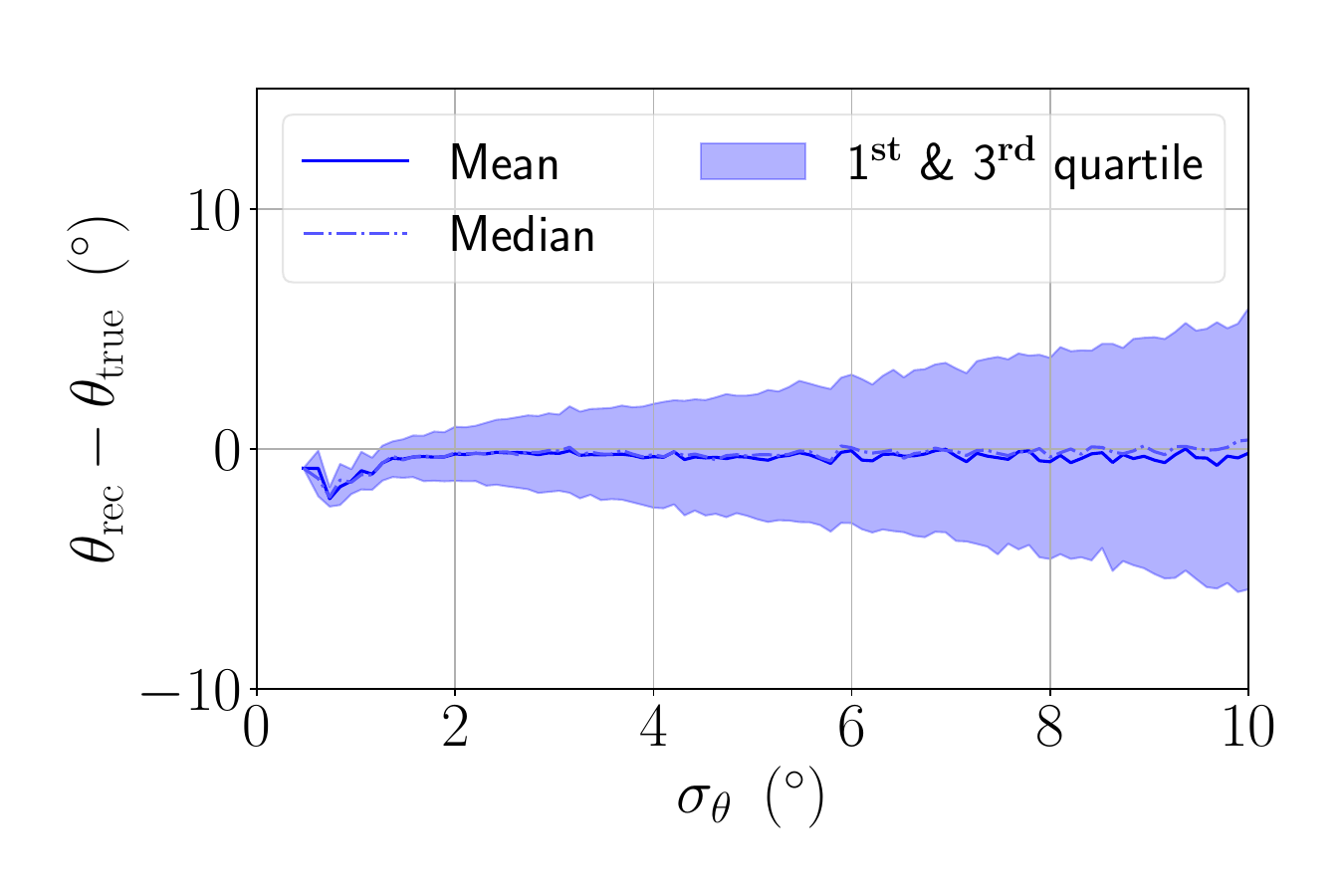}
	\includegraphics[width=.48\textwidth]{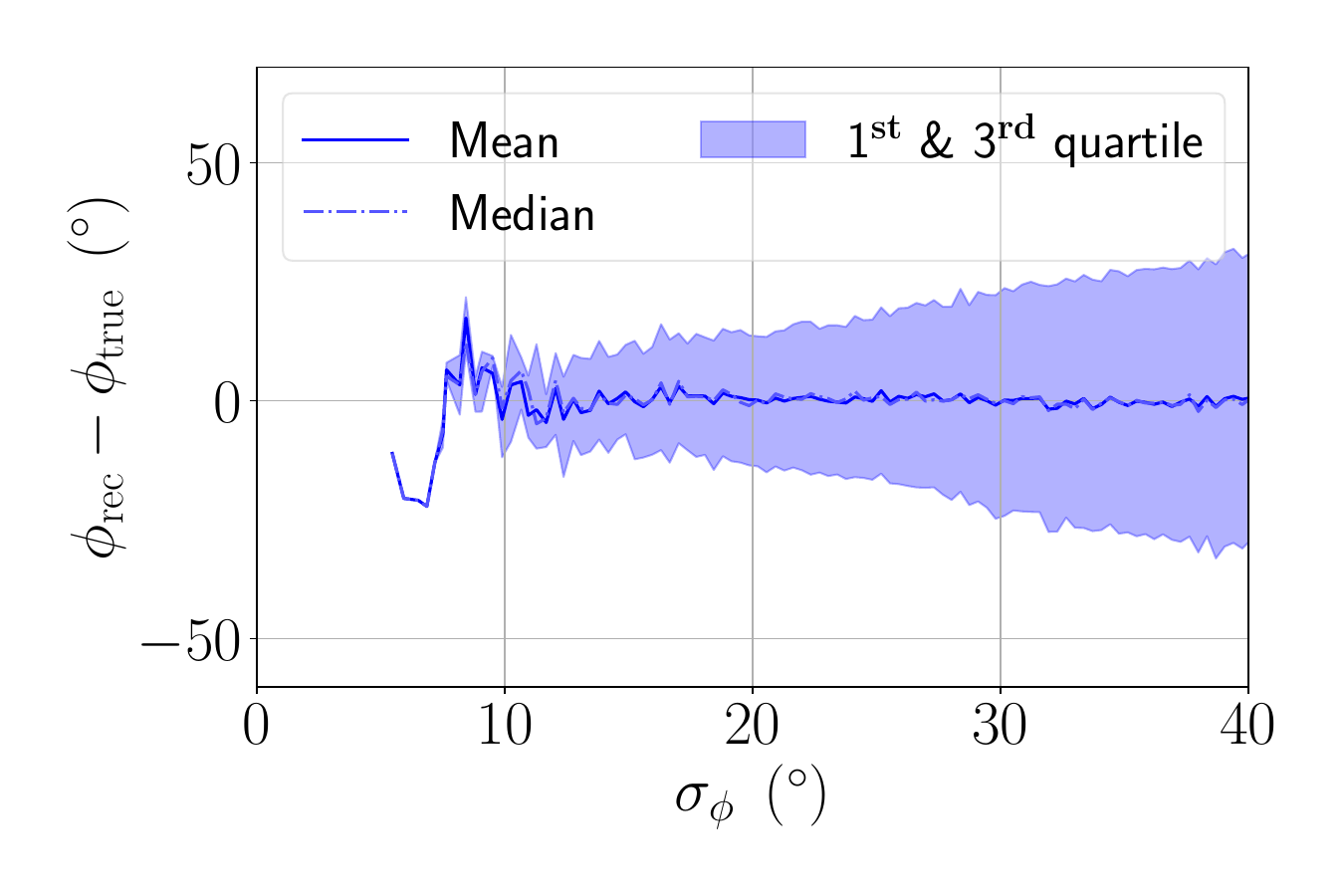}
	\caption{\label{fig:sigma}The error on angles $\theta$ (left) and $\phi$ (right) as a function of the predicted uncertainty for the track branch. The mean and median error stay close to zero with no significant bias. The first and third quartiles behave as expected for a Gaussian distribution, especially for $\theta$.}
\end{figure}

Note that the uncertainty estimation assumes a Gaussian distribution. This means that the distribution of errors, although centred, should become wider with growing $\sigma$ values. For most of the error ranges this is precisely the case, especially for $\theta$ (\autoref{fig:sigma} and \autoref{fig:sigma_sh}). As for the uncertainty of the direction reconstruction ($\sigma_{\Omega}$),  we used the following expression derived from the solid angle definition:
\begin{equation}
\label{eq:sigma_omega}
\sigma_\Omega = \sqrt{\sin^2(\theta)\sigma_\phi^2 + \sigma_\theta^2} \,.
\end{equation}

\subsubsection{Closest point and interaction vertex}
\label{subsubsec:closest}

In addition to the reconstruction of direction, estimating the energy of neutrino events is fundamental for physics analyses. Energy and distance to the detector are, however, intermingled in SL events: distant high-energy events may appear as SL as well as near low-energy events. Then, to improve the energy estimation, we included in $N$-fit the reconstruction of the \emph{closest point} of track events to the ANTARES detector line, and the \emph{interaction vertex} position of shower events, since these two magnitudes are characterized by the horizontal distance in meters ($R_c$ for closest point of tracks, $R_v$ for interaction vertex of showers) and their vertical position ($Z_c$ for tracks, $Z_v$ for showers) defined in the ANTARES frame of reference.

\begin{figure}[htbp]
	\centering
	\includegraphics[scale=0.73]{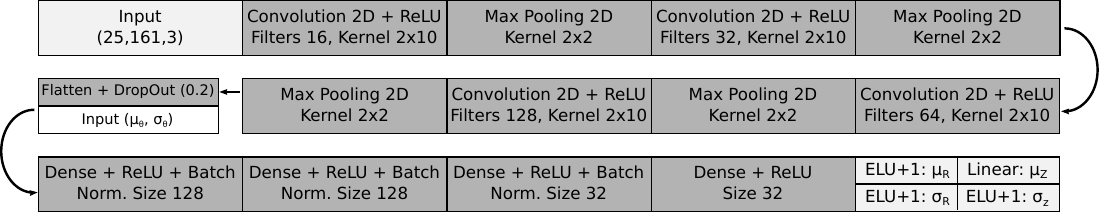}
	\caption{\label{fig:network_RZ}Details of the closest point and interaction vertex network architecture.}
\end{figure}

Independent networks (\autoref{fig:network_RZ}) were used for the horizontal distance ($R$) and vertical coordinate ($Z$). Their architecture was that of the direction reconstruction without further tuning. Differences lay in input and output of the networks to fulfil the characteristics of these reconstructions. Thus, events were not centred across images in these network models to ease the reconstruction of $Z$. In addition, these networks also incorporated the reconstruction of $\theta$ as input, which was introduced in parallel to the convolutional layers. Lastly, the loss function was adapted in consonance with the outputs of these networks: 
\begin{equation}
\label{eq:loss_RZ}
\begin{split}
\mathcal{L} = &\ln (\sqrt{2\pi}\sigma_{R}) + \frac{1}{2}\frac{(R_{t}-\mu_{R} )^2}{\sigma_{R}^2}+\\
+ &\ln (\sqrt{2\pi}\sigma_{Z}) + \frac{1}{2}\frac{(Z_{t}-\mu_{Z} )^2}{\sigma_{Z}^2} \,.
\end{split}
\end{equation}

\subsubsection{Energy}
\label{subsubsec:energy}

As described earlier, estimating the energy is particularly challenging for SL events, especially for track events, given their physical characteristics and the limited information available by a single line of the detector. Shower events are better suited for energy reconstruction because of their physical topology. Moreover, the neutrino energy is directly inferred by $N$-fit in shower events, whereas for track events, the reconstructed energy by $N$-fit is limited to that of the secondary muon. This is due to the stochastic energy loss in neutrino interactions producing track events. The neutrino energy of track events for physics analyses can, however, be inferred indirectly, by combining $N$-fit energy reconstruction and the statistical properties of the interactions.

As a first approximation, we applied the same model architecture of the direction reconstruction to infer energy without further tuning. The energy reconstructions by such model presented very low accuracy. Reasons underlying the poor performance include that the secondary muon can escape the detector in track events, or that events could be very close or far away from the detector line in shower events. To better guide training in the $N$-fit energy reconstruction, we preselected events that had good direction reconstruction (i.e., 50\% of the events with the lowest predicted $\sigma_\theta$), and that were close to the line (according to the $\{R,Z\}$ reconstruction). The specific cuts appear in expression (\ref{eq:cuts_tr}) for tracks, and in expression (\ref{eq:cuts_sh}) for showers.

\begin{equation}
\label{eq:cuts_tr}
\begin{gathered}
	R_c \leq 50\,m~~~~Z_c \in (-150, 150)\,m~~~~\sigma_\theta\leq8.6^\circ\\
	\sigma_{R_c} \leq 4.3\,m~~~~\sigma_{Z_c} \leq 6.2\,m \,,
\end{gathered}
\end{equation}

\begin{equation}
\label{eq:cuts_sh}
\begin{gathered}
R_v \leq 50\,m~~~~Z_v \in (-150, 150)\,m~~~~\sigma_\theta\leq 14.9^\circ\\
\sigma_{R_v} \leq 5.6\,m~~~~\sigma_{Z_v} \leq 3.3\,m \,.
\end{gathered}
\end{equation}

After training with these cuts, energy inference slightly improved. We considered this model as the \textit{benchmark} to evaluate further improvements, particularly that from indirect transfer learning through PCA-based knowledge distillation (see \autoref{subsubsec:TL}). In such an approach, we aimed to exploit the relationship between the energy of the neutrino and all other physical parameters of the event processed by $N$-fit direction and distance models. We assumed that internal representations from those models could then benefit energy predictions. To test the hypothesis, neuron activations in hidden layers from the $\theta$ and $\{R,Z\}$ networks were taken as feature dimensions from which to infer the energy of the events. We disregarded, however, activations from the $\phi$ network for two main reasons: first due to the symmetry in $\{X,Y\}$, and second because of the low accuracy of $\phi$ estimations in SL events, compared to $\theta$ and $\{R,Z\}$ estimations.

\begin{figure}[htbp]
	\centering
	\includegraphics[scale=0.73]{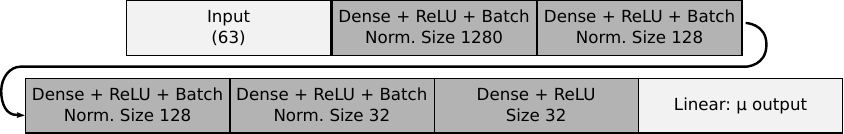}
	\caption{\label{fig:network_e}Details of the energy network architecture.}
\end{figure}

We linearly transformed all feature dimensions, using the PCA, to rank components according to their variability. We then selected those most relevant features as inputs to a FFN trained to infer the energy as point predictions (\autoref{fig:network_e}). The final number of components, 63 features for the track branch and 68 for the shower branch, was determined applying the elbow rule, i.e., stop incorporating more features once the cumulated variance reached a \textit{plateau}. Concurrently, we realized that inferring $\log (E)$ performed better than estimating $E$ directly, due to the large energy range of neutrino events (from 5 GeV up to 20 TeV).

In a subsequent phase aimed to further improve energy predictions for physics analyses, we designed a specialized FFN by feeding the model during training with only the 50\% best reconstructed events from the original model shown in \autoref{fig:network_e}. In addition, this specialized FFN incorporated a MDN output layer to provide uncertainty estimation of energy predictions, as well as loss weighting factors to balance the impact of the non-uniform energy distribution. 

\subsubsection{Classification}
\label{subsubsec:class}

$N$-fit includes a neural network for track vs. shower classification, which leverages TL from the specialized track and shower branches. The output is a track classifier score ($S$), with shower-likeness defined as $1-S$. To train, validate and test the classifier, we used a combined dataset made of 200,000 MC events, half of them representing track events ($\numuCC$) and the other half representing shower events ($\numuNC$, $\nueNC$, $\nueCC$). This new dataset, as well as the previous used in the training phase, are selected according to the $\chi^2$-fit SL criteria with no further refinement. All events were randomly sampled in a uniform manner covering the full period in which ANTARES collected data.

\begin{figure}[htbp]
	\centering 
	\includegraphics[scale=0.73]{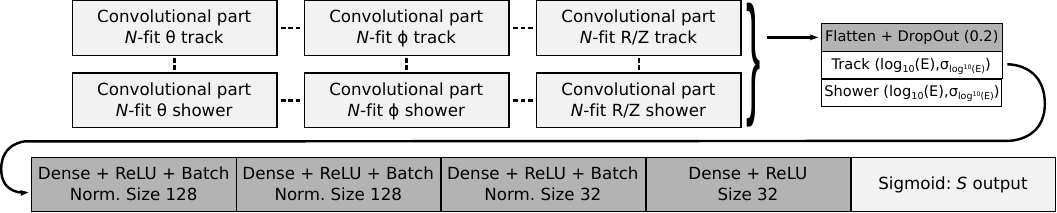}
	\caption{\label{fig:network_class}Details of the classifier network architecture. All convolutional blocks are connected in parallel to the feed-forward part.}
\end{figure}

The \emph{control} network model for classification consisted of the same model architecture of the direction reconstruction without further tuning. Only the output was adjusted: a single neuron represented the track classifier score, using the sigmoid as its activation function. Given that track and shower events typically present different trace characteristics and that these traits were presumably present as internal representations in the convolutional layers of $N$-fit direction and distance reconstruction models, we decided to exploit this knowledge in a second, more sophisticated classifier endowed with TL. Specifically, we incorporated the convolutional blocks of the $\theta$, $\phi$ and $\{R,Z\}$ network models from both $N$-fit branches as frozen components that were plugged in parallel into a FFN. This network model also incorporated the reconstruction of the energy and its uncertainty from the networks specialized in track and shower events. The complete architecture is shown in \autoref{fig:network_class}.

\subsubsection{Code implementation and computational efficiency}
\label{subsubsec:code}

All neural network models were implemented using the TensorFlow 2.4.1 framework \cite{TensorFlow}, with model evaluation and analysis conducted in Python 3.9.15. The implementation relied on standard scientific computing libraries, including \texttt{NumPy}~1.23.4, \texttt{Scikit-learn}~1.1.3, and \texttt{Pandas}~1.5.1, along with several auxiliary packages.

Training times varied depending on the network architecture. Specifically, direction and position networks required approximately 12 hours each to train, leveraging GPU acceleration via TensorFlow. The PCA, applied separately to each branch (track and shower), required roughly 15 minutes per branch. The subsequent training of the energy reconstruction networks took between 5 and 15 minutes, being much faster than the spatial models primarily due to their lack of convolutional layers, which are typically the most computationally demanding components. Training the track vs. shower classifier, which reuses frozen convolutional layers transferred from the specialized branches, was comparatively lightweight, requiring only about 20 minutes. GPU-accelerated training was conducted on a workstation running KDE Neon 24.04, equipped with two NVIDIA Quadro RTX 8000 GPUs (TU102GL architecture, 64-bit interface), each of them having a 48 GB GDDR6 memory and supporting 672 GB/s memory bandwidth. 

The final application of the $N$-fit algorithm to ANTARES data was organized into three main phases: (i) reading the ANTARES data files, (ii) processing the raw data into the standardized $N$-fit input format, and (iii) performing the actual reconstruction. The first phase is common to any reconstruction pipeline and does not affect the evaluation of $N$-fit's computational performance. The image generation stage, where detector hits are transformed into 2D representations for network input, took an average of 0.2 seconds per event. The complete reconstruction process required approximately 0.1 seconds per event. These estimates are based on the averaged processing times over 10,000 events, using conservative rounding to ensure upper-bound accuracy. In both cases, model parameters and dependencies were preloaded into memory cache, eliminating the overhead of repeated I/O operations during batch processing.

The whole application of $N$-fit to the ANTARES data was executed on the high-throughput computing (HTC) partition of the IN2P3 Computing Center, managed via the SLURM workload manager. The HTC partition consists of heterogeneous CPU-only nodes running Red Hat Enterprise Linux release 9.6 (64-bit). Most nodes are equipped with AMD EPYC 7302 16-core, EPYC 7453 28-core, or EPYC 9334 32-core processors, with memory per node ranging from 192~GB to over 1.2~TB. For our runs, a single CPU core and less than 2~GB of RAM were sufficient for efficient reconstruction inference. GPU resources were not required for deployment, although training leveraged TensorFlow's GPU acceleration when available.

\section{Results}
\label{sec:results}

This section presents the main performance results of the $N$-fit algorithm. All evaluations are performed on the corresponding test subsets associated with each reconstruction branch, as defined in \autoref{subsec:data}, except for the classifier performance, which is tested with its own dataset. Finally, additional validation studies -- addressing the algorithm robustness with respect to training data variations, resilience to noise, and applicability to real ANTARES data -- are briefly presented.


\subsection{Direction reconstruction improvement}
\label{subsec:track}

\begin{figure}[htbp]
    \centering
    
    \begin{subfigure}{\textwidth}
        \centering
        \subcaption{\label{fig:err_dist_A}} 

        \makebox[0.5cm][c]{\raisebox{2.3cm}{\small\textbf{(1)}}}~\includegraphics[width=.45\textwidth]{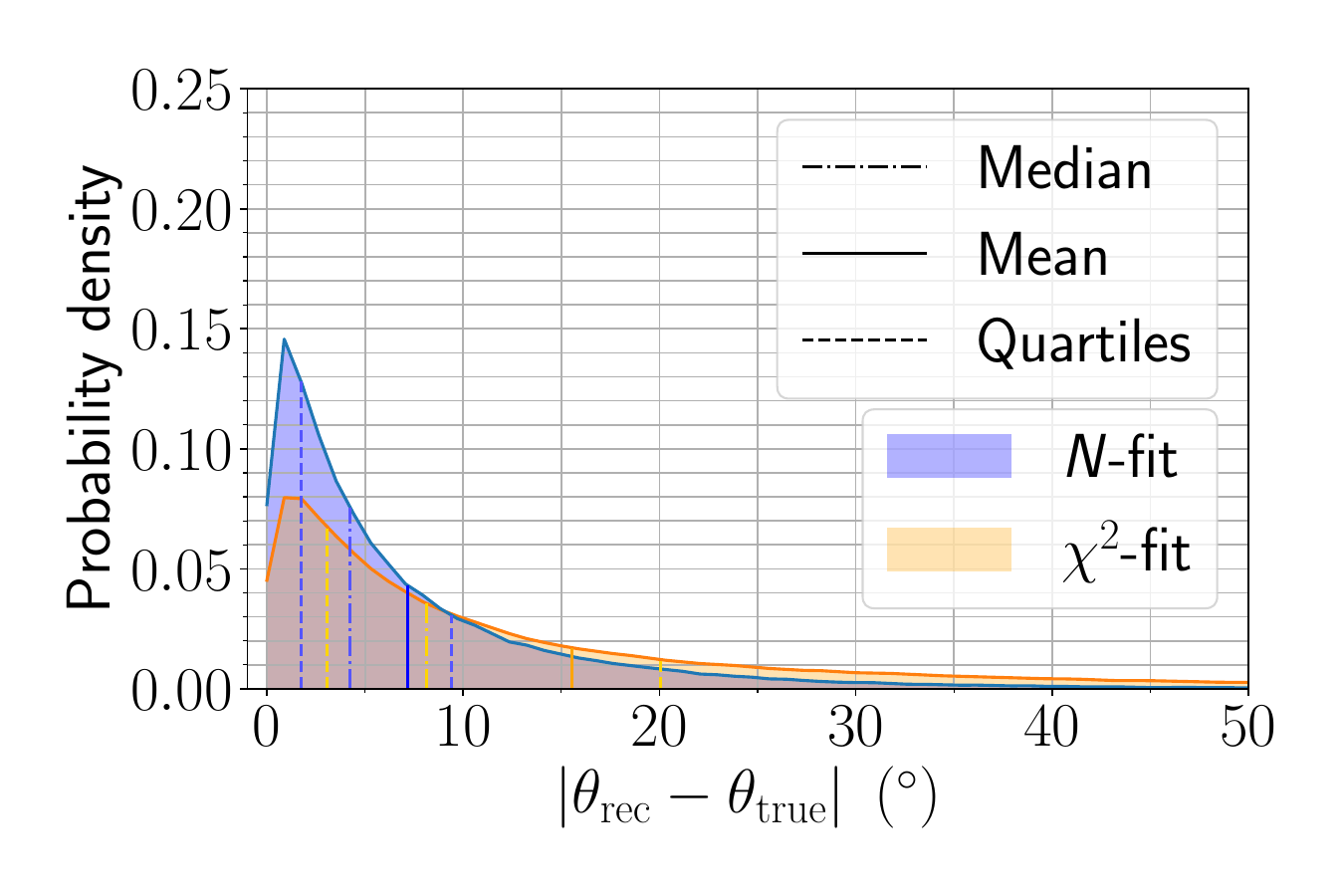}\hfill
        \includegraphics[width=.45\textwidth]{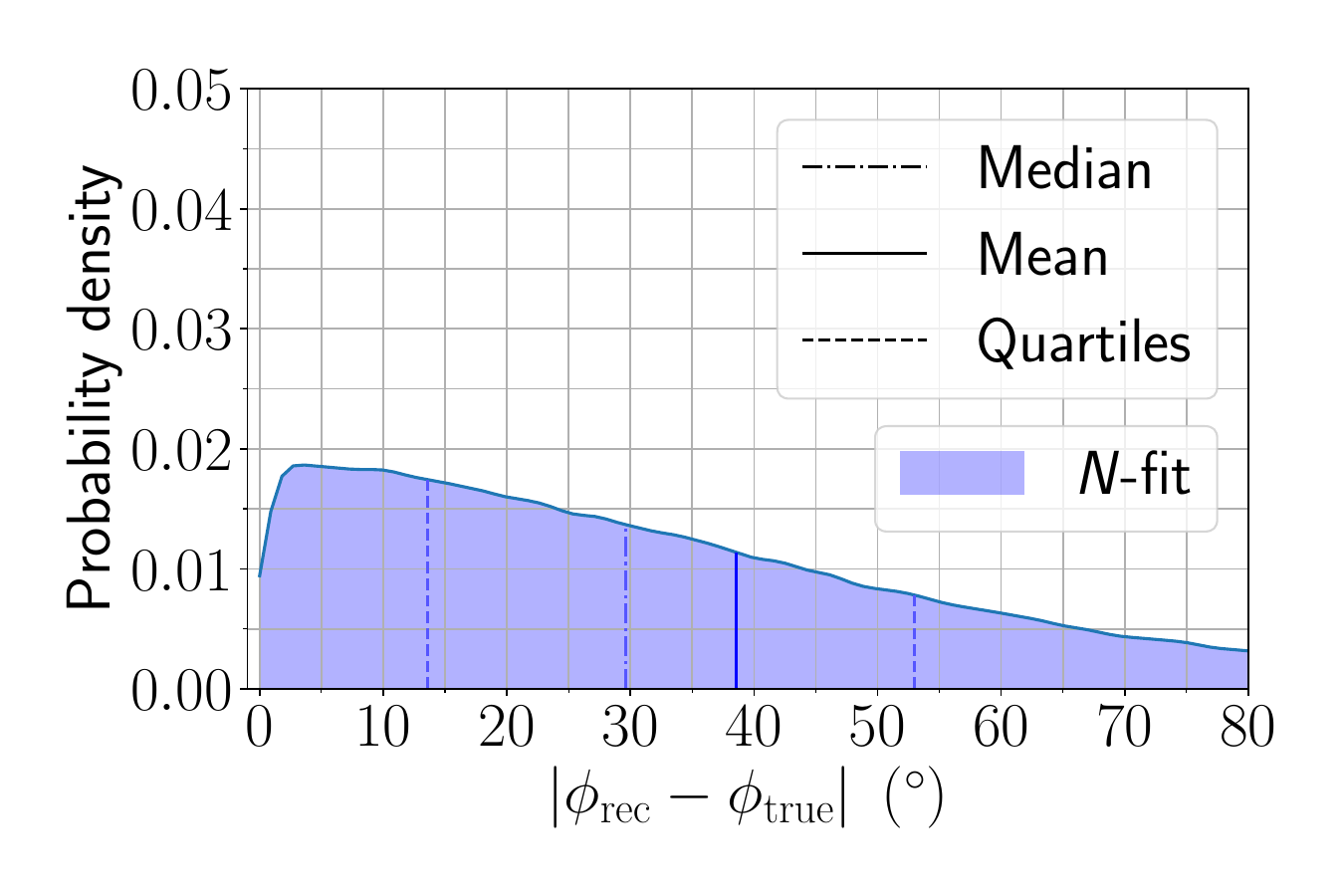}

        \makebox[0.5cm][c]{\raisebox{1.7cm}{\small\textbf{(2)}}}\hfill
        \includegraphics[width=.32\textwidth]{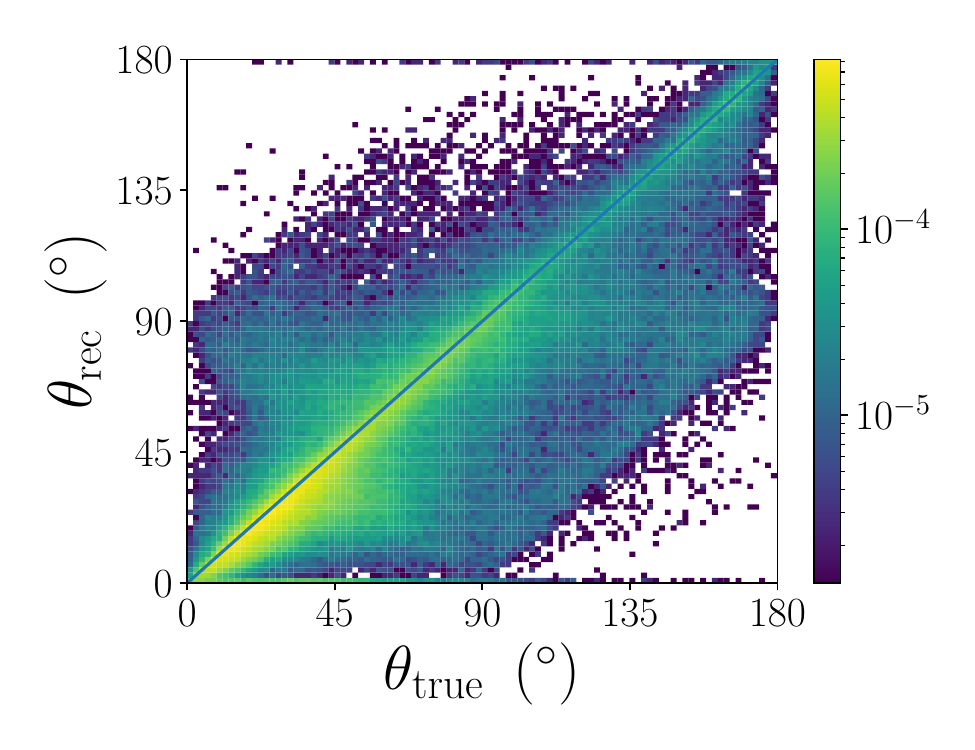}\hfill
        \vline\hfill
        \includegraphics[width=.32\textwidth]{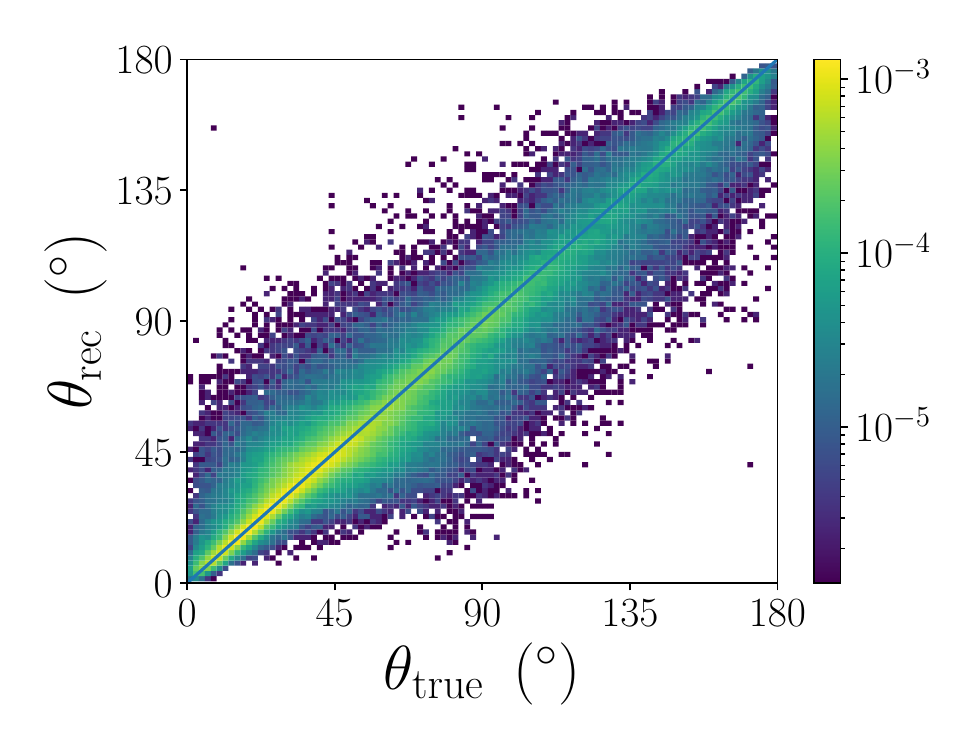}\hfill
        \includegraphics[width=.32\textwidth]{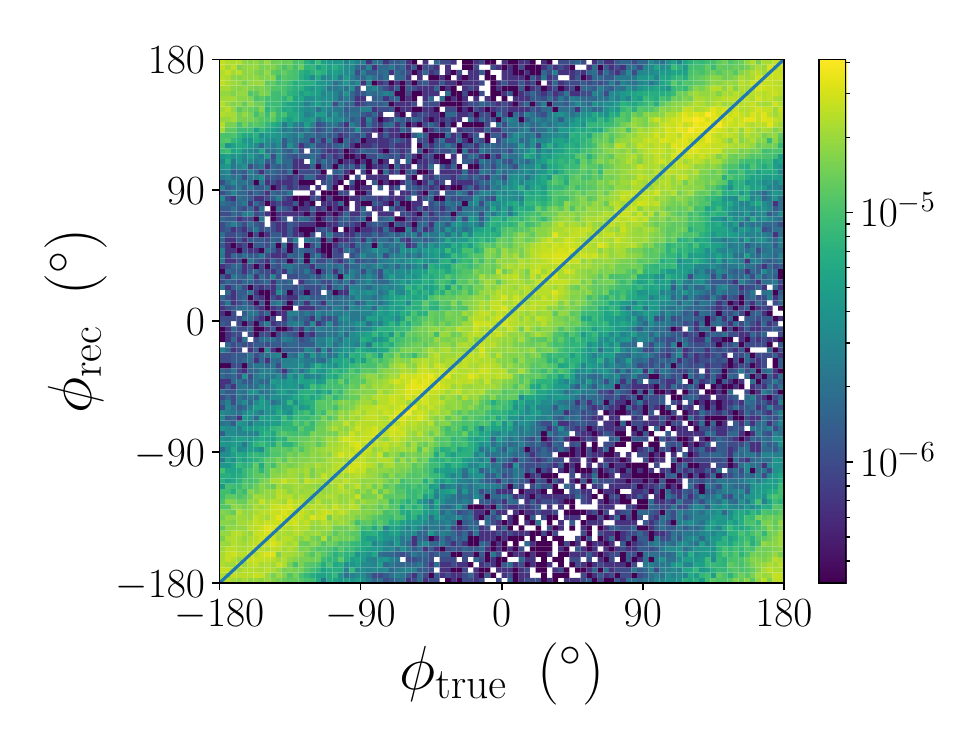}
    \end{subfigure}

    \vskip 4pt
    \hrule
    \vskip 4pt

    \begin{subfigure}{\textwidth}
        \centering
        \subcaption{\label{fig:err_dist_B}}
        
        \makebox[0.5cm][c]{\raisebox{2.3cm}{\small\textbf{(1)}}}~\includegraphics[width=.45\textwidth]{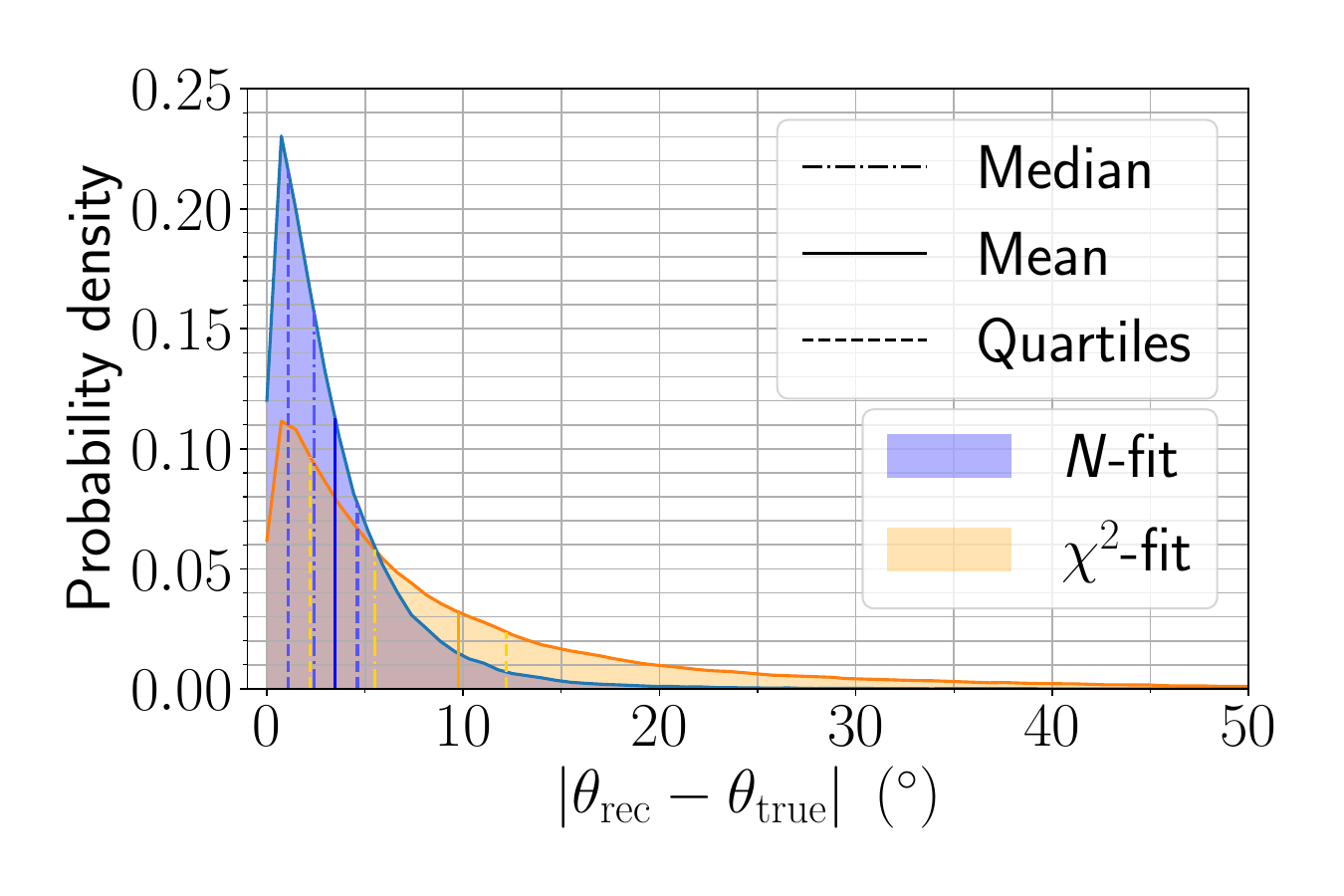}\hfill
        \includegraphics[width=.45\textwidth]{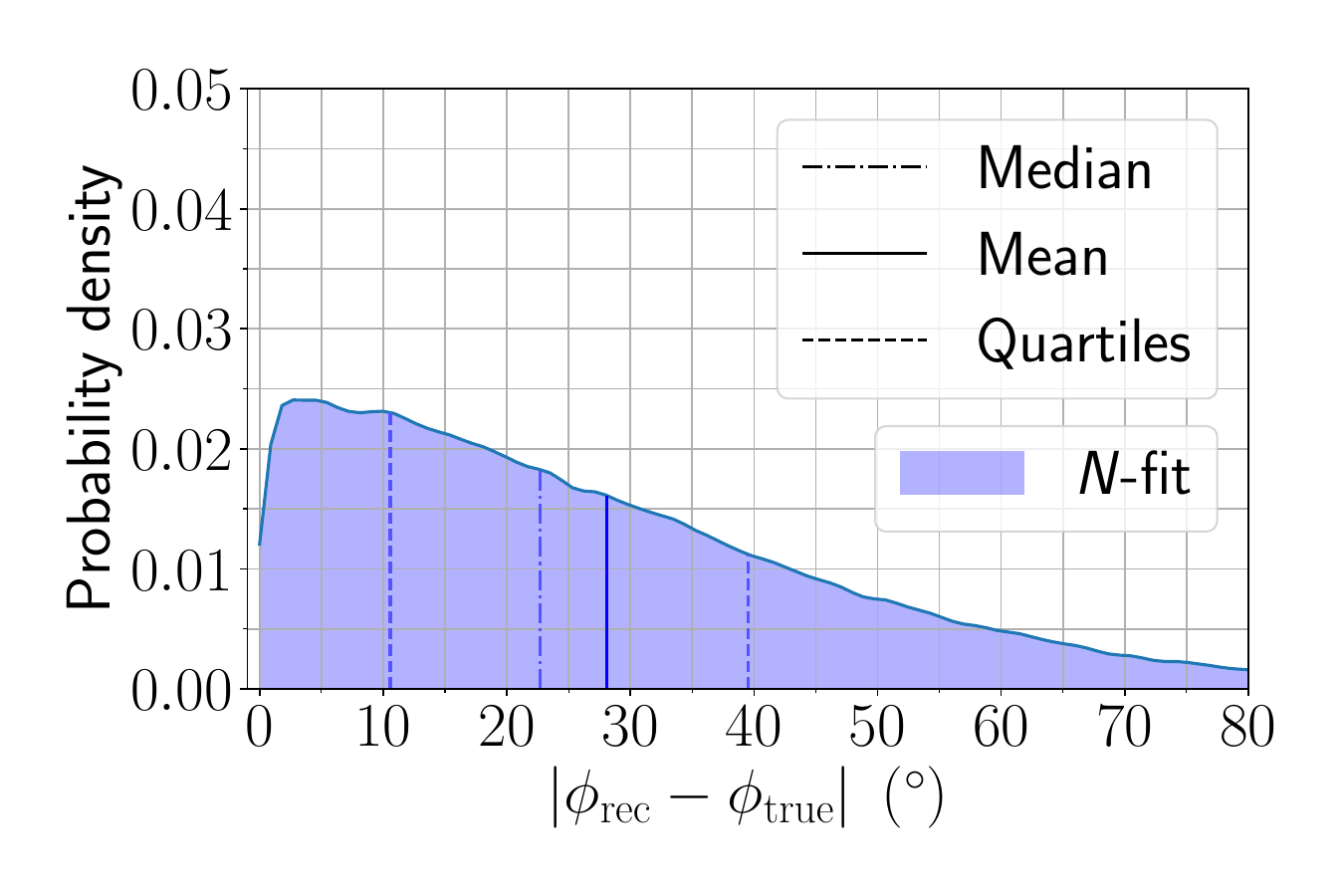}

        \makebox[0.5cm][c]{\raisebox{1.7cm}{\small\textbf{(2)}}}\hfill
        \includegraphics[width=.32\textwidth]{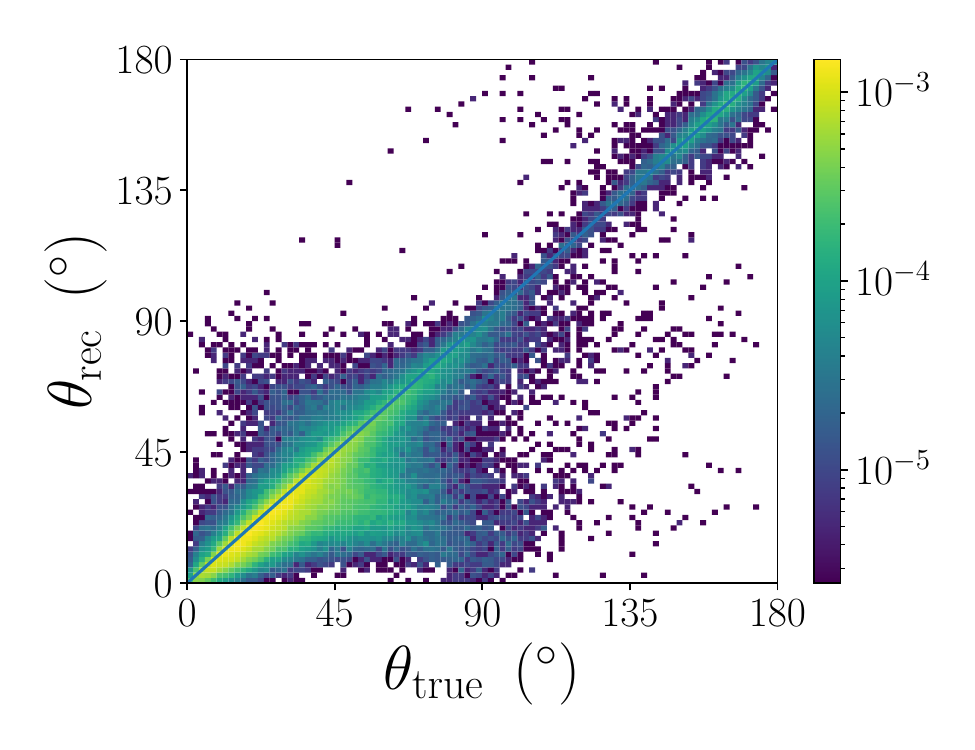}\hfill
        \vline\hfill
        \includegraphics[width=.32\textwidth]{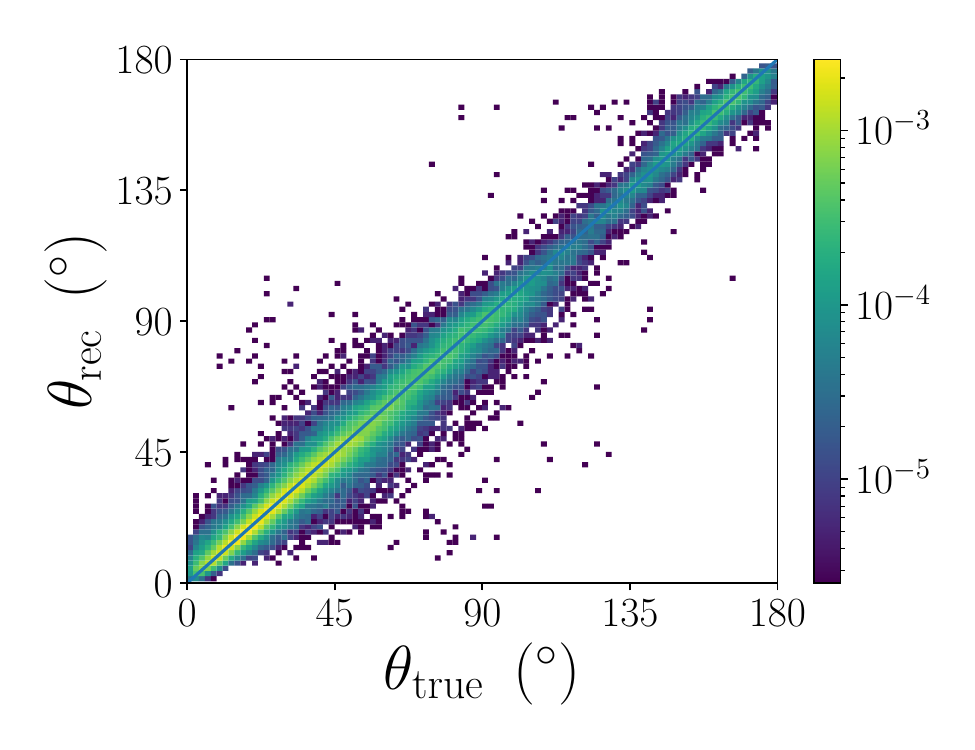}\hfill
        \includegraphics[width=.32\textwidth]{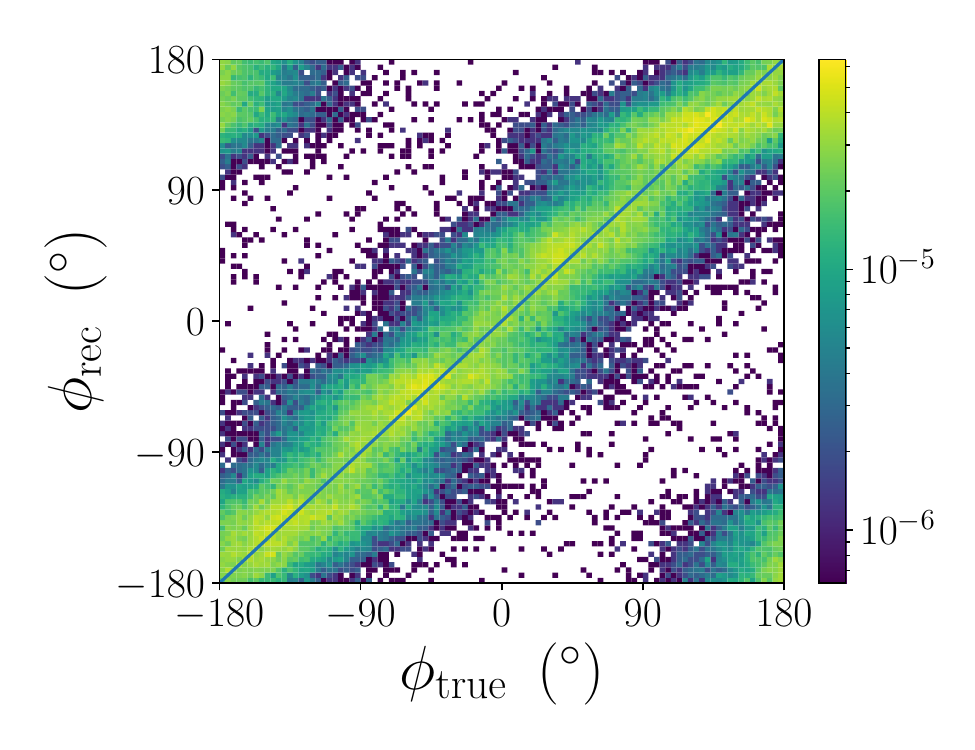}
    \end{subfigure}

    \caption{\label{fig:err_dist}Track branch results. \textbf{(a)} Whole test dataset. \textbf{(b)} 50\% of the test dataset with lowest values of $\sigma$ or $\chi^2$, respectively. \textbf{(1)}: Absolute error distributions of angles $\theta$ and $\phi$. \textbf{(2)}: 2D density histograms of true vs. reconstructed angles for $\chi^2$-fit (left) and $N$-fit (right). Note that $\phi$ points at the off-diagonal corners represent in fact good predictions due to periodicity.}
\end{figure}

For the track branch of $N$-fit, statistics of the error distribution for the reconstructed angles can be seen in \autoref{tab:results}. Errors are computed as the absolute values of the difference between the reconstructed ($\mu_\alpha$) and the simulated ($\alpha_t$) angles for the $\numuCC$ MC simulations. The error distributions are shown in \autoref{fig:err_dist_A}. The $N$-fit algorithm clearly outperforms the $\chi^2$-fit SL reconstruction method: results show a significantly larger proportion of low errors for the $\theta$ angle and a first estimation of the $\phi$ angle, which was previously missing for SL events. Remarkably, our approach allows to focus on best predictions by selecting the events with the lowest values of $\sigma$ provided by $N$-fit. Applying a selection criterion on $\sigma$ improves results  (see \autoref{tab:results} and \autoref{fig:err_dist_B}). These plots for the shower branch --where $\numuNC$, $\nueNC$ and $\nueCC$ MC simulations are employed-- are shown in the supplementary material, \autoref{fig:err_dist_sh}, as well as the statistics of the error for the reconstructed angles, \autoref{tab:results_sh}.

%
%

\begin{table}[htbp]
    \centering
    \renewcommand{\arraystretch}{1.2} 
    \setlength{\tabcolsep}{5pt} 

    \begin{tabular}{>{\centering\arraybackslash}m{2.7cm}  >{\centering\arraybackslash}m{0.9cm} >{\centering\arraybackslash}m{0.9cm}  >{\centering\arraybackslash}m{0.9cm} >{\centering\arraybackslash}m{0.9cm}  >{\centering\arraybackslash}m{0.9cm} >{\centering\arraybackslash}m{0.9cm}  >{\centering\arraybackslash}m{0.9cm} >{\centering\arraybackslash}m{0.9cm}}  
        \hline
        Mean/Median 
        & \multicolumn{2}{c}{$\chi^2$-fit} 
        & \multicolumn{2}{c}{$N$-fit} 
        & \multicolumn{2}{c}{$\chi^2$-fit (50\%)} 
        & \multicolumn{2}{c}{$N$-fit (50\%)} \\  
        \hline \hline
        $\theta$ & 15.5$^\circ$ & 8.5$^\circ$  & 7.4$^\circ$  & 4.4$^\circ$ & 9.7$^\circ$  & 5.5$^\circ$ & 3.7$^\circ$  & 2.5$^\circ$  \\
        $\phi$    & - & - & 41.4$^\circ$ & 31.5$^\circ$ & - & - & 29.2$^\circ$ & 23.6$^\circ$ \\ 
        $\Omega$  & - & - & 28.3$^\circ$ & 22.7$^\circ$ & - & - & 18.7$^\circ$ & 13.7$^\circ$ \\ 
        \hline
    \end{tabular}
    
    \caption{\label{tab:results}Mean and median absolute error of reconstructed neutrino angles $\theta$, $\phi$, and total deviation, $\Omega$, for the track branch. The 50\% best events are selected according to the lowest values of the parameters $\sigma$ for $N$-fit and $\chi^2$ for the $\chi^2$-fit.}
\end{table}

%
%

\begin{figure}[htbp]
    \centering
    
    \begin{subfigure}{\textwidth}
        \centering
        \subcaption{\label{fig:tot_err_dist_A}} 

        \makebox[0pt][c]{\raisebox{2.3cm}{\textbf{(1)}} \quad} 
        \includegraphics[width=.48\textwidth]{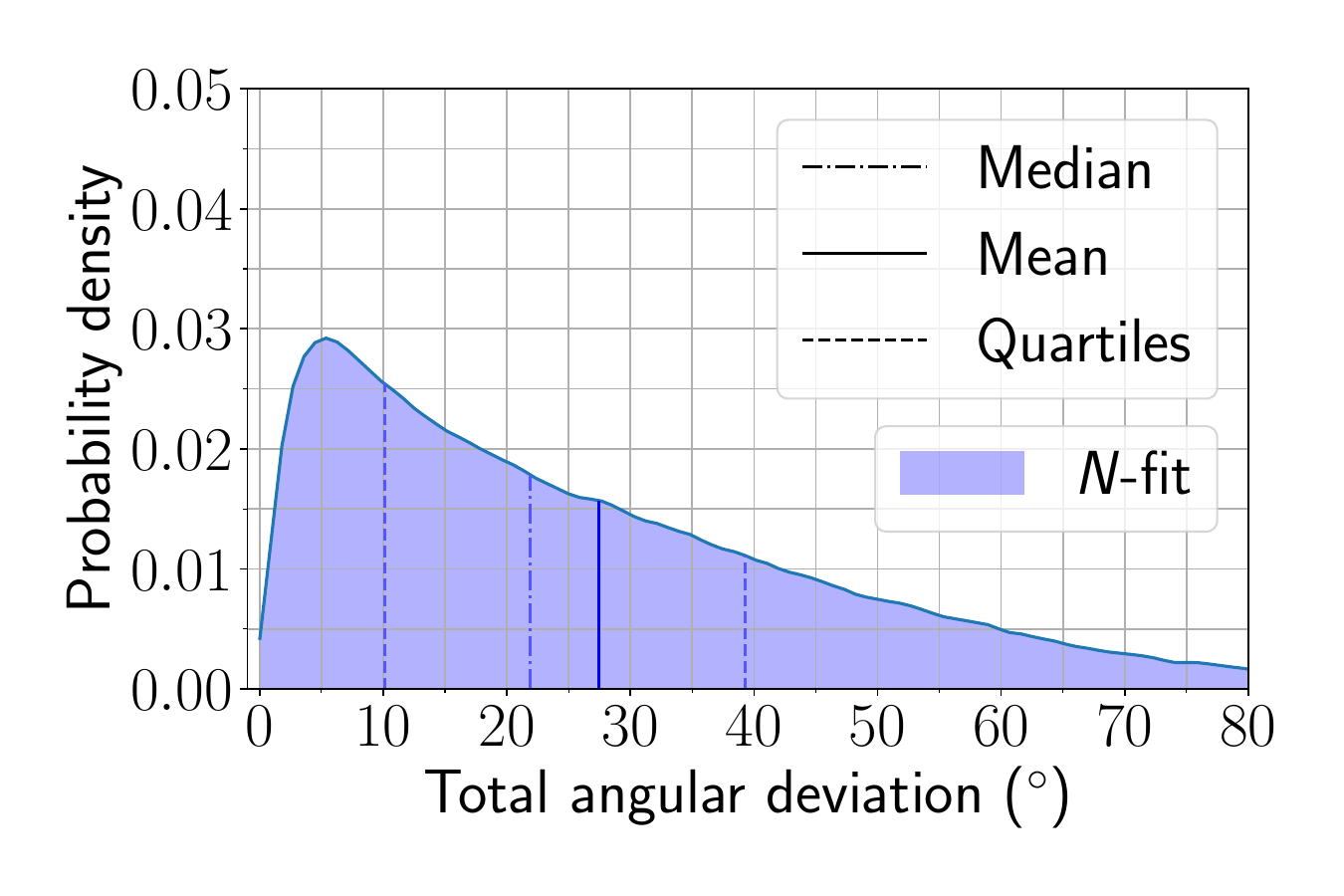}
	    \includegraphics[width=.48\textwidth]{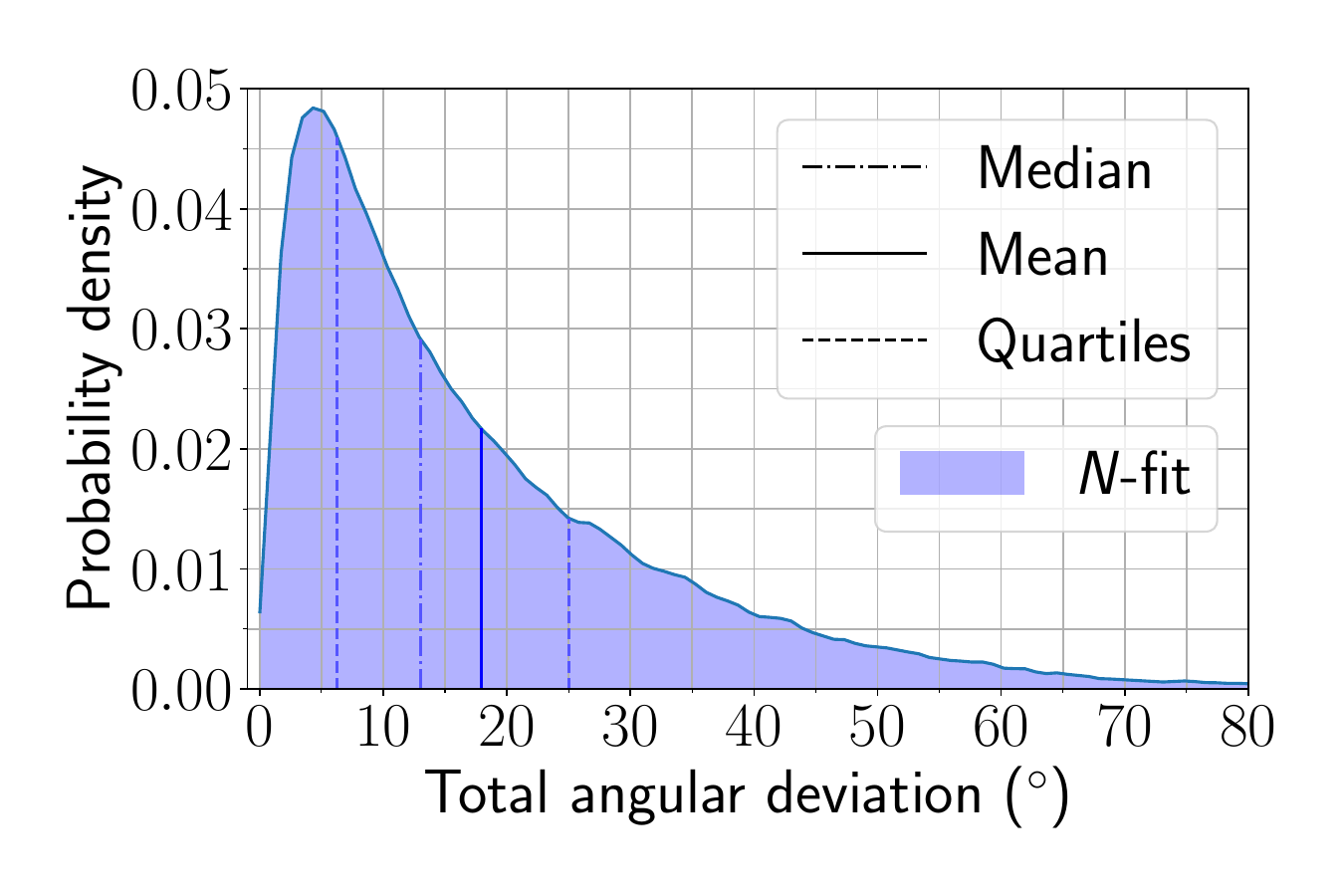}

        \makebox[0pt][c]{\raisebox{2.3cm}{\textbf{(2)}} \quad} 
        \includegraphics[width=.48\textwidth]{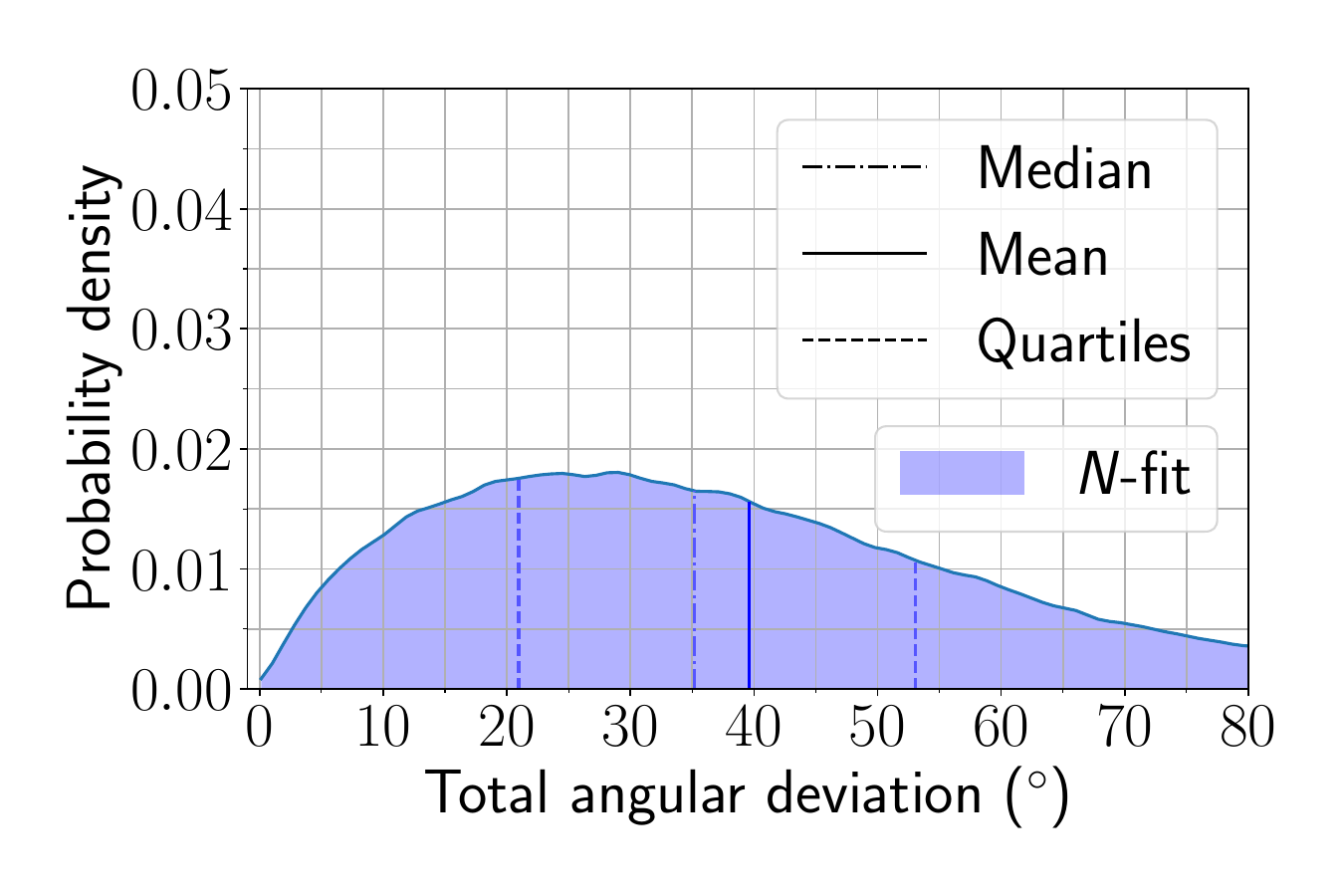}
	    \includegraphics[width=.48\textwidth]{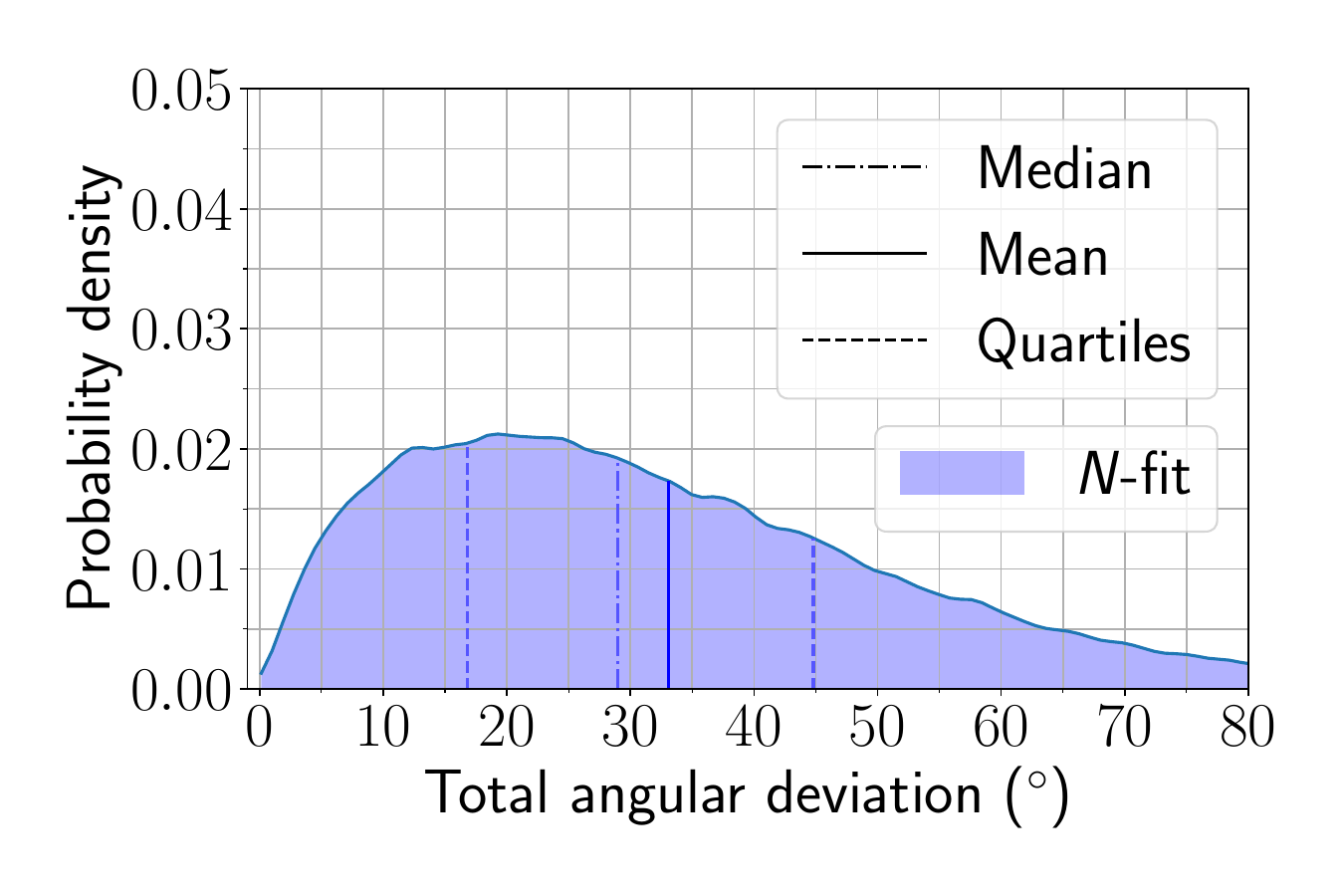}

    \end{subfigure}

    \vskip 5pt
    \hrule
    \vskip 5pt

    \begin{subfigure}{\textwidth}
        \centering
        \subcaption{\label{fig:tot_err_dist_B}}
        
        \includegraphics[width=.48\textwidth]{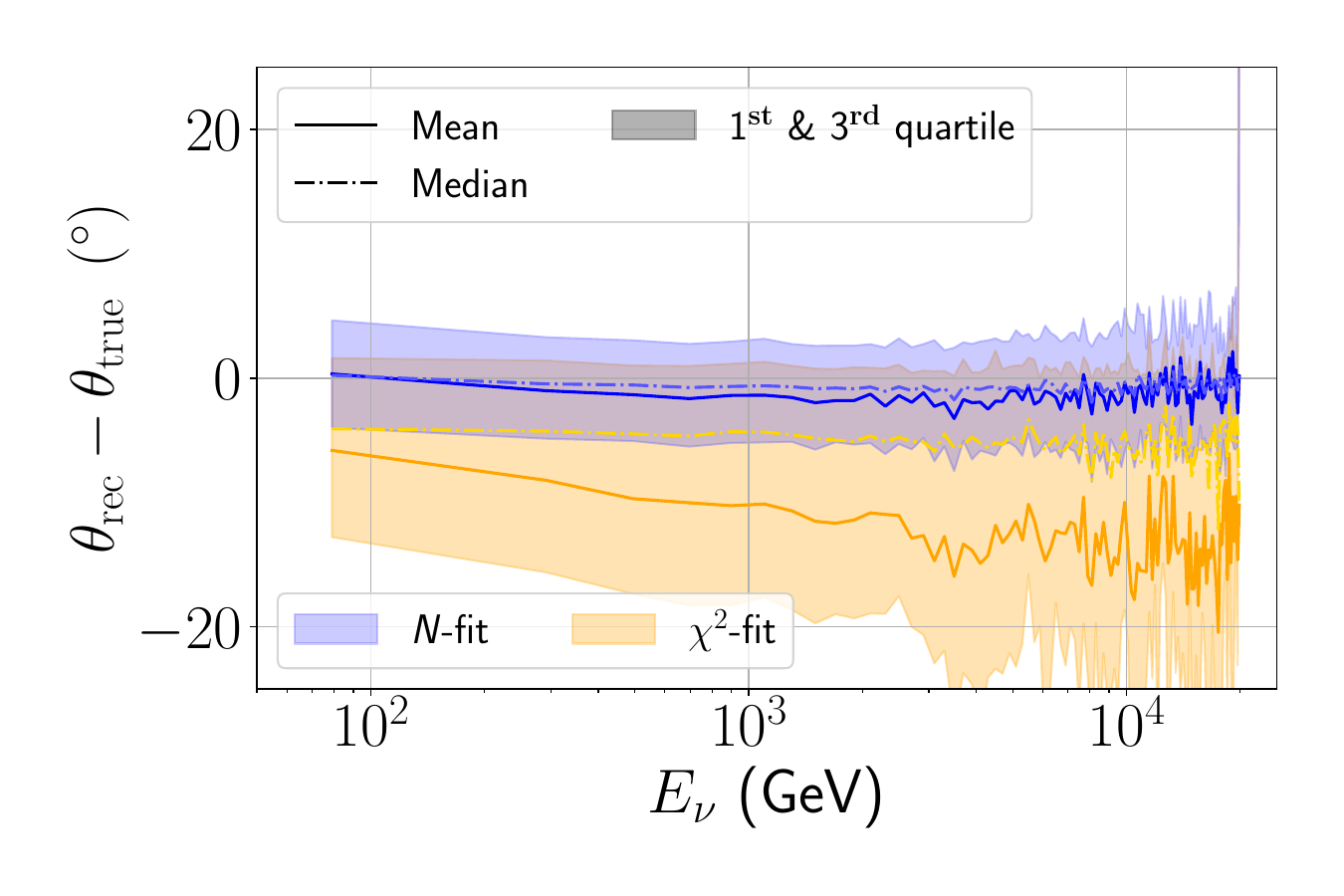}
        \includegraphics[width=.48\textwidth]{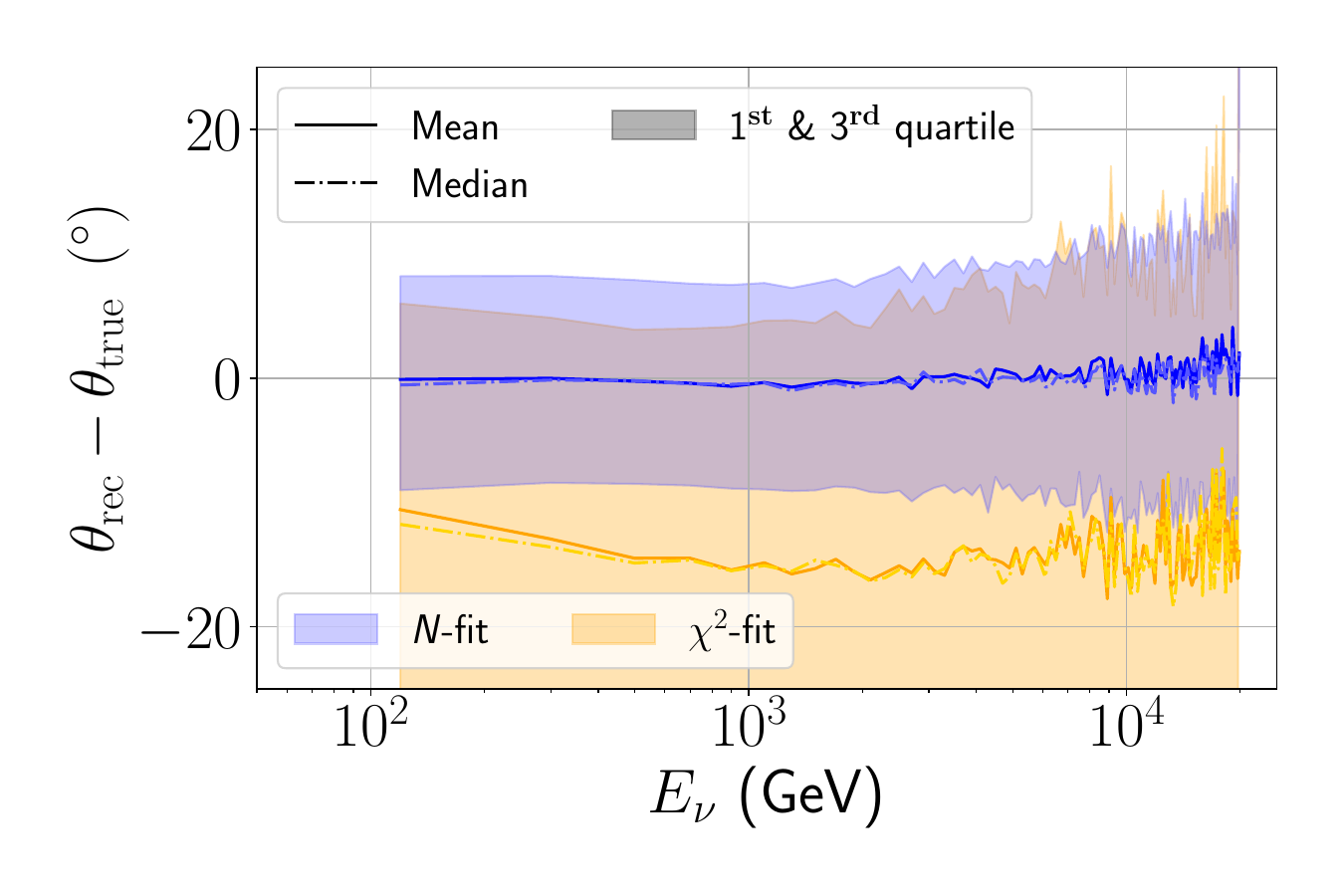}

    \end{subfigure}

    \caption{\label{fig:tot_err_dist}\textbf{(a)} Total angular deviation distribution for tracks (1) and showers (2) of all test dataset events (left) and of the 50\% events with lowest value of $\sigma_{\Omega}$ (right). \textbf{(b)} Angle $\theta$ error as a function of the true neutrino energy for tracks (left) and showers (right).}
\end{figure}

By using expression (\ref{eq:sigma_omega}), we can also apply the selection criterion to the total angular deviation for both track and shower branches (\autoref{fig:tot_err_dist_A}). The direction reconstruction for showers is worse than for tracks due to the physical geometry of the shower signature. It is easier to determine the direction of a source emitting light following a straight path than from an almost point-like source, typical of showers. This issue, however, does not prevent the use of $N$-fit and the estimated uncertainty can still be used to choose the best reconstructed events.

Finally, we checked that $N$-fit direction reconstructions show no bias for the whole MC simulated energy range of neutrinos, in contrast to the $\chi^2$-fit (\autoref{fig:tot_err_dist_B}). This is particularly relevant for using this method in physics analyses that involve low energy events.
 

%
%

\subsection{Transfer learning applications}

\subsubsection{Energy reconstruction}

\begin{figure}[htbp]
    \centering
    
    \begin{subfigure}{\textwidth}
        \centering
        \subcaption{\label{fig:err_dist_RZ_A}} 

        \includegraphics[width=.32\textwidth]{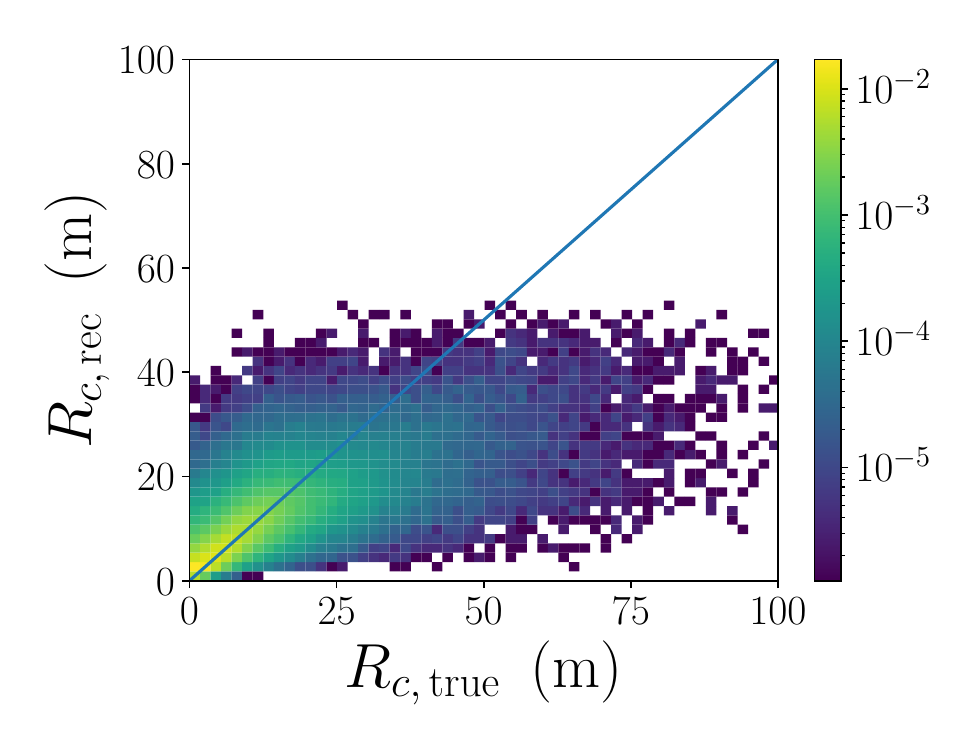}  
        \includegraphics[width=.32\textwidth]{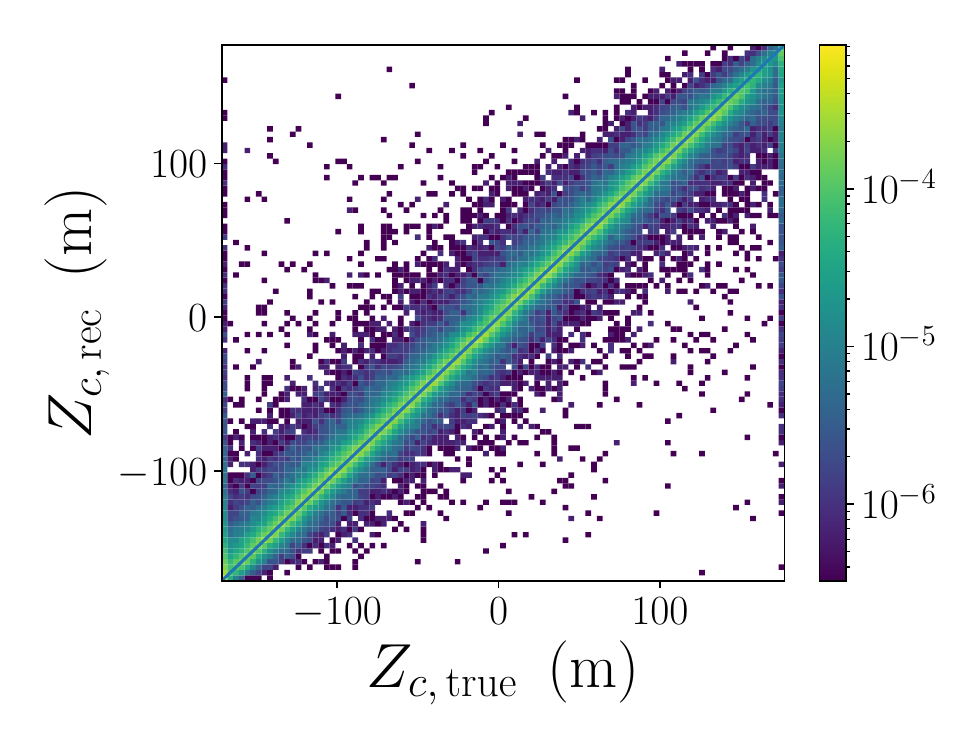}
    
    \end{subfigure}

    \vskip 5pt
    \noindent\rule{0.7\textwidth}{0.4pt}    
    \vskip 5pt

    \begin{subfigure}{\textwidth}
        \centering
        \subcaption{\label{fig:err_dist_RZ_B}}
        
        \includegraphics[width=.32\textwidth]{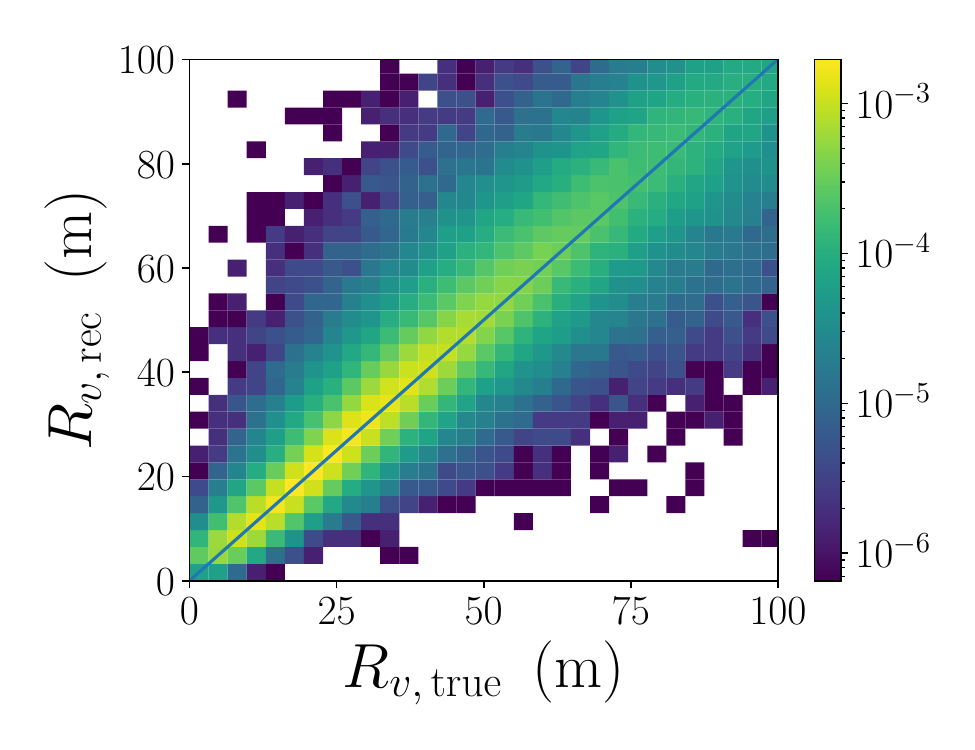}
    	\includegraphics[width=.32\textwidth]{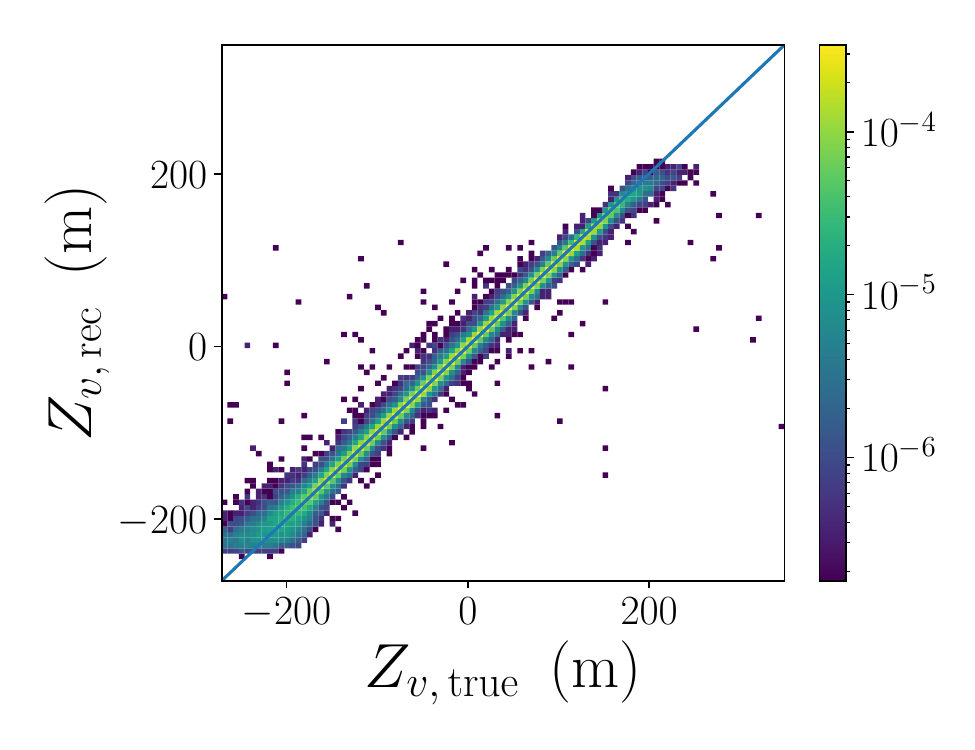}
    
    \end{subfigure}

    \caption{\label{fig:2D_dist_RZ}2D density histograms of reconstructed vs. true horizontal distance ($R$) and vertical position ($Z$) of the closest point for tracks (a) and of the interaction vertex for showers (b).}
\end{figure}

As mentioned in \autoref{subsubsec:energy}, the energy was difficult to reconstruct due to its dependency on the distance of the neutrino interaction and its physical properties. Thus, we first reconstructed the closest point of tracks to the detector line and the interaction vertex position of showers. These reconstructions were very satisfactory, as can be seen in \autoref{fig:2D_dist_RZ}. Supplementary plots regarding these reconstructions can be seen in \autoref{fig:err_dist_RZ}, \autoref{fig:sigma_RZ}, \autoref{fig:err_dist_RZ_sh} and \autoref{fig:sigma_RZ_sh}. After this, the reconstruction of the energy was addressed using transfer learning with the PCA, as explained in \autoref{subsubsec:energy}.

\begin{figure}[htbp]
    \centering
    
    \begin{subfigure}{\textwidth}
        \centering
        \subcaption{\label{fig:energy_2D_A}} 

        \makebox[0pt][c]{\raisebox{1.7cm}{\textbf{(1)}} \quad}\includegraphics[width=.32\textwidth]{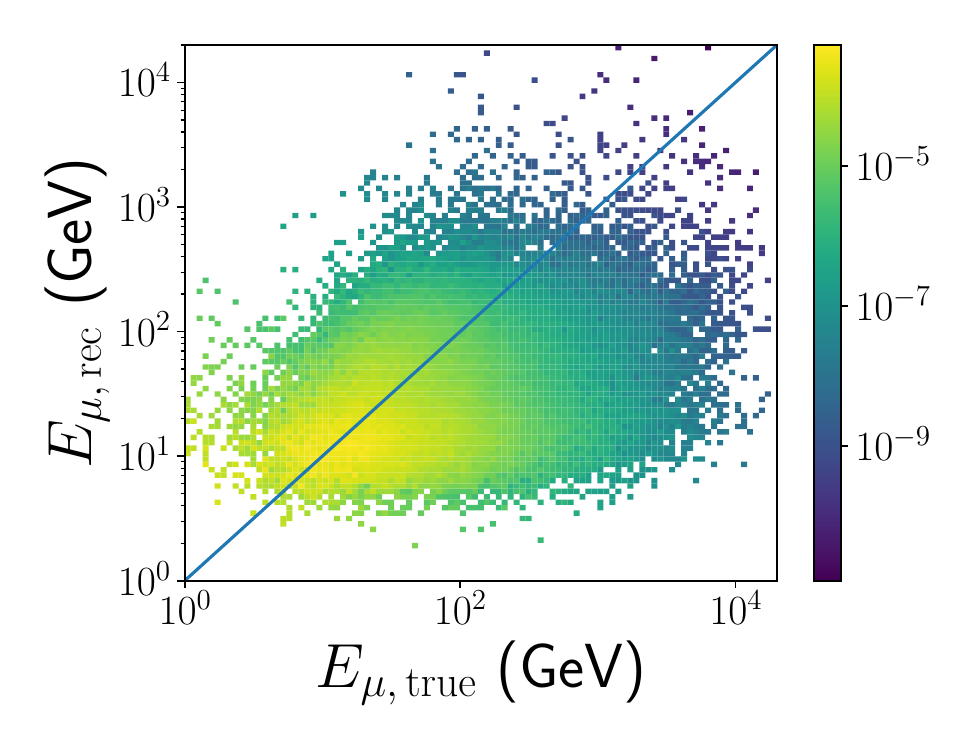}~
		\includegraphics[width=.32\textwidth]{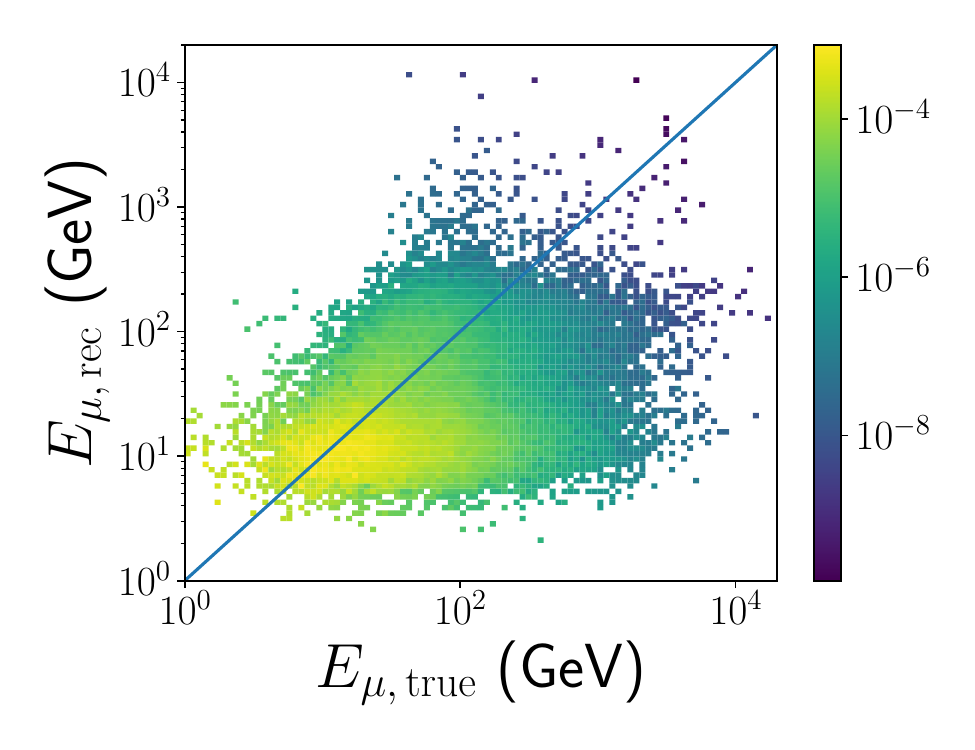}~
		\includegraphics[width=.32\textwidth]{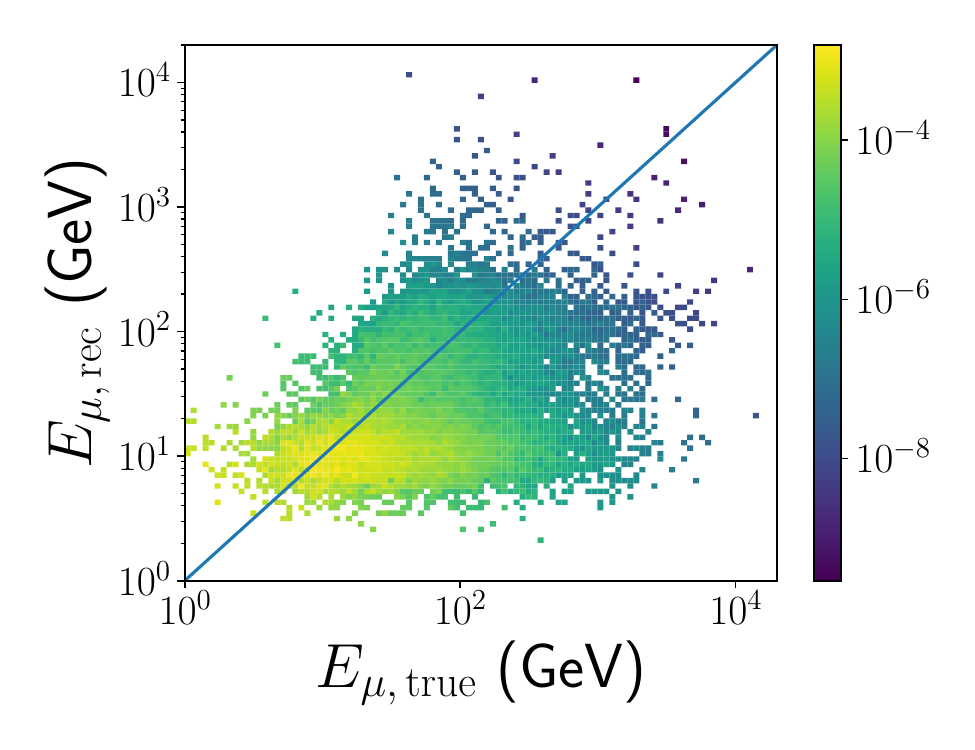}

        \makebox[0pt][c]{\raisebox{1.7cm}{\textbf{(2)}} \quad}\includegraphics[width=.32\textwidth]{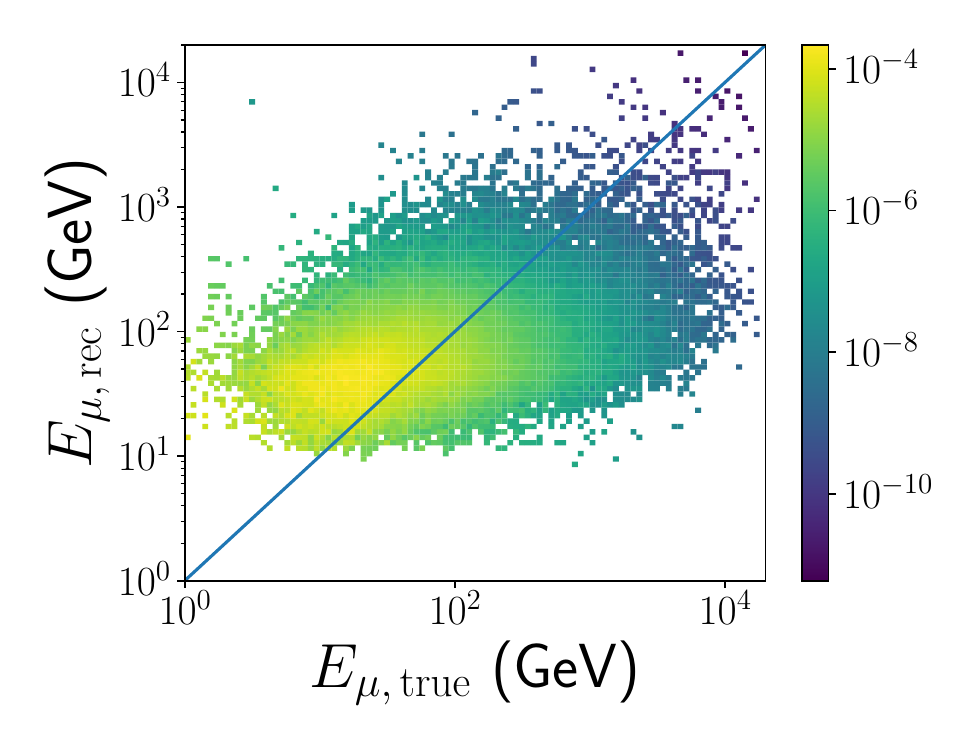}~
		\includegraphics[width=.32\textwidth]{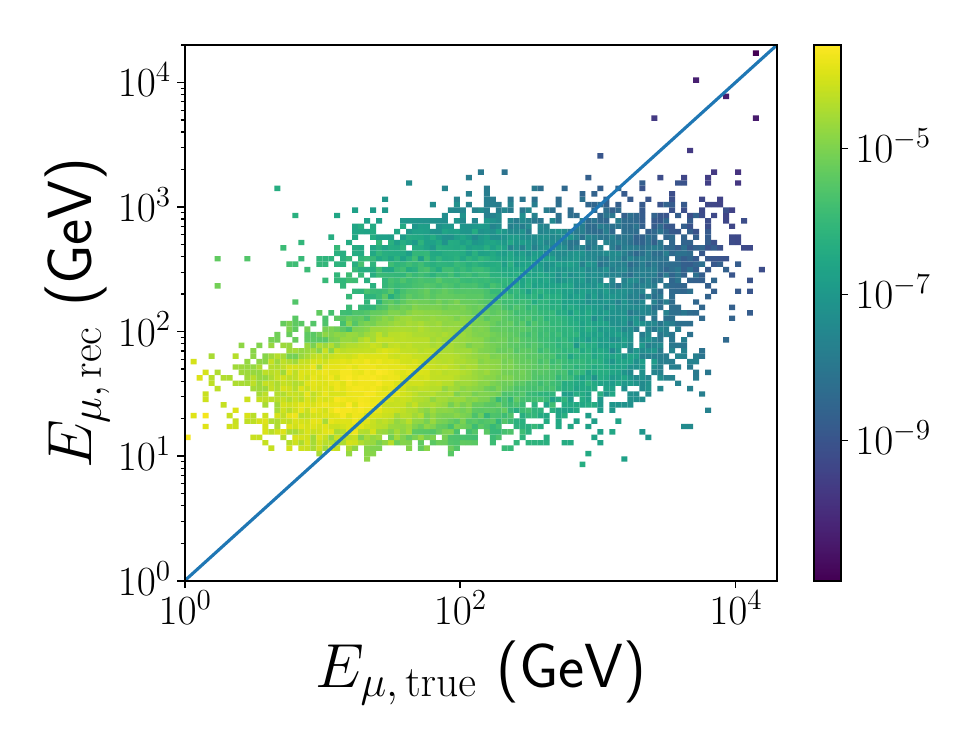}~
		\includegraphics[width=.32\textwidth]{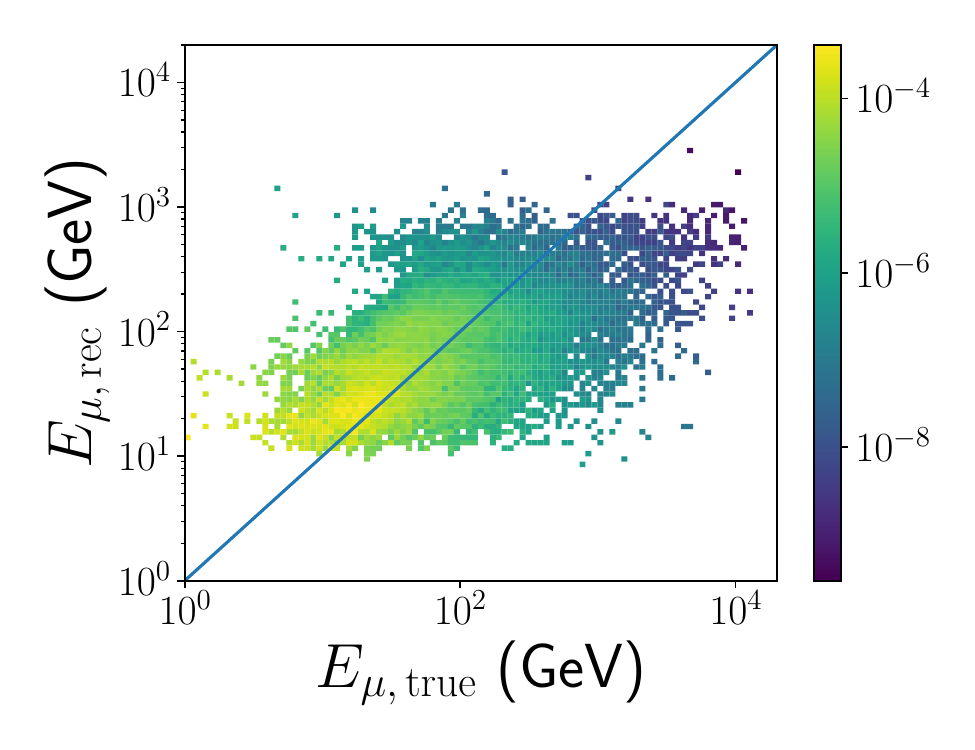}

    \end{subfigure}

    \vskip 5pt
    \hrule
    \vskip 5pt

    \begin{subfigure}{\textwidth}
        \centering
        \subcaption{\label{fig:energy_2D_B}}
        
        \makebox[0pt][c]{\raisebox{1.7cm}{\textbf{(1)}} \quad}\includegraphics[width=.32\textwidth]{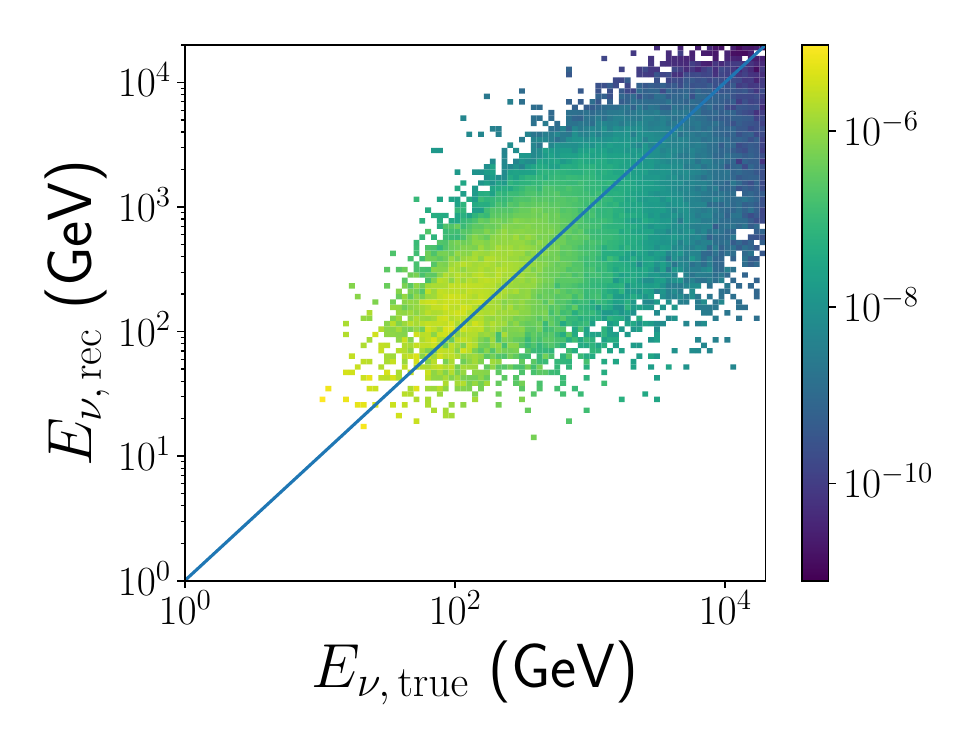}~
		\includegraphics[width=.32\textwidth]{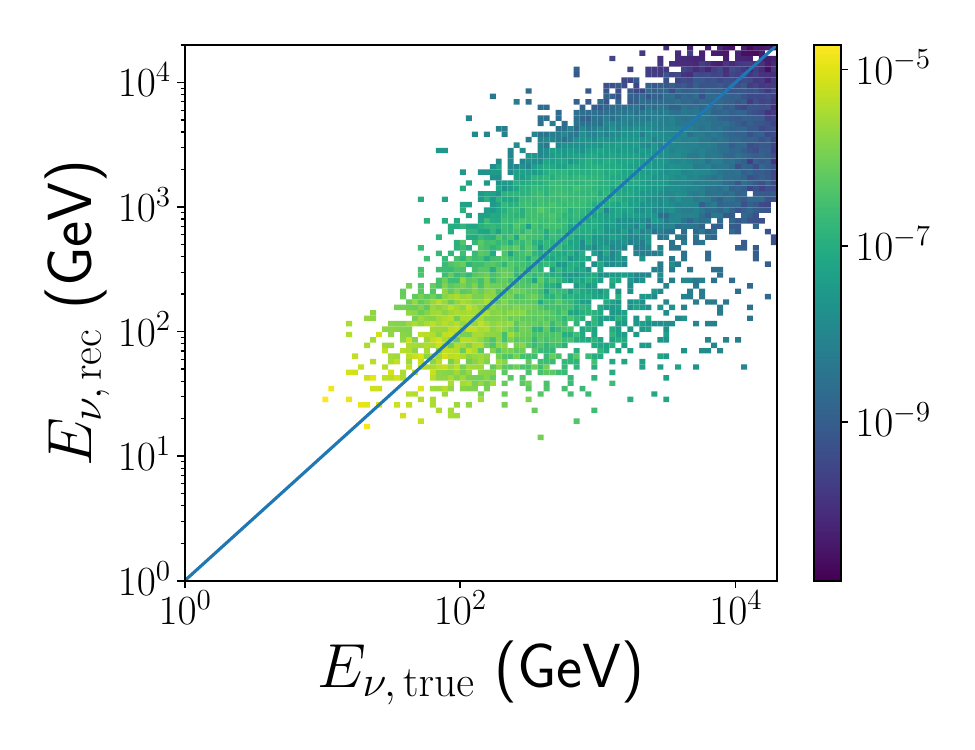}~
		\includegraphics[width=.32\textwidth]{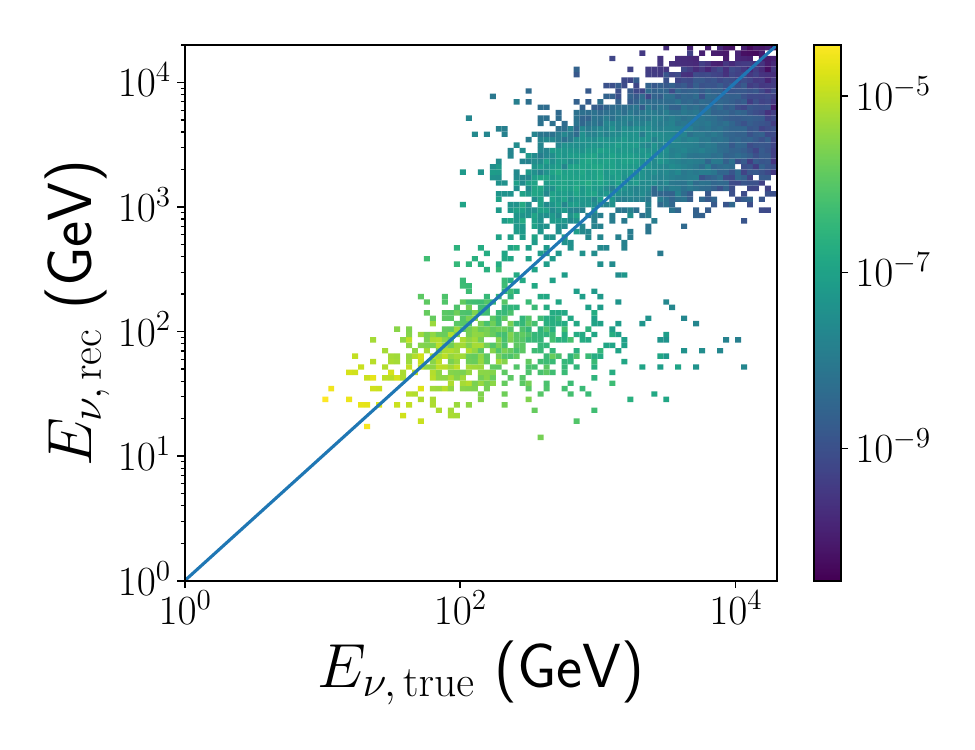}

        \makebox[0pt][c]{\raisebox{1.7cm}{\textbf{(2)}} \quad}\includegraphics[width=.32\textwidth]{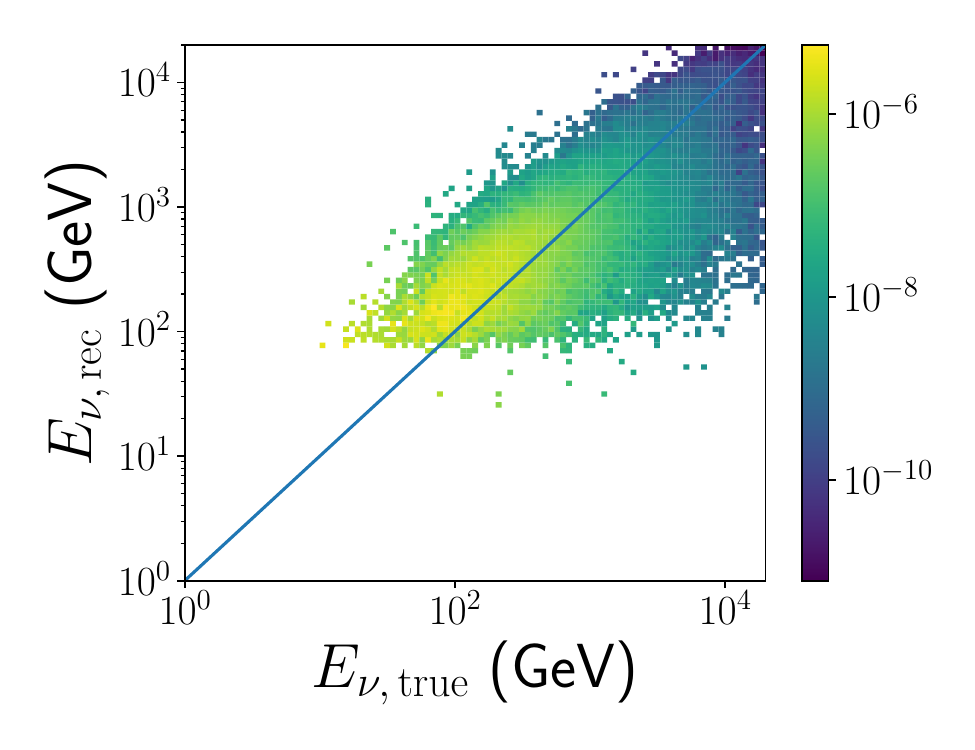}~
		\includegraphics[width=.32\textwidth]{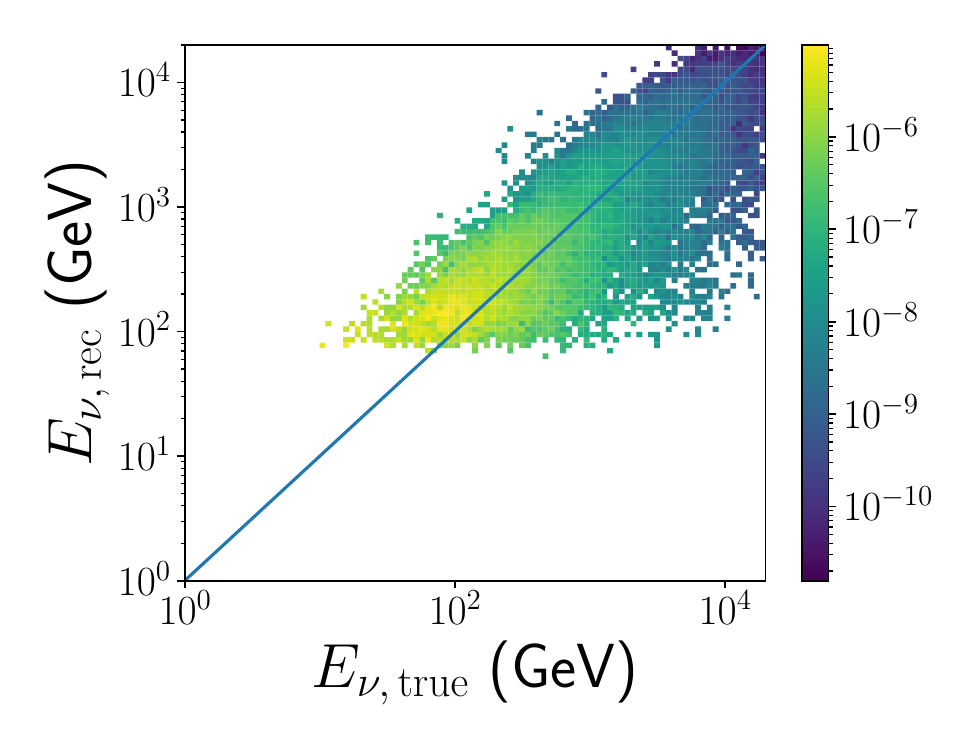}~
		\includegraphics[width=.32\textwidth]{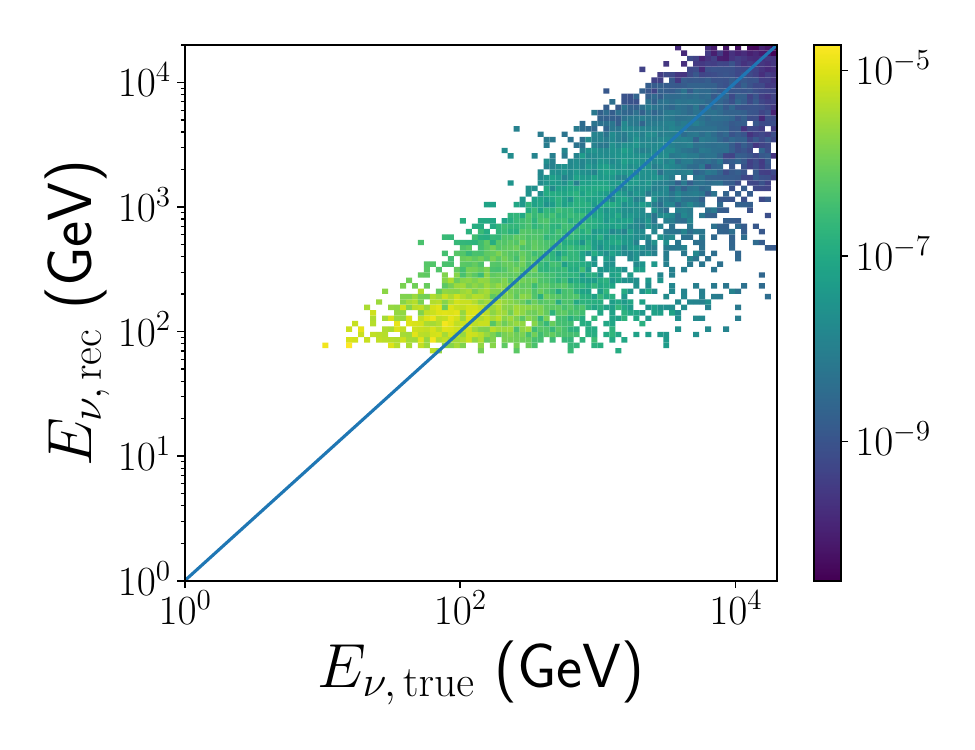}

    \end{subfigure}

    \caption{\label{fig:energy_2D} 2D density histograms of reconstructed vs. true energy values for the pre-selected test dataset (left), the 50\% lowest values of $\sigma$ (mid) and the 25\% (right). \textbf{(a)} Track branch. \textbf{(b)} Shower branch. \textbf{(1)} Benchmark reconstruction. \textbf{(2)} Final reconstruction approach. The diagonal represents the ideal fit.}
\end{figure}

Results for the pre-selected test dataset according to expressions (\ref{eq:cuts_tr}) and (\ref{eq:cuts_sh}) can be seen in \autoref{fig:energy_2D}, showing the utility of the $\sigma_{\log_{10}(E)}$ in selecting the best predictions. Compared to the benchmark reconstruction, the results from the track fit show an improvement when using inheritance from previous networks, even if the results are still moderately good due to the physical complications. The results for showers are slightly better because they are better suited for energy reconstructions due to its topology. Compared to the benchmark, we can see that the uncertainty parameter improves considerably in order to select the best reconstructed events.

In order to quantify the reconstruction resolution, we have computed the median relative error as a function of the true energy:

\begin{equation}
\left| \frac{E_{\mathrm{rec}}-E_{\mathrm{true}}}{E_{\mathrm{true}}} \right| \,,
\end{equation}
for 25\% of events with the best energy reconstruction according to the uncertainty prediction. For tracks, this quantity has a stable value of 0.55 between a true muon energy of 20 and 200 GeV, worsening considerably outside this range. For showers, the error is stable for true neutrino energies higher than 100 GeV, with a value of 0.50 and worsening significantly for lower energies.



\subsubsection{Classifier performance}
\label{subsec:class}

To measure the performance of the classifier, we computed the overall accuracy as well as the recall and precision for both type of events of its dedicated test dataset. The accuracy is defined as the fraction of correctly classified events from the total number of events. The recall is a similar measurement but taking into account only events that truly are from a specific class. Finally, the precision is the fraction of events correctly classified of a class from the total number of events classified in that class.

\begin{figure}[htbp]
	\centering{
	\includegraphics[width=.8\textwidth]{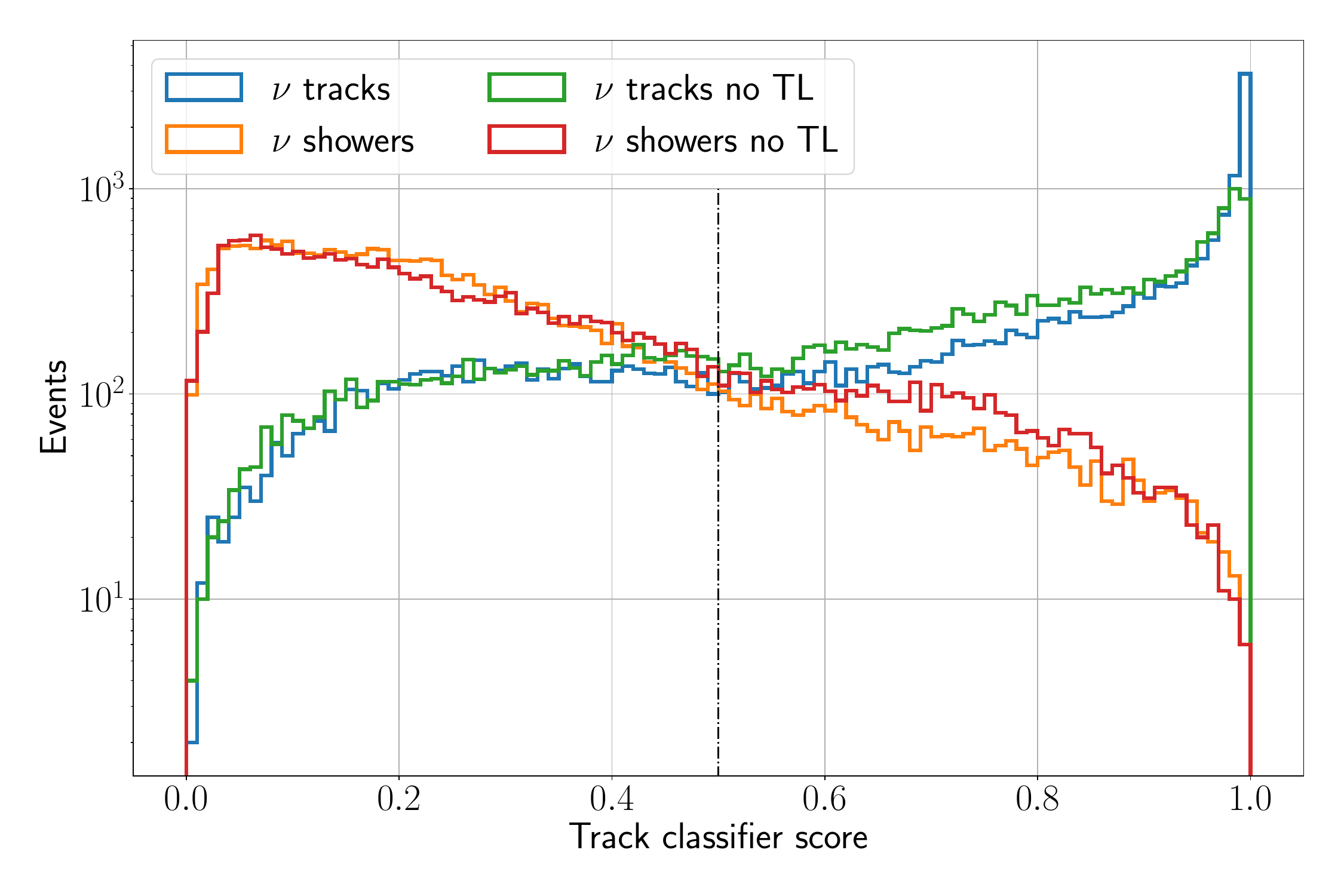}
	\caption{\label{fig:class}Distributions of the track score according to the $N$-fit classifier, with and without Transfer Learning (TL), for tracks and showers.}}
\end{figure}

We get an accuracy of around 80\%, with a recall of 75\% for tracks and 85\% for showers. The precision is 81\% for tracks and 77\% for showers. This means that 25\% of tracks are mis-classified as showers, but that we have a confidence of 81\% that events classified as tracks are truly of that kind. \autoref{fig:class} illustrates the distributions of the track classifier score for track-like and shower-like events, both from the final transfer learning approach and from the benchmark without it (``\textit{no TL}'' tag). There is a higher peak for tracks near to the score equal to one for the TL approach. Also, more showers are classified as tracks without the TL. The overall accuracy reduces to 77\%, with a recall of 73\% and 81\% and precision of 80\% and 75\% for tracks and showers, respectively.

\subsection{Robustness tests}
\label{subsec:robust}

To ensure that the $N$-fit algorithm is stable, unbiased, and suitable for real physics analyses, a series of robustness checks were conducted. These tests aimed to assess the model dependence on the training dataset, its behaviour under extreme or unphysical inputs, and its capability to reproduce realistic results when applied to experimental data. The analyses were carried out using the track branch of $N$-fit and are described in detail in \ref{app:robust}.

The first test consisted of a K-fold cross-validation, a standard method to evaluate the stability of ML models against data partitioning. The dataset used in the track branch development was divided into five equally sized folds, cyclically used for training, validation, and testing. The resulting mean absolute errors remained consistent across folds, indicating that the network’s performance does not depend on a particular training subset. Additionally, a date-sorted K-fold test was performed to verify robustness over the operational history of the ANTARES detector. Despite small fluctuations related to detector conditions at different periods, the results confirmed that $N$-fit can handle data from various stages of the experiment without degradation in accuracy. See \ref{subsubsec:kfold} for details.

The second test addressed the model’s response to \textit{background noise}. By feeding the networks with background-only images, we evaluated whether $N$-fit could mistakenly reconstruct a direction in the absence of a signal. As expected, the reconstructed directions followed a random uniform distribution, confirming that the model does not infer spurious correlations from pure noise. Moreover, the predicted uncertainties were appropriately large, reflecting the low confidence of the algorithm when the input lacks physical features of neutrino events. See \ref{subsubsec:background} for details.

The last test compared the model’s predictions on ANTARES data with those obtained from MC simulations. Since $N$-fit was trained exclusively on MC samples, this validation step was essential in verifying its applicability to experimental data. Using runs not employed during training, and ensuring the availability of the corresponding simulation files for a \textit{run-by-run} comparison, we obtained distributions of reconstructed angles where simulations were in close agreement with the data. After applying quality cuts based on the estimated angular uncertainty ($\sigma_{\Omega}$), most background and atmospheric muon events were rejected for up-going events, and the agreement further improved. This demonstrates that the $N$-fit method is reliable for use in real physics analyses. See \ref{subsubsec:dataMC} for details.



\section{Physics application example}
\label{sec:physics}

The $N$-fit track branch has been already used for different physics analyses. A binned methodology has been applied for an indirect Dark Matter search towards the Sun \cite{ICRC23} and a final unbinned analysis is under development using the whole ANTARES dataset \cite{DMneutrino}. Also, the potential of the method has been tested with a point source search following an IceCube alert, as shown in this section.

The IceCube Collaboration reported a neutrino track-like event with a moderate probability of being of astrophysical origin that occurred on December 8, 2021 at 20:02:51.1 UT \cite{Alert}. The reported position, in equatorial coordinates referred to J2000 and with 90\% point-spread function (PSF) containment, was:

$$
RA = 114.52^\circ ~^{+2.82^\circ}_{-2.50^\circ}\quad \quad
\delta = 15.56^\circ ~^{+1.81^\circ}_{-1.39^\circ}.
$$

The IceCube Collaboration pointed to two gamma-ray sources listed in the 4FGL-DR2 Fermi-LAT catalog and located within the error region presented for the event: 4FGL J0738.4+1539 and 4FGL J0743.1+1713. Hence, the Collaboration encouraged additional follow-up by ground and space-based instruments. The Fermi-LAT Collaboration inspected the vicinity of the neutrino event with all-sky survey data from the Large Area Telescope (LAT) \cite{AlertFermi}. They did not find any significant photon detection from the aforementioned sources nor did they find a new transient source compatible with the best-fit direction of the event. However, outside the 90\% error region, at a distance of $2.1^\circ$, the cataloged gamma-ray source 4FGL J0738.1+1742 was significantly detected. This source is associated with the BL Lac object PKS 0735+17, that is a blazar located at $RA = 114.5^\circ$ and $\delta = 17.7^\circ$.

The ANTARES Collaboration performed a follow-up of the blazar PKS 0735+17 in a sensible time window around the time of the IceCube alert. Two types of analyses were carried out: one binned analysis and another one unbinned \cite{Analisis}. Both of them covered a wide neutrino energy range above $300$ GeV. The $N$-fit track branch is designed to be sensitive to even lower energies, so we performed a binned analysis for the follow-up of the blazar PKS 0735+17 using $N$-fit reconstructed parameters as a complementary study to that already performed. This study served as the first real physics analysis for the new reconstruction method.

The methodology followed here is based on reference \cite{Method}. It is based on comparing the measured data with the expected background inside a determined solid angle region --or Region of Interest (RoI)-- and then computing the significance of the signal excess \cite{Method}. Classically, a circular RoI over the sphere is used, defined by its radius. However, in contrast to traditional methods analyzing multi-line events, our technique is asymmetric for the direction angles, to take into account the better reconstruction of the angle $\theta$. Thus, we decided to modify the methodology by using an asymmetrical RoI, i.e., a rectangle in ($\theta$, $\phi$) coordinates defined by the half-length of the sides: $R_{\theta}$ and $R_{\phi}$. In addition, to optimize the set of cuts, typically the equatorial coordinates are used, for which the right ascension is scrambled (blinded). In our case, we scrambled the azimuthal coordinate, which better suited our $N$-fit reconstruction algorithm.

After completing the optimization, the set of parameters found for the study are:

\begin{equation}
\begin{split}
R_{\theta} = 5^\circ \quad & \quad R_{\phi} =   70^\circ \\
\sigma_{\Omega, \max} &=  21^\circ \,,
\end{split}
\end{equation}
where $\sigma_{\Omega, \max}$ is the maximum angular error estimation allowed for the event selection. We did not find any signal in this region for the given date of the study. Therefore, we present the 90\% confidence level (CL) upper limits on the neutrino fluence ($\mathcal{F}$).

\begin{figure}[htbp]
	\centering
	\includegraphics[width=0.8\textwidth]{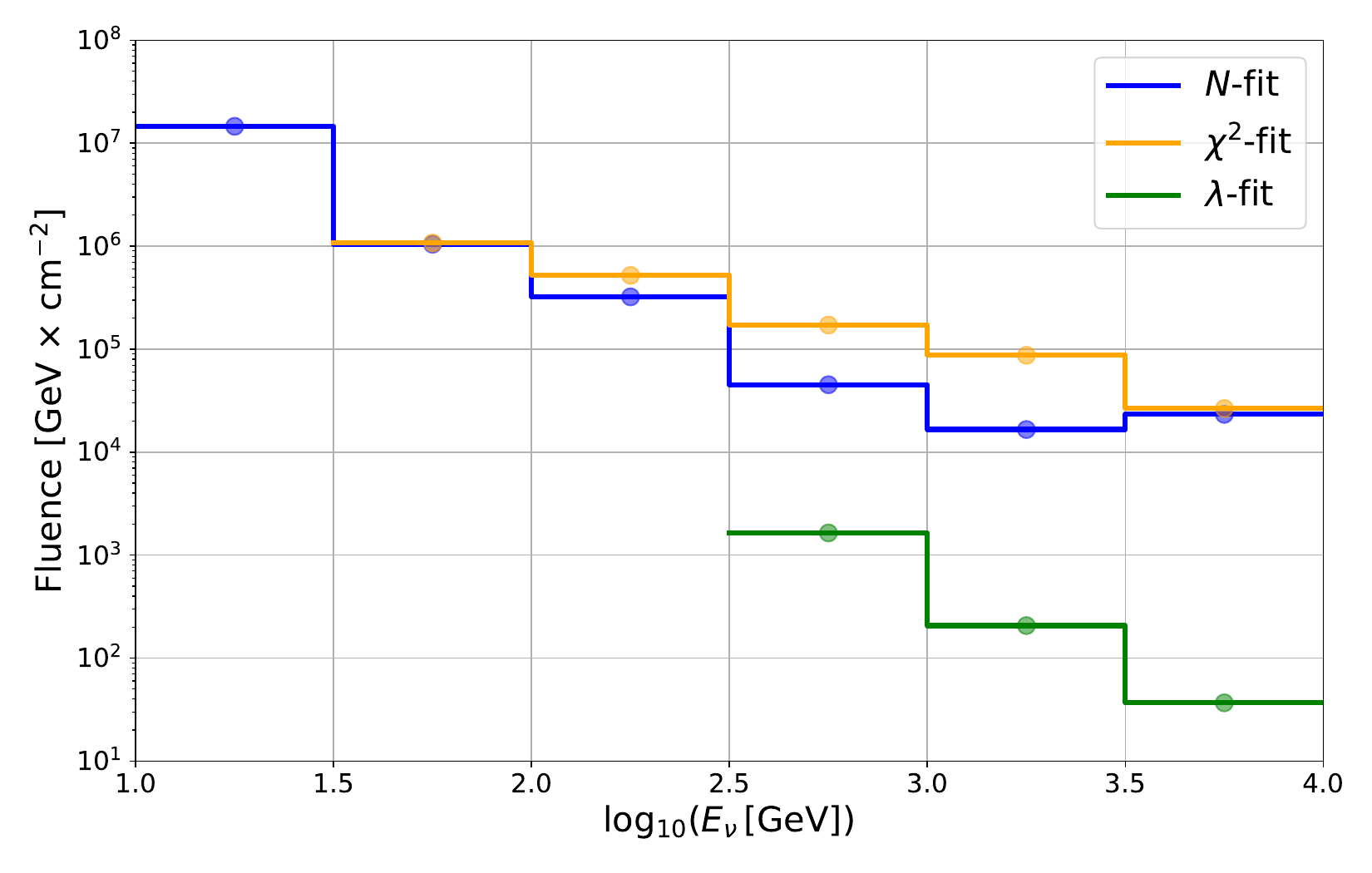}
	\caption{\label{fig:fluence}90\% confidence upper limits on the neutrino fluence as a function of the energy for $N$-fit, $\chi^2$-fit and $\lambda$-fit from PKS 0735+17 over $\pm1$ day around 20:02:51.1 UT on December 8, 2021. The values of the study performed with the $\lambda$-fit reconstruction are taken from \cite{Analisis}, while the values for $\chi^2$-fit --using only SL events-- were computed just for comparison purposes.}
\end{figure}

Using the fact that the 90\% CL upper limit compatible with no event detection is $n_s^{90\%}=2.44$, according to the Feldman and Cousins method \cite{FC} and assuming an $E^{-2}$ energy spectrum, the upper limit for the fluence is:

\begin{equation}
\mathcal{F}^{90\%} = \frac{2.44}{Acc} \cdot \log(E_{\max} / E_{\min}) \cdot T \quad [\mathrm{GeV}/\mathrm{cm}^2] \,,
\end{equation}
where $Acc$ is the acceptance of the detector for our reconstruction technique, and $T$ is the search time ($\sim$ 2 days). This computation is performed for a discrete histogram, in such a way that for each bin we have that $\log(E_{\max} / E_{\min}) = 0.5\cdot \log(10)$. The same procedure was applied for the $\chi^2$-fit reconstructions, just skipping the steps that include a computation with the azimuthal angle, since its reconstruction is missing. The RoI for the $\chi^2$-fit method was simply a full zenithal band characterized by its width. The results are shown in \autoref{fig:fluence}, along with the ones of the previous study carried out with the usual $\lambda$-fit reconstruction method.

The $N$-fit study is sensitive to low energies, being complementary to the previous $\lambda$-fit analysis in this sense. Only $N$-fit is able to retain enough statistics to compute the limits below 30 GeV. The limits of the $N$-fit method are slightly better to those of $\chi^2$-fit using SL events in the energy range between 30 and 300 GeV. Above 300 GeV, where all methods can be compared, the $\lambda$-fit provides more stringent limits than $N$-fit, as expected.



\section{Discussion and conclusions}
\label{sec:conclusion}

In this work, we have developed and presented a new reconstruction and classification approach based on Deep Learning for single-line events in the ANTARES telescope. The exploration to achieve the best possible reconstruction has led us to a novel combination of machine learning techniques, such as the use of deep convolutional layers, a mixture density output layer, and transfer learning. The reconstruction of the spatio-temporal parameters of the SL event, such as the incoming neutrino direction and location, can directly be obtained using deep convolutional neural networks. Combining a deep convolutional network with a mixture density output layer allowed us to estimate the quality of the prediction. Other more complex tasks, however, such as energy estimation or event classification, benefitted from using transfer learning. For the energy reconstruction, this study presents an innovative \emph{indirect} transfer learning approach based on Principal Component Analysis knowledge distillation. Using this dimensionality reduction technique, we selected the most informative neuron activations from networks performing spatial tasks, and used them as input features for training a feed-forward network in energy reconstruction. This method provided results that can be exploited in physics analyses. \emph{Direct} transfer learning was applied to classification tasks, by freezing the convolutional layers of pretrained network models specialized in event-type-specific spatial information. 

Thus, we developed $N$-fit as a repertoire of specialized modules that together fully characterized SL neutrino events in terms of energy, spatial properties and type of event, track or shower (\autoref{sec:results}). Compared to previous techniques applied to SL events in ANTARES, $N$-fit outperforms the $\theta$ angle reconstruction of SL events provided by the $\chi^2$-fit. For track-like events, the mean $\theta$ error decreased from $\sim 9.7^\circ$ (standard $\chi^2$-fit) to $\sim 3.7^\circ$ for the best 50\% reconstructed events. Moreover, $N$-fit provides a sufficiently accurate estimation for the $\phi$ angle --with a mean error of $\sim 29^\circ$ for the best 50\% track events-- that is useful in physics analyses. Such information was not attempted in $\chi^2$-fit due to the approximations made in the method. Specifically for track events, $N$-fit provides a very precise reconstruction of the closest point of the track event to the ANTARES detector line and a first approximation of the muon energy reconstruction, taking into account the physical difficulties in this endeavour, particularly in tracks. Regarding shower events, their interaction vertex reconstruction was very precisely estimated in $N$-fit. The neutrino energy in shower events was reconstructed with a 0.50 relative error, compared to a 0.55 relative error for muon energy in track events. Lastly, discrimination between the two types of neutrino events by the $N$-fit classifier reached a high accuracy level, of around $80\%$. When comparing data and MC simulations, $N$-fit demonstrates a good agreement in the track branch, heavily reducing the background noise and atmospheric muons of the up-going sample when quality cutoffs are applied.

$N$-fit modules characterizing track events have already been used in phy\-sics analyses as shown in \autoref{sec:physics}. We have a promising improvement in the low energy range for dark matter WIMP searches towards the Sun \cite{ICRC23}. Also, we have demonstrated that the new methodology is able to complement classical reconstruction methods in point sources neutrino searches. In addition, we are actively applying $N$-fit reconstructions in different analyses considering the complete ANTARES dataset, for example, in the search for dark matter in the Sun \cite{DMneutrino}.

More generally, the repertoire and combination of machine learning techniques used in $N$-fit may be adapted to reconstructions in other particle detectors, such as KM3NeT, as well as inspiring new applications in other areas of computational physics.



\hide{
\appendix
\section{Example Appendix Section}
\label{app1}

Appendix text.
}






\clearpage
\bibliographystyle{elsarticle-num}
\bibliography{paper_bib}

@misc{TensorFlow,
title={ {TensorFlow}: Large-Scale Machine Learning on Heterogeneous Systems},
url={https://www.tensorflow.org/},
note={Software available from tensorflow.org},
author={
    Mart\'{i}n~Abadi and
    Ashish~Agarwal and
    Paul~Barham and
    Eugene~Brevdo and
    Zhifeng~Chen and
    Craig~Citro and
    Greg~S.~Corrado and
    Andy~Davis and
    Jeffrey~Dean and
    Matthieu~Devin and
    Sanjay~Ghemawat and
    Ian~Goodfellow and
    Andrew~Harp and
    Geoffrey~Irving and
    Michael~Isard and
    Yangqing Jia and
    Rafal~Jozefowicz and
    Lukasz~Kaiser and
    Manjunath~Kudlur and
    Josh~Levenberg and
    Dandelion~Man\'{e} and
    Rajat~Monga and
    Sherry~Moore and
    Derek~Murray and
    Chris~Olah and
    Mike~Schuster and
    Jonathon~Shlens and
    Benoit~Steiner and
    Ilya~Sutskever and
    Kunal~Talwar and
    Paul~Tucker and
    Vincent~Vanhoucke and
    Vijay~Vasudevan and
    Fernanda~Vi\'{e}gas and
    Oriol~Vinyals and
    Pete~Warden and
    Martin~Wattenberg and
    Martin~Wicke and
    Yuan~Yu and
    Xiaoqiang~Zheng},
  year={2015},
}

@Article{Transfer,
  author   = {Zhuang, Fuzhen and Qi, Zhiyuan and Duan, Keyu and Xi, Dongbo and Zhu, Yongchun and Zhu, Hengshu and Xiong, Hui and He, Qing},
  journal  = {Proceedings of the IEEE},
  title    = {{A Comprehensive Survey on Transfer Learning}},
  year     = {2021},
  number   = {1},
  pages    = {43-76},
  volume   = {109},
  doi      = {10.1109/JPROC.2020.3004555},
}

@Article{BBfit,
  author   = {{ANTARES Collaboration}},
  journal  = {Astroparticle Physics},
  title    = {A fast algorithm for muon track reconstruction and its application to the {ANTARES} neutrino telescope},
  year     = {2011},
  issn     = {0927-6505},
  month    = apr,
  number   = {9},
  pages    = {652--662},
  volume   = {34},
  doi      = {10.1016/j.astropartphys.2011.01.003},
  language = {en},
  url      = {https://www.sciencedirect.com/science/article/pii/S0927650511000053},
  urldate  = {2022-07-05},
}

@PhdThesis{AAfit,
  author = {Heijboer, Aart},
  school = {Amsterdam U.},
  title  = {Track reconstruction and point source searches with {ANTARES}},
  year   = {2004},
  type   = {{PhD} {Thesis}},
}

@Article{Antares,
  author     = {{ANTARES Collaboration}},
  journal    = {Nuclear Instruments and Methods in Physics Research Section A: Accelerators, Spectrometers, Detectors and Associated Equipment},
  title      = {{ANTARES}: {The} first undersea neutrino telescope},
  year       = {2011},
  issn       = {0168-9002},
  month      = nov,
  number     = {1},
  pages      = {11--38},
  volume     = {656},
  doi        = {10.1016/j.nima.2011.06.103},
  language   = {en},
  shorttitle = {{ANTARES}},
  url        = {https://www.sciencedirect.com/science/article/pii/S0168900211013994},
  urldate    = {2022-07-05},
}

@InCollection{PCA,
  author    = {Kherif, Ferath and Latypova, Adeliya},
  booktitle = {Machine Learning},
  publisher = {Academic Press},
  title     = {{Chapter 12 - Principal component analysis}},
  year      = {2020},
  editor    = {Andrea Mechelli and Sandra Vieira},
  isbn      = {978-0-12-815739-8},
  pages     = {209-225},
  doi       = {https://doi.org/10.1016/B978-0-12-815739-8.00012-2},
  url       = {https://www.sciencedirect.com/science/article/pii/B9780128157398000122},
}

@Article{perceptron,
  author     = {Rosenblatt, Frank},
  journal    = {Psychological Review},
  title      = {The perceptron: {A} probabilistic model for information storage and organization in the brain},
  year       = {1958},
  issn       = {1939-1471},
  number     = {6},
  pages      = {386--408},
  volume     = {65},
  address    = {US},
  doi        = {10.1037/h0042519},
  publisher  = {American Psychological Association},
  shorttitle = {The perceptron},
}

@TechReport{ini_weights,
  author   = {Kumar, Siddharth Krishna},
  title    = {On weight initialization in deep neural networks},
  year     = {2017},
  month    = may,
  note     = {arXiv:1704.08863 [cs]},
  doi      = {10.48550/arXiv.1704.08863},
  school   = {arXiv},
  url      = {http://arxiv.org/abs/1704.08863},
  urldate  = {2022-07-05},
}

@TechReport{Adam,
  author     = {Kingma, Diederik P. and Ba, Jimmy},
  title      = {Adam: {A} {Method} for {Stochastic} {Optimization}},
  year       = {2017},
  month      = jan,
  note       = {arXiv:1412.6980 [cs]},
  doi        = {10.48550/arXiv.1412.6980},
  school     = {arXiv},
  shorttitle = {Adam},
  url        = {http://arxiv.org/abs/1412.6980},
  urldate    = {2022-07-05},
}

@TechReport{SGD,
  author   = {Ruder, Sebastian},
  title    = {An overview of gradient descent optimization algorithms},
  year     = {2017},
  month    = jun,
  note     = {arXiv:1609.04747 [cs]},
  doi      = {10.48550/arXiv.1609.04747},
  school   = {arXiv},
  url      = {http://arxiv.org/abs/1609.04747},
  urldate  = {2022-07-05},
}

@InCollection{EarlyStop,
  author    = {Prechelt, Lutz},
  publisher = {Springer},
  title     = {Early {Stopping} - {But} {When}?},
  year      = {1998},
  address   = {Berlin, Heidelberg},
  editor    = {Orr, Genevieve B. and Müller, Klaus-Robert},
  isbn      = {9783540494300},
  pages     = {55--69},
  series    = {Lecture {Notes} in {Computer} {Science}},
  doi       = {10.1007/3-540-49430-8_3},
  language  = {en},
  url       = {https://doi.org/10.1007/3-540-49430-8_3},
  urldate   = {2022-07-05},
}

@Article{CNN,
  author   = {Gu, Jiuxiang and Wang, Zhenhua and Kuen, Jason and Ma, Lianyang and Shahroudy, Amir and Shuai, Bing and Liu, Ting and Wang, Xingxing and Wang, Gang and Cai, Jianfei and Chen, Tsuhan},
  journal  = {Pattern Recognition},
  title    = {Recent advances in convolutional neural networks},
  year     = {2018},
  issn     = {0031-3203},
  month    = may,
  pages    = {354--377},
  volume   = {77},
  doi      = {10.1016/j.patcog.2017.10.013},
  language = {en},
  url      = {https://www.sciencedirect.com/science/article/pii/S0031320317304120},
  urldate  = {2022-07-05},
}

@Article{MDN,
  author  = {Bishop, Christopher M.},
  journal = {NCRG/94/004},
  title   = {Mixture density networks},
  year    = {1994},
  url     = {https://research.aston.ac.uk/en/publications/mixture-density-networks},
  urldate = {2022-07-05},
}

@Article{MCsim,
  author    = {{ANTARES Collaboration}},
  journal   = {Journal of Cosmology and Astroparticle Physics},
  title     = {Monte {Carlo} simulations for the {ANTARES} underwater neutrino telescope},
  year      = {2021},
  issn      = {1475-7516},
  month     = jan,
  number    = {01},
  pages     = {064--064},
  volume    = {2021},
  doi       = {10.1088/1475-7516/2021/01/064},
  language  = {en},
  publisher = {IOP Publishing},
  url       = {https://doi.org/10.1088/1475-7516/2021/01/064},
  urldate   = {2022-07-05},
}

@InProceedings{Glorot,
  author    = {Glorot, Xavier and Bengio, Yoshua},
  title     = {Understanding the difficulty of training deep feedforward neural networks},
  year      = {2010},
  month     = mar,
  pages     = {249--256},
  publisher = {JMLR Workshop and Conference Proceedings},
  issn      = {1938-7228},
  language  = {en},
  url       = {https://proceedings.mlr.press/v9/glorot10a.html},
  urldate   = {2022-07-05},
}

@InProceedings{He,
  author    = {He, Kaiming and Zhang, Xiangyu and Ren, Shaoqing and Sun, Jian},
  booktitle = {Proceedings of the IEEE International Conference on Computer Vision (ICCV)},
  title     = {Delving {Deep} into {Rectifiers}: {Surpassing} {Human-Level} {Performance} on {ImageNet} {Classification}},
  year      = {2015},
  month     = {December},
}

@InProceedings{KCV,
  author       = {Anguita, Davide and Ghelardoni, Luca and Ghio, Alessandro and Oneto, Luca and Ridella, Sandro},
  booktitle    = {20th European Symposium on Artificial Neural Networks, Computational Intelligence and Machine Learning (ESANN)},
  title        = {The ‘{K}’in {K}-fold cross validation},
  year         = {2012},
  organization = {i6doc. com publ},
  pages        = {441--446},
}

@PhdThesis{weights,
  author   = {Colnard, Claudine M. M.},
  title    = {Ultra-{High} {Energy} {Neutrino} {Simulations}},
  year     = {2009},
  month    = jun,
  language = {en},
  url      = {https://inspirehep.net/literature/823582},
  urldate  = {2022-07-05},
}

@Article{ICRC23,
  author  = {García-Méndez, Juan and Ardid, Salva and Ardid, Miguel},
  journal = {PoS},
  title   = {{Dark matter search towards the Sun using Machine Learning reconstructions of single-line events in ANTARES}},
  year    = {2023},
  pages   = {1443},
  volume  = {ICRC2023},
  doi     = {10.22323/1.444.1443},
}

@Misc{DMneutrino,
  author    = {Poirè, Chiara},
  month     = aug,
  title     = {Indirect dark matter searches towards the {Sun} using the full {ANTARES} data set},
  year      = {2024},
  doi       = {10.5281/zenodo.13350925},
  publisher = {Zenodo},
  url       = {https://doi.org/10.5281/zenodo.13350925},
}

@Electronic{Alert,
  author       = {{IceCube Collaboration}},
  note         = {\url{https://gcn.gsfc.nasa.gov/gcn/gcn3/31191.gcn3}},
  organization = {IceCube Collaboration},
  title        = {{GCN} circular \#31191},
  url          = {https://gcn.gsfc.nasa.gov/gcn/gcn3/31191.gcn3},
  year         = {2021},
}

@Electronic{AlertFermi,
  author       = {{Fermi-{LAT} Collaboration}},
  note         = {\url{https://gcn.gsfc.nasa.gov/gcn3/31194.gcn3}},
  organization = {Fermi-LAT Collaboration},
  title        = {{GCN} circular \#31194},
  url          = {https://gcn.gsfc.nasa.gov/gcn3/31194.gcn3},
  year         = {2021},
}

@Misc{Analisis,
  author   = {Alves, Sergio and on behalf of the ANTARES collaboration},
  month    = jul,
  title    = {{ANTARES} follow-up of {IceCube} event {IC211208A} coincident with flaring blazar {PKS} 0735+17},
  year     = {2022},
  doi      = {10.5281/zenodo.6785295},
  url      = {https://zenodo.org/record/6785295},
  urldate  = {2022-07-08},
}

@InProceedings{Method,
  author    = {Alves Garre, Sergio and Versari, Federico and Avrorin, A. D. and Dzhlkibaev, Zhan-Arys M. and Shelepov, M. D. and Suvorova, O. V. and on behalf of the ANTARES collaboration},
  booktitle = {37th {International} {Cosmic} {Ray} {Conference}},
  title     = {{ANTARES} offline study of three alerts after {Baikal}-{GVD} follow-up found coincident cascade neutrino events},
  year      = {2021},
  address   = {Berlin, Germany},
  month     = jul,
  pages     = {1121},
  volume    = {ICRC2021},
  doi       = {10.22323/1.395.1121},
  url       = {https://hal.archives-ouvertes.fr/hal-03373262},
  urldate   = {2022-07-05},
}

@Article{FC,
  author    = {Feldman, Gary J. and Cousins, Robert D.},
  journal   = {Physical Review D},
  title     = {Unified approach to the classical statistical analysis of small signals},
  year      = {1998},
  month     = apr,
  number    = {7},
  pages     = {3873--3889},
  volume    = {57},
  doi       = {10.1103/PhysRevD.57.3873},
  publisher = {American Physical Society},
  url       = {https://link.aps.org/doi/10.1103/PhysRevD.57.3873},
  urldate   = {2022-07-05},
}

@article{Psihas,
author = {Psihas, Fernanda and Groh, Micah and Tunnell, Christopher and Warburton, Karl},
title = {A review on machine learning for neutrino experiments},
journal = {International Journal of Modern Physics A},
volume = {35},
number = {33},
pages = {2043005},
year = {2020},
doi = {10.1142/S0217751X20430058},

URL = {   
        https://doi.org/10.1142/S0217751X20430058
    
},
}

@Article{Dropout,
  author    = {Srivastava, Nitish and Hinton, Geoffrey and Krizhevsky, Alex and Sutskever, Ilya and Salakhutdinov, Ruslan},
  journal   = {The journal of machine learning research},
  title     = {Dropout: a simple way to prevent neural networks from overfitting},
  year      = {2014},
  number    = {1},
  pages     = {1929--1958},
  volume    = {15},
  publisher = {JMLR. org},
  url       = {https://jmlr.org/papers/v15/srivastava14a.html},
}

@article{moslemi2024,
  title={A Survey on Knowledge Distillation: Recent Advancements},
  author={Moslemi, Amir and Briskina, Anna and Dang, Zubeka and Li, Jason},
  journal={Machine Learning with Applications},
  pages={100605},
  year={2024},
  publisher={Elsevier}
}

@Article{IC_DL,
author={Bukhari, Habib
and Chakraborty, Dipam
and Eller, Philipp
and Ito, Takuya
and Shugaev, Maxim V.
and {\O}rs{\o}e, Rasmus},
title={IceCube -- Neutrinos in Deep Ice},
journal={The European Physical Journal C},
year={2024},
month={Jun},
day={25},
volume={84},
number={6},
pages={646},
issn={1434-6052},
doi={10.1140/epjc/s10052-024-12977-2},
url={https://doi.org/10.1140/epjc/s10052-024-12977-2}
}

@article{Km3_DL,
doi = {10.1088/1748-0221/16/10/C10011},
url = {https://dx.doi.org/10.1088/1748-0221/16/10/C10011},
year = {2021},
month = {oct},
publisher = {IOP Publishing},
volume = {16},
number = {10},
pages = {C10011},
author = {Reck, S. and Guderian, D. and Vermariën, G. and Domi, A. and on behalf of the KM3NeT collaboration},
title = {{Graph neural networks for reconstruction and classification in KM3NeT}},
journal = {Journal of Instrumentation}
}

@Article{Km3_DL2,
  author = {{KM3NeT Collaboration}},
  journal  = {Journal of Instrumentation},
  title    = {{Event reconstruction for KM3NeT/ORCA using convolutional neural networks}},
  year     = {2020},
  month    = {oct},
  number   = {10},
  pages    = {P10005},
  volume   = {15},
  doi      = {10.1088/1748-0221/15/10/P10005},
  url      = {https://dx.doi.org/10.1088/1748-0221/15/10/P10005},
}

@Article{Jones2023,
  author   = {Jones, Anne and Kuehnert, Julian and Fraccaro, Paolo and Meuriot, Oph{\'e}lie and Ishikawa, Tatsuya and Edwards, Blair and Stoyanov, Nikola and Remy, Sekou L. and Weldemariam, Kommy and Assefa, Solomon},
  journal  = {npj Climate and Atmospheric Science},
  title    = {{AI} for climate impacts: applications in flood risk},
  year     = {2023},
  issn     = {2397-3722},
  month    = {Jun},
  number   = {1},
  pages    = {63},
  volume   = {6},
  day      = {08},
  doi      = {10.1038/s41612-023-00388-1},
  url      = {https://doi.org/10.1038/s41612-023-00388-1},
}

@Article{Higgs,
  author        = {Glaysher, Paul},
  journal       = {PoS},
  title         = {{BDTs in the Search for $t\bar{t}H$ Production with Higgs Decays to $b\bar{b}$ at ATLAS}},
  year          = {2018},
  pages         = {698},
  volume        = {EPS-HEP2017},
  collaboration = {ATLAS},
  doi           = {10.22323/1.314.0698},
  editor        = {Checchia, Paolo and others},
  reportnumber  = {ATL-PHYS-PROC-2017-214},
}

@Article{relu,
  author   = {Daubechies, I. and DeVore, R. and Foucart, S. and Hanin, B. and Petrova, G.},
  journal  = {Constructive Approximation},
  title    = {{Nonlinear Approximation and (Deep) $\mathrm{ReLU}$ Networks}},
  year     = {2022},
  issn     = {1432-0940},
  month    = {Feb},
  number   = {1},
  pages    = {127-172},
  volume   = {55},
  day      = {01},
  doi      = {10.1007/s00365-021-09548-z},
  url      = {https://doi.org/10.1007/s00365-021-09548-z},
}

@Article{Tantra,
  author        = {{ANTARES Collaboration}},
  journal       = {The Astronomical Journal},
  title         = {An {A}lgorithm for the {R}econstruction of {N}eutrino-induced {S}howers in the {ANTARES} {N}eutrino {T}elescope},
  year          = {2017},
  month         = dec,
  number        = {6},
  pages         = {275},
  volume        = {154},
  collaboration = {The ANTARES Collaboration},
  doi           = {10.3847/1538-3881/aa9709},
  publisher     = {The American Astronomical Society},
  url           = {https://dx.doi.org/10.3847/1538-3881/aa9709},
}

@Article{AAfit2,
  author    = {{ANTARES Collaboration}},
  journal   = {The Astrophysical Journal},
  title     = {SEARCH FOR COSMIC NEUTRINO POINT SOURCES WITH FOUR YEARS OF DATA FROM THE ANTARES TELESCOPE},
  year      = {2012},
  month     = {nov},
  number    = {1},
  pages     = {53},
  volume    = {760},
  doi       = {10.1088/0004-637X/760/1/53},
  publisher = {The American Astronomical Society},
  url       = {https://doi.org/10.1088/0004-637X/760/1/53},
}

@Article{legacy,
  author   = {{ANTARES Collaboration}},
  journal  = {Physics Reports},
  title    = {The {ANTARES} detector: {T}wo decades of neutrino searches in the {M}editerranean {S}ea},
  year     = {2025},
  issn     = {0370-1573},
  pages    = {1-46},
  volume   = {1121-1124},
  doi      = {https://doi.org/10.1016/j.physrep.2025.04.001},
  url      = {https://www.sciencedirect.com/science/article/pii/S0370157325001450},
}

@Article{Duev2019,
  author   = {Duev, Dmitry A and Mahabal, Ashish and Masci, Frank J and Graham, Matthew J and Rusholme, Ben and Walters, Richard and Karmarkar, Ishani and Frederick, Sara and Kasliwal, Mansi M and Rebbapragada, Umaa and Ward, Charlotte},
  journal  = {Monthly Notices of the Royal Astronomical Society},
  title    = {Real-bogus classification for the {Z}wicky {T}ransient {F}acility using deep learning},
  year     = {2019},
  issn     = {0035-8711},
  month    = {08},
  number   = {3},
  pages    = {3582-3590},
  volume   = {489},
  doi      = {10.1093/mnras/stz2357},
  eprint   = {https://academic.oup.com/mnras/article-pdf/489/3/3582/30029533/stz2357.pdf},
  url      = {https://doi.org/10.1093/mnras/stz2357},
}


\section*{Acknowledgements}
\addcontentsline{toc}{section}{Acknowledgements}
\begin{sloppypar}
The authors acknowledge the financial support of the funding agencies:
Centre National de la Recherche Scientifique (CNRS), Commissariat \`a
l'\'ener\-gie atomique et aux \'energies alternatives (CEA),
Commission Europ\'eenne (FEDER fund and Marie Curie Program),
LabEx UnivEarthS (ANR-10-LABX-0023 and ANR-18-IDEX-0001),
R\'egion Alsace (contrat CPER), R\'e\-gion Provence-Alpes-C\^ote d'Azur,
D\'e\-par\-tement du Var and Ville de La
Seyne-sur-Mer, France;
Bundesministerium f\"ur Bildung und Forschung
(BMBF), Germany; 
Istituto Nazionale di Fisica Nucleare (INFN), Italy;
Nederlandse organisatie voor Wetenschappelijk Onderzoek (NWO), the Netherlands;
Ministry of Education and Scientific Research, Romania;
MCIN for PID2024-156285NB-C41, -C42, -C43, PID2021-124591NB-C41, -C42, -C43, PID2020-120037GA-I00,  and PDC2023-145913-I00 funded by MICIU/AEI/10.13039/501100011033 and by “ERDF A way of making Europe”, for ASFAE/2022/014 and ASFAE/2022 /023 with funding from the EU NextGenerationEU (PRTR-C17.I01) and Generalitat Valenciana, for Grant AST22\_6.2 with funding from Consejer\'{\i}a de Universidad, Investigaci\'on e Innovaci\'on and Gobierno de Espa\~na and European Union - NextGenerationEU, for CSIC-INFRA23013 and for CNS2023-144099, Generalitat Valenciana for CIDEGENT/2020/049, CIDEGENT/2021/23, CIDEIG/2023/20, ESGENT/2024/24, CIPROM/2023/51, GRISOLIAP/2021/192 and INNVA1/2024/110 (IVACE+i), Spain;
Ministry of Higher Education, Scientific Research and Innovation, Morocco, and the Arab Fund for Economic and Social Development, Kuwait.
We also acknowledge the technical support of Ifremer, AIM and Foselev Marine
for the sea operation and the CC-IN2P3 for the computing facilities.
\end{sloppypar}

\pagebreak
\appendix
\section{Supplementary material}
\label{app:figures}

\begin{table}[htbp]
    \centering
    \renewcommand{\arraystretch}{1.2} 
    \setlength{\tabcolsep}{5pt} 

    \begin{tabular}{>{\centering\arraybackslash}m{2.7cm}  >{\centering\arraybackslash}m{0.9cm} >{\centering\arraybackslash}m{0.9cm}  >{\centering\arraybackslash}m{0.9cm} >{\centering\arraybackslash}m{0.9cm}  >{\centering\arraybackslash}m{0.9cm} >{\centering\arraybackslash}m{0.9cm}  >{\centering\arraybackslash}m{0.9cm} >{\centering\arraybackslash}m{0.9cm}}  
        \hline
        Mean/Median 
        & \multicolumn{2}{c}{$\chi^2$-fit} 
        & \multicolumn{2}{c}{$N$-fit} 
        & \multicolumn{2}{c}{$\chi^2$-fit (50\%)} 
        & \multicolumn{2}{c}{$N$-fit (50\%)} \\  
        \hline \hline
        $\theta$ & 27.4$^\circ$ & 22.8$^\circ$ & 12.3$^\circ$ & 8.9$^\circ$ & 24.9 & 20.5 & 7.9$^\circ$ & 5.8$^\circ$  \\ 
        $\phi$    & - & - & 47.0$^\circ$ & 37.7$^\circ$ & - & -  &34.2$^\circ$ & 28.8$^\circ$ \\
        $\Omega$  & - & - & 39.4$^\circ$ & 34.9$^\circ$ & - & -  & 32.2$^\circ$ & 27.8$^\circ$ \\  
        \hline
    \end{tabular}
    
    \caption{\label{tab:results_sh}Mean and median absolute error of reconstructed neutrino angles $\theta$, $\phi$, and total deviation, $\Omega$, for the shower branch. The 50\% best events are selected according to the lowest values of the parameters $\sigma$ for $N$-fit and $\chi^2$ for the $\chi^2$-fit.}
\end{table}

\begin{figure}[htbp]
	\centering
	\includegraphics[width=.48\textwidth]{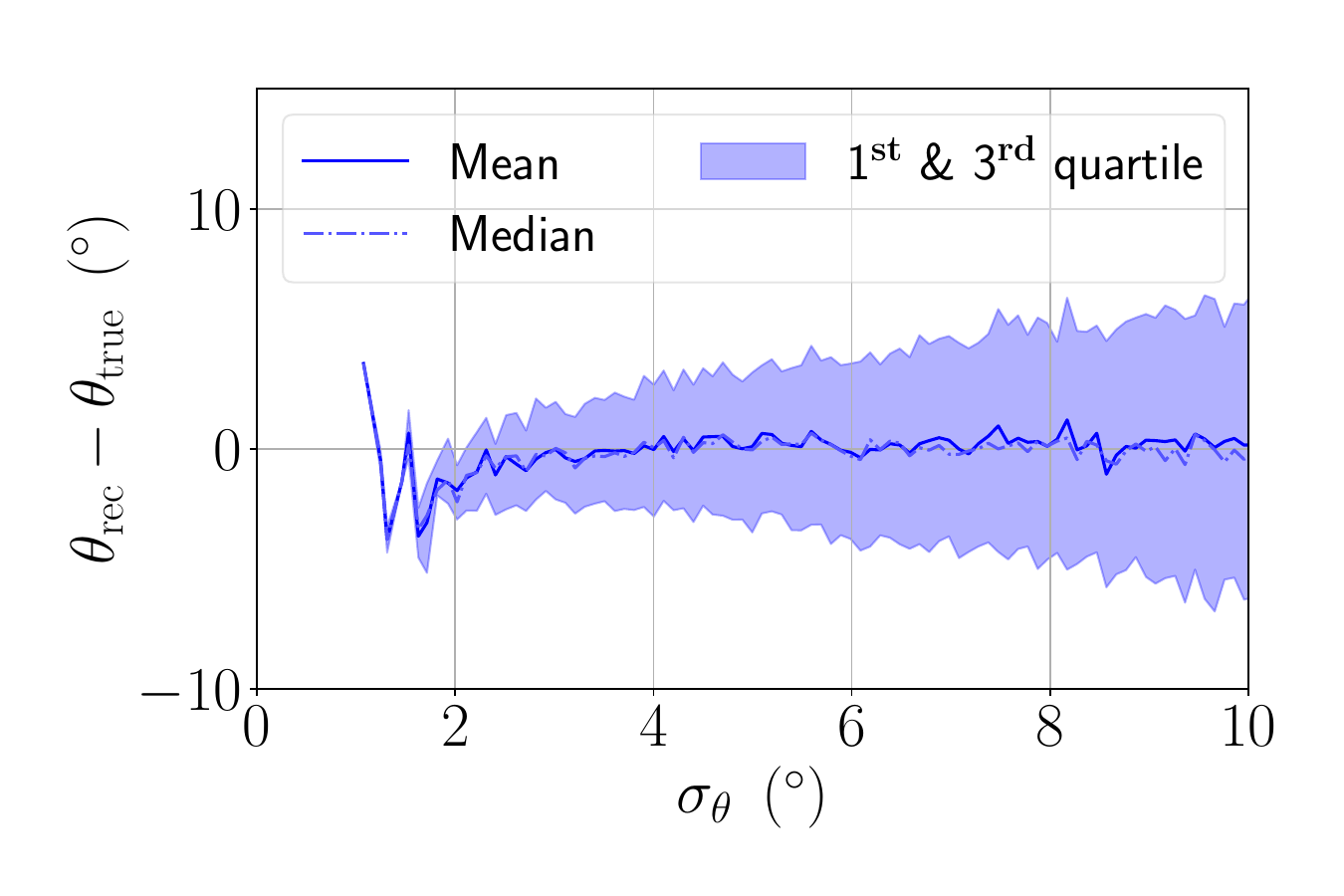}
	\includegraphics[width=.48\textwidth]{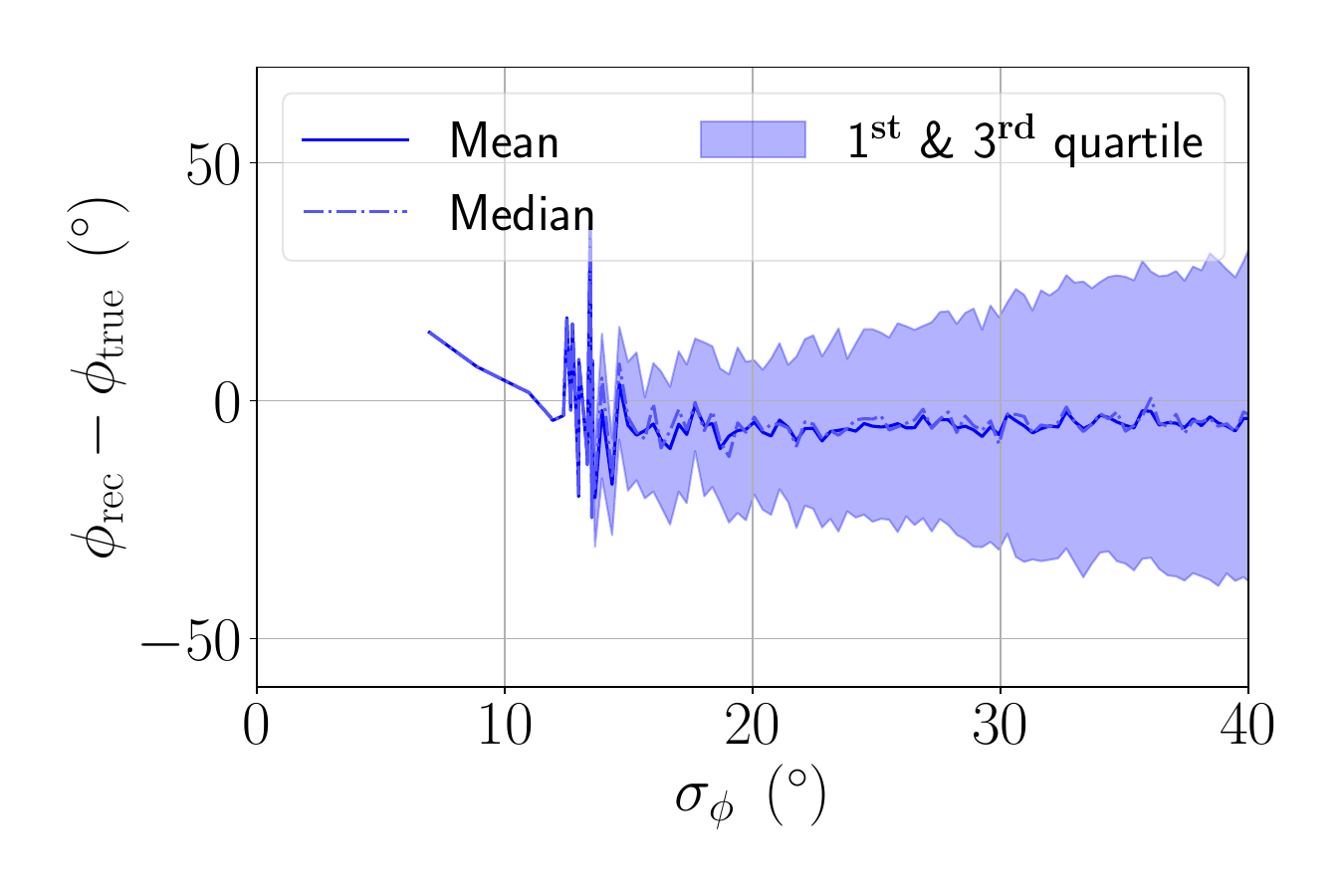}
	\caption{\label{fig:sigma_sh}The error on angles $\theta$ (left) and $\phi$ (right) as a function of the predicted uncertainty for the shower branch. The mean and median error stay close to zero with no significant bias. The first and third quartiles behave as expected for a Gaussian distribution, especially for $\theta$.}
\end{figure}

%
%

\begin{figure}[htbp]
    \centering
    
    \begin{subfigure}{\textwidth}
        \centering
        \subcaption{\label{fig:err_dist_sh_A}} 

        \makebox[0.5cm][c]{\raisebox{2.3cm}{\small\textbf{(1)}}}~\includegraphics[width=.45\textwidth]{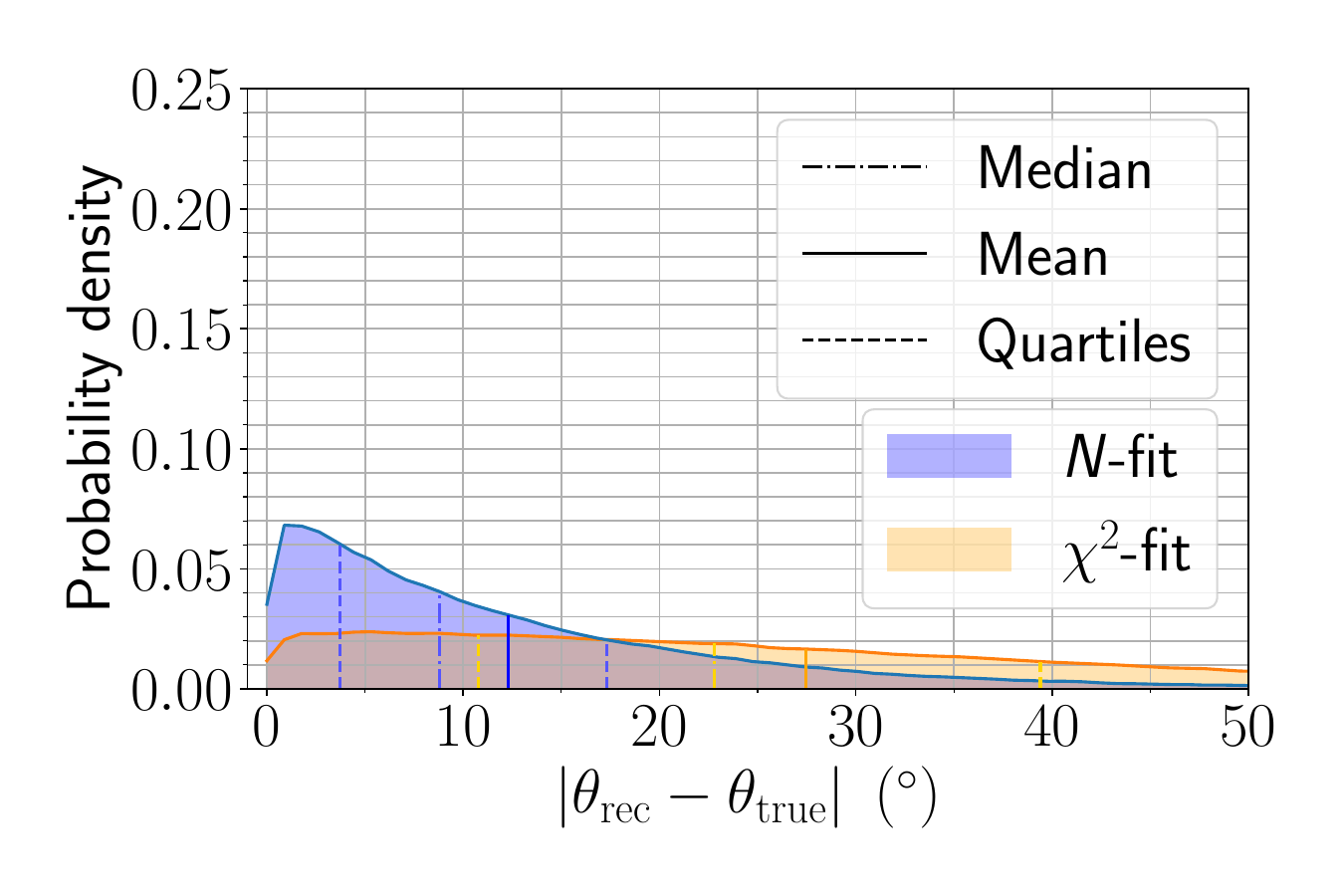}\hfill
        \includegraphics[width=.45\textwidth]{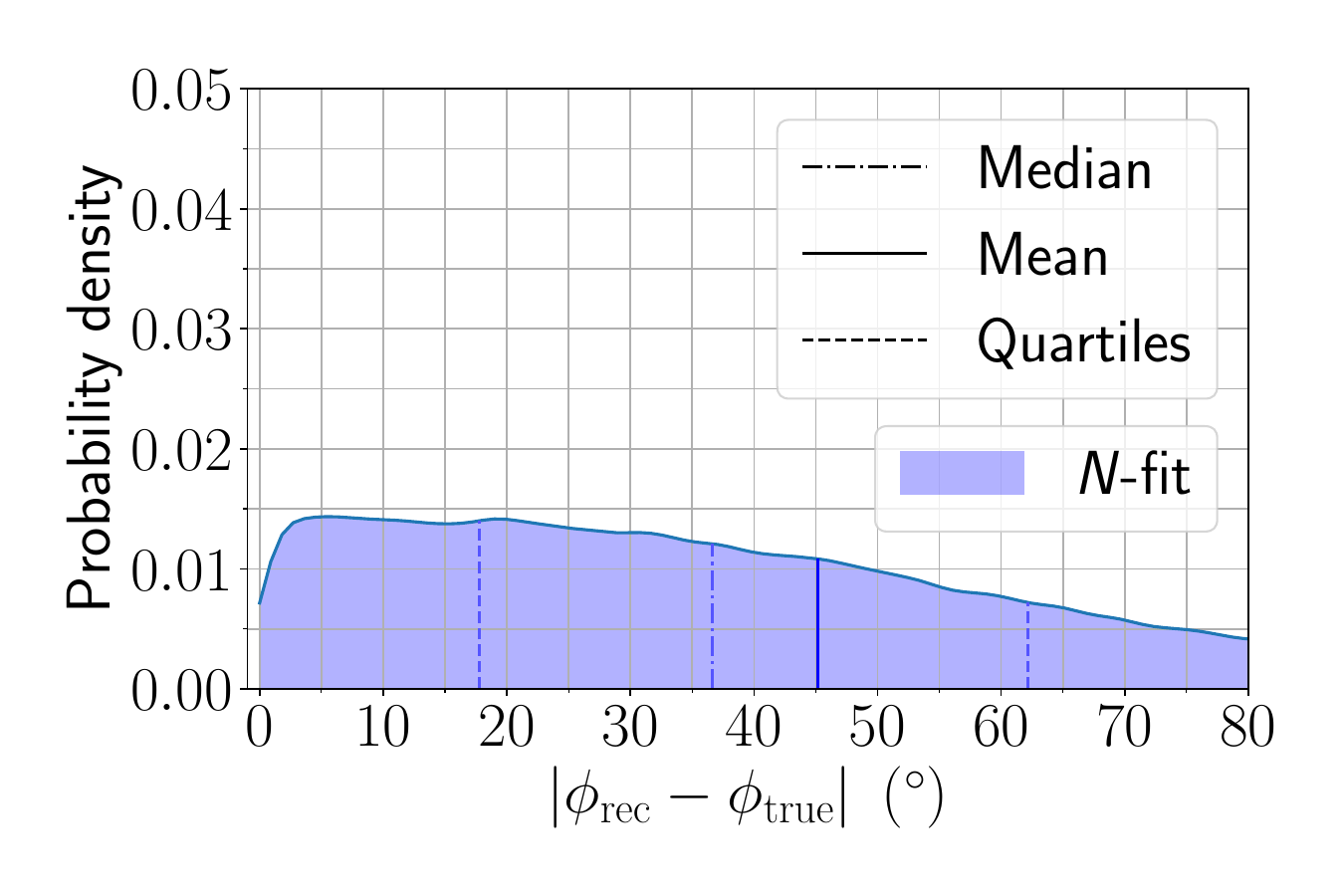}

        \makebox[0.5cm][c]{\raisebox{1.7cm}{\small\textbf{(2)}}}\hfill
        \includegraphics[width=.32\textwidth]{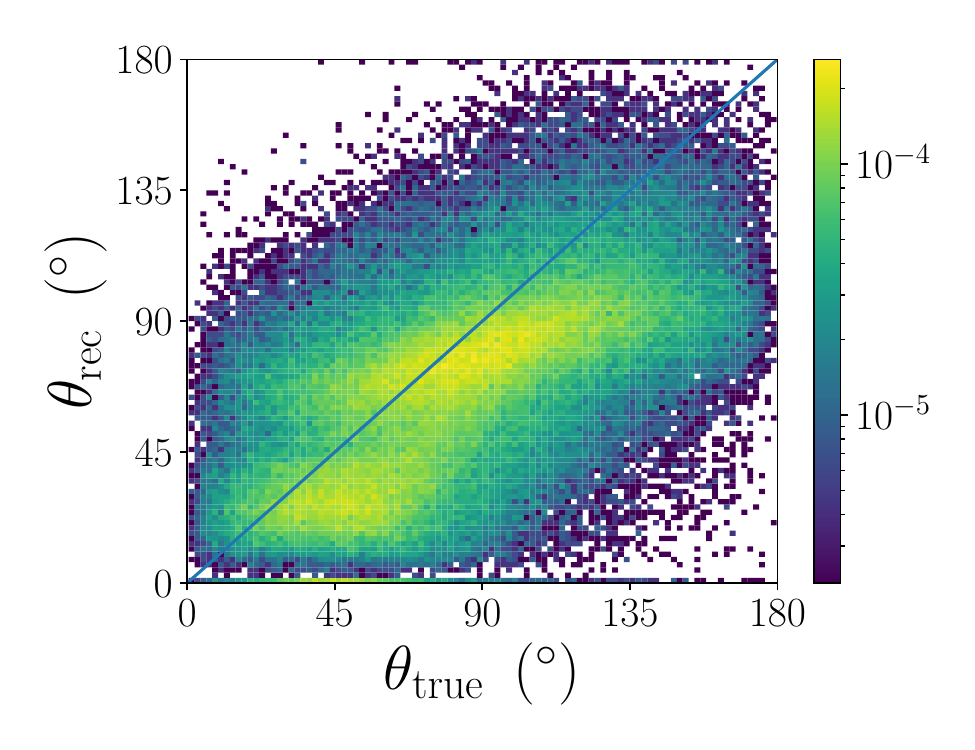}\hfill
        \vline\hfill
        \includegraphics[width=.32\textwidth]{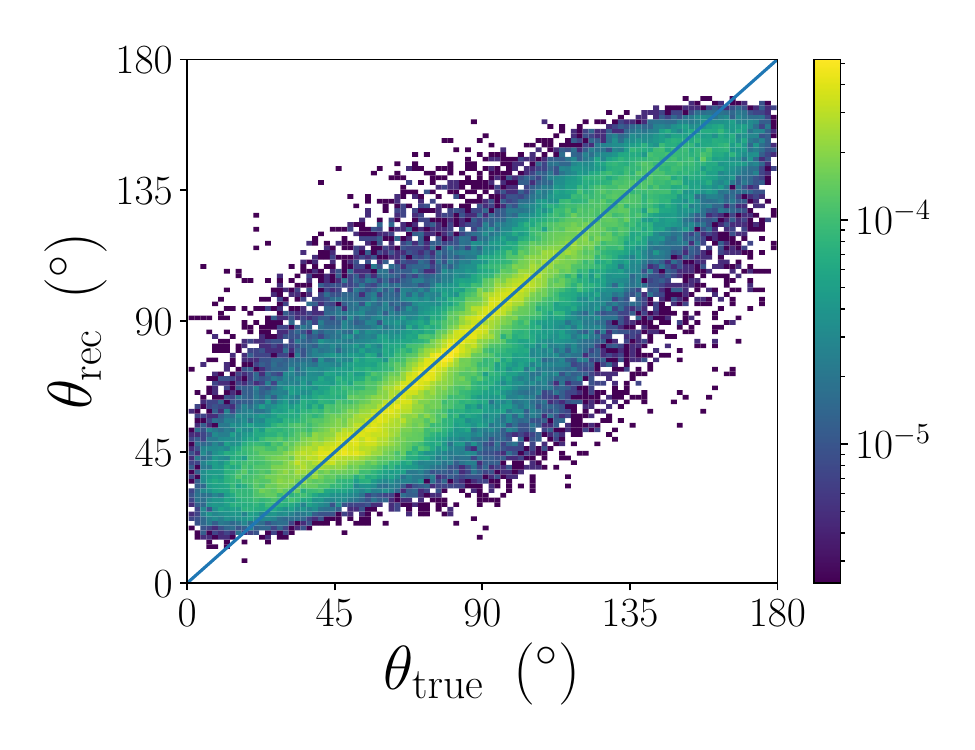}\hfill
        \includegraphics[width=.32\textwidth]{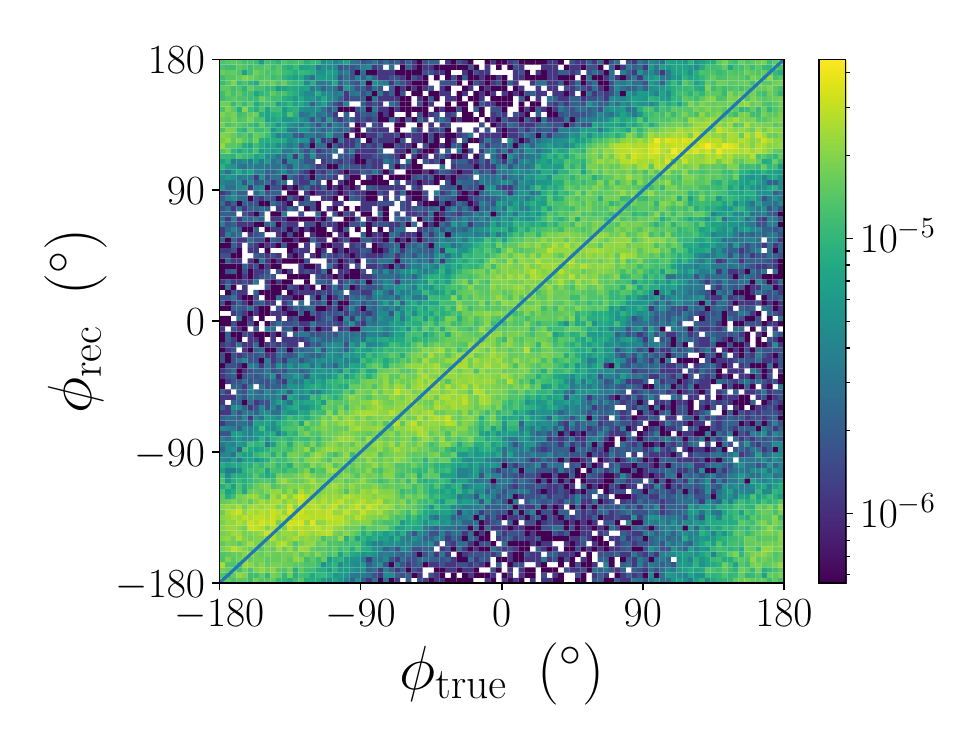}
    \end{subfigure}

    \vskip 4pt
    \hrule
    \vskip 4pt

    \begin{subfigure}{\textwidth}
        \centering
        \subcaption{\label{fig:err_dist_sh_B}}
        
        \makebox[0.5cm][c]{\raisebox{2.3cm}{\small\textbf{(1)}}}~\includegraphics[width=.45\textwidth]{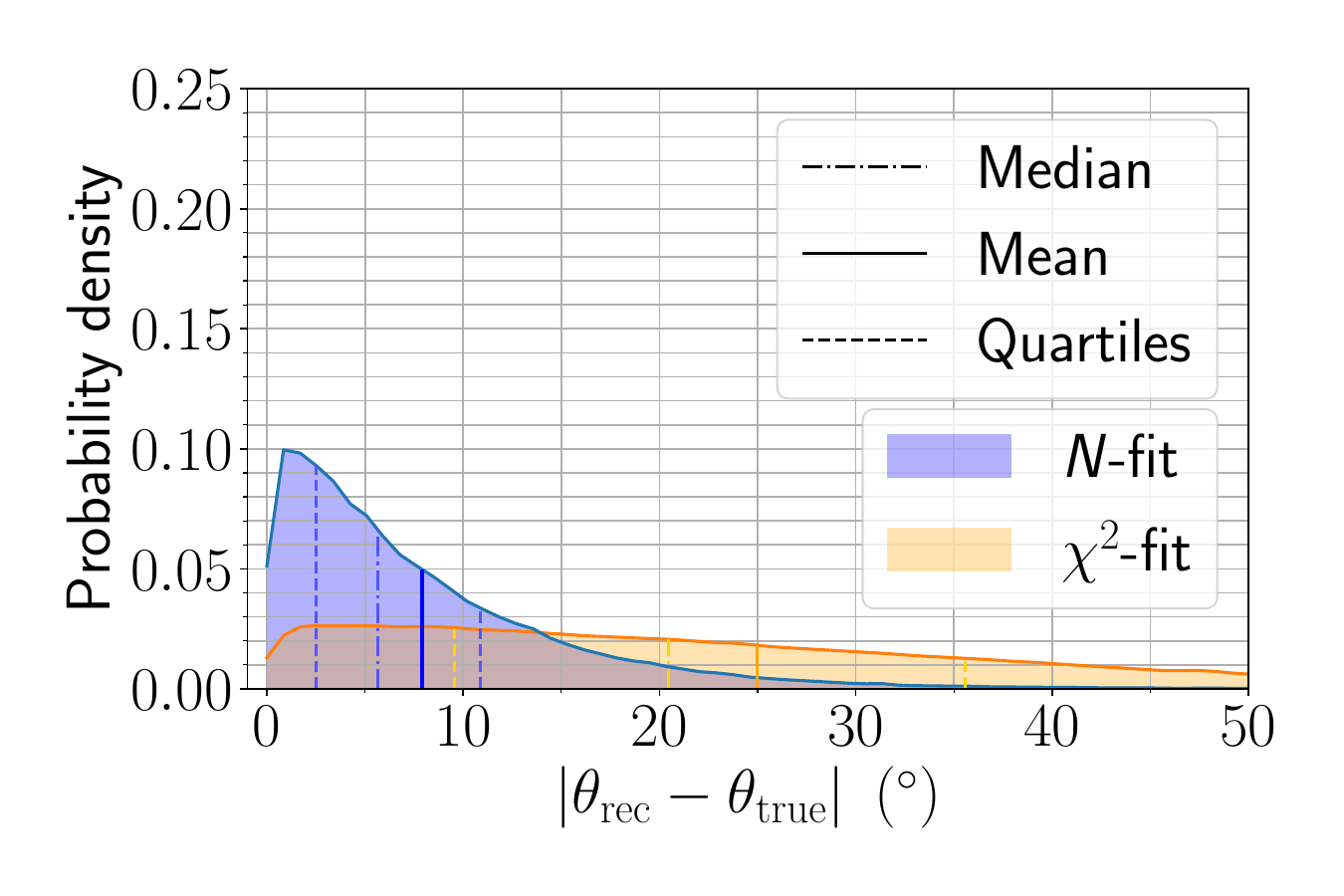}\hfill
        \includegraphics[width=.45\textwidth]{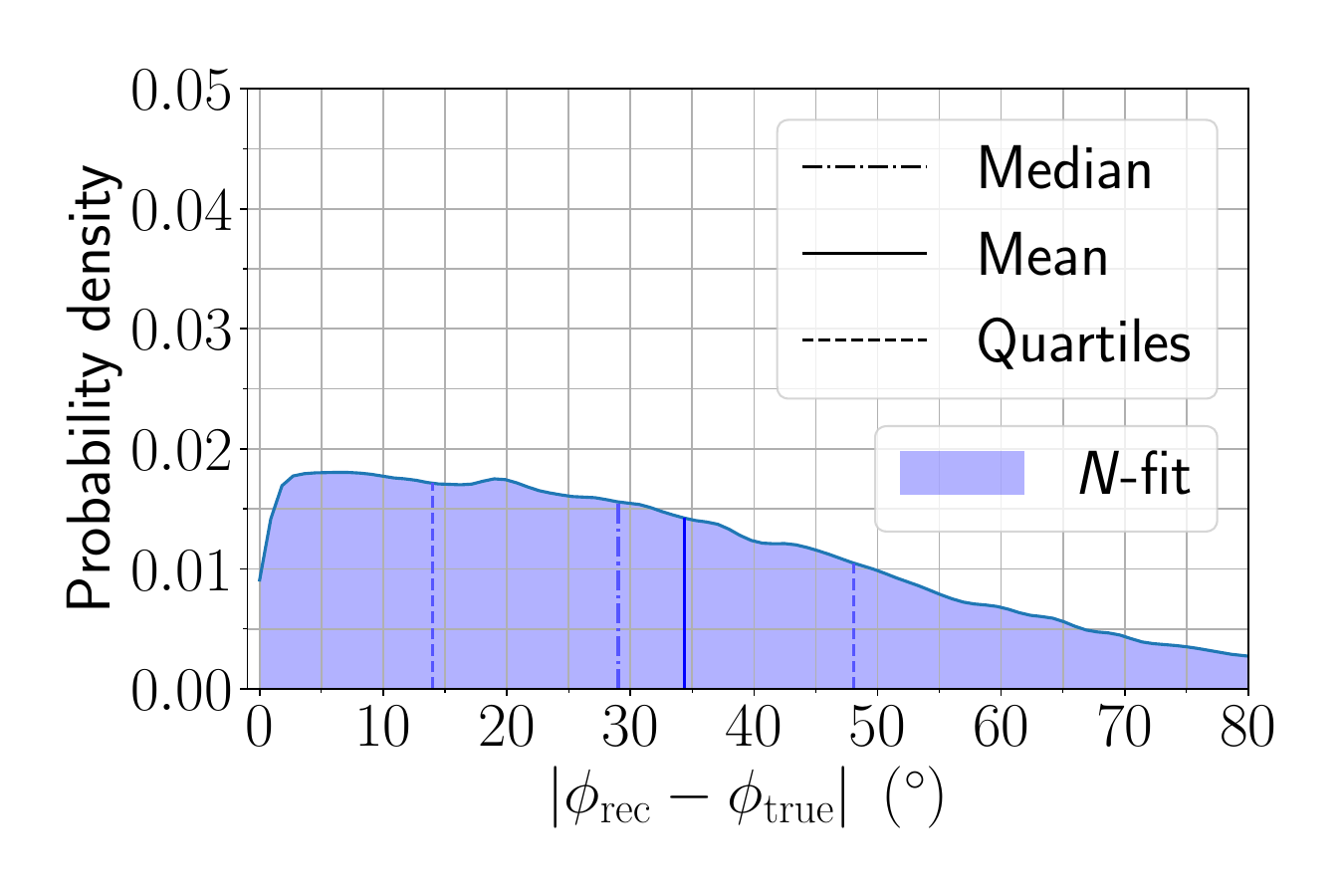}

        \makebox[0.5cm][c]{\raisebox{1.7cm}{\small\textbf{(2)}}}\hfill
        \includegraphics[width=.32\textwidth]{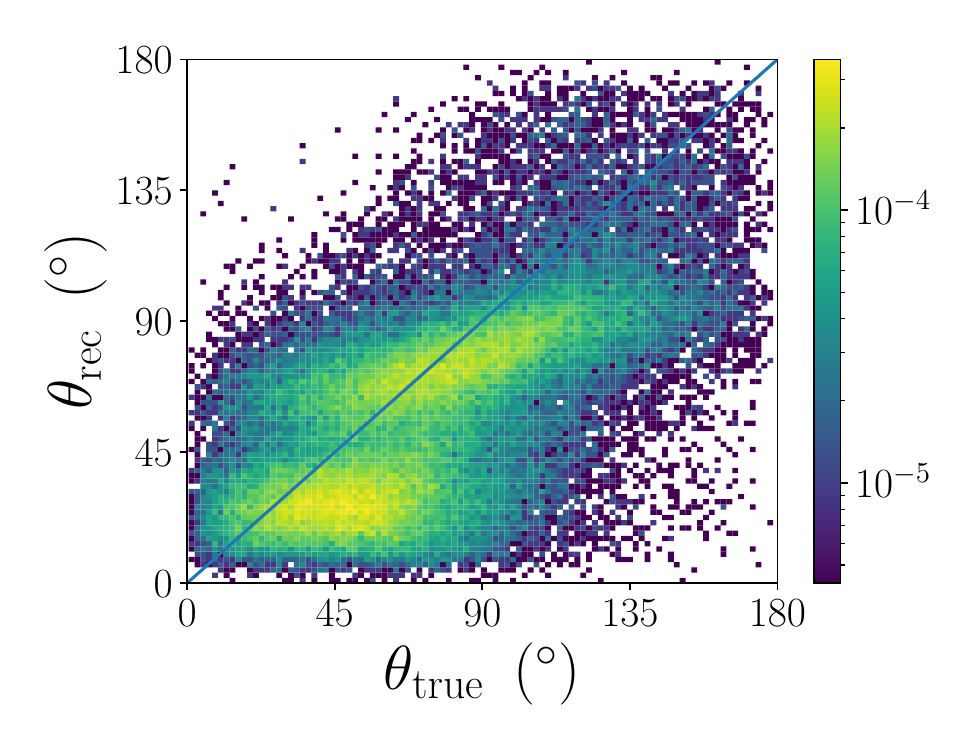}\hfill
        \vline\hfill
        \includegraphics[width=.32\textwidth]{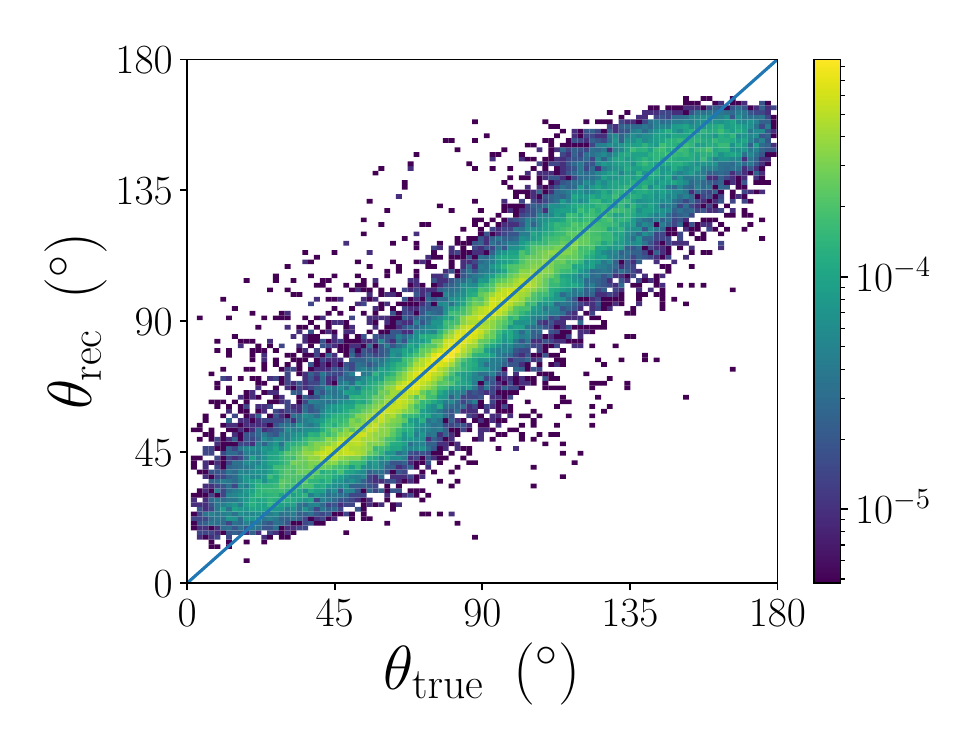}\hfill
        \includegraphics[width=.32\textwidth]{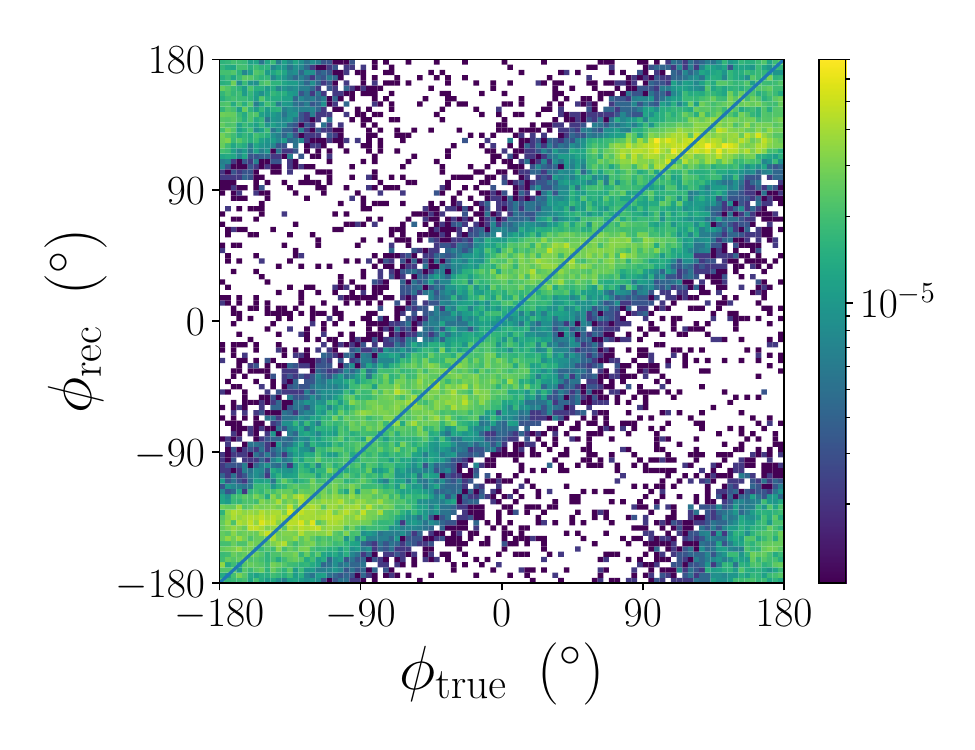}
    \end{subfigure}

    \caption{\label{fig:err_dist_sh}Shower branch results. \textbf{(a)} Whole test dataset. \textbf{(b)} 50\% of the test dataset with lowest values of $\sigma$ or $\chi^2$, respectively. \textbf{(1)}: Absolute error distributions of angles $\theta$ and $\phi$. \textbf{(2)}: 2D density histograms of true vs. reconstructed angles for $\chi^2$-fit (left) and $N$-fit (right). Note that $\phi$ points at the off-diagonal corners represent in fact good predictions due to periodicity.}
\end{figure}

\begin{figure}[htbp]
	\centering
	\includegraphics[width=.48\textwidth]{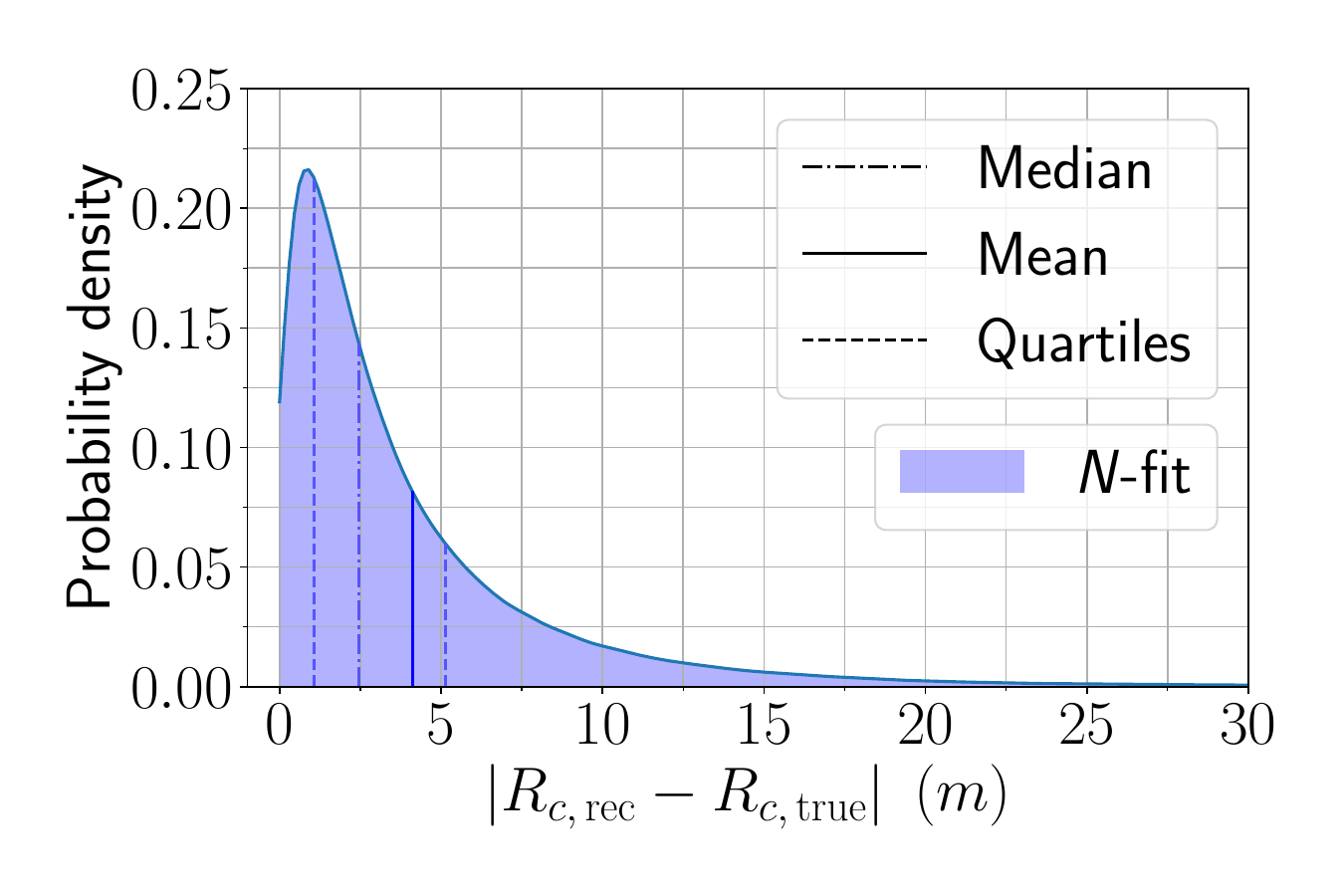}
	\includegraphics[width=.48\textwidth]{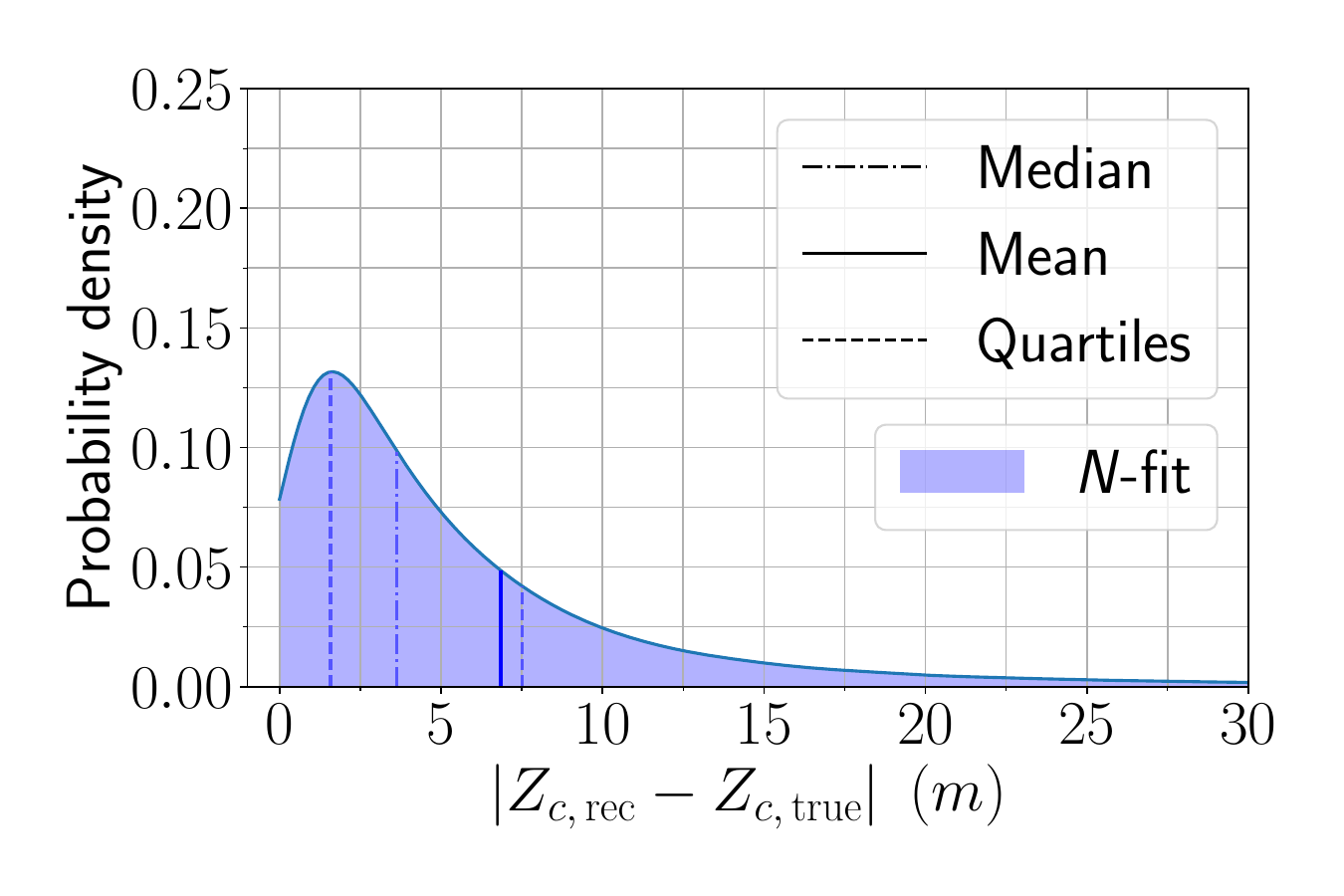}
	\caption{\label{fig:err_dist_RZ}$R_c$ and $Z_c$ absolute error distributions.}
\end{figure}

\begin{figure}[htbp]
	\centering
	\includegraphics[width=.48\textwidth]{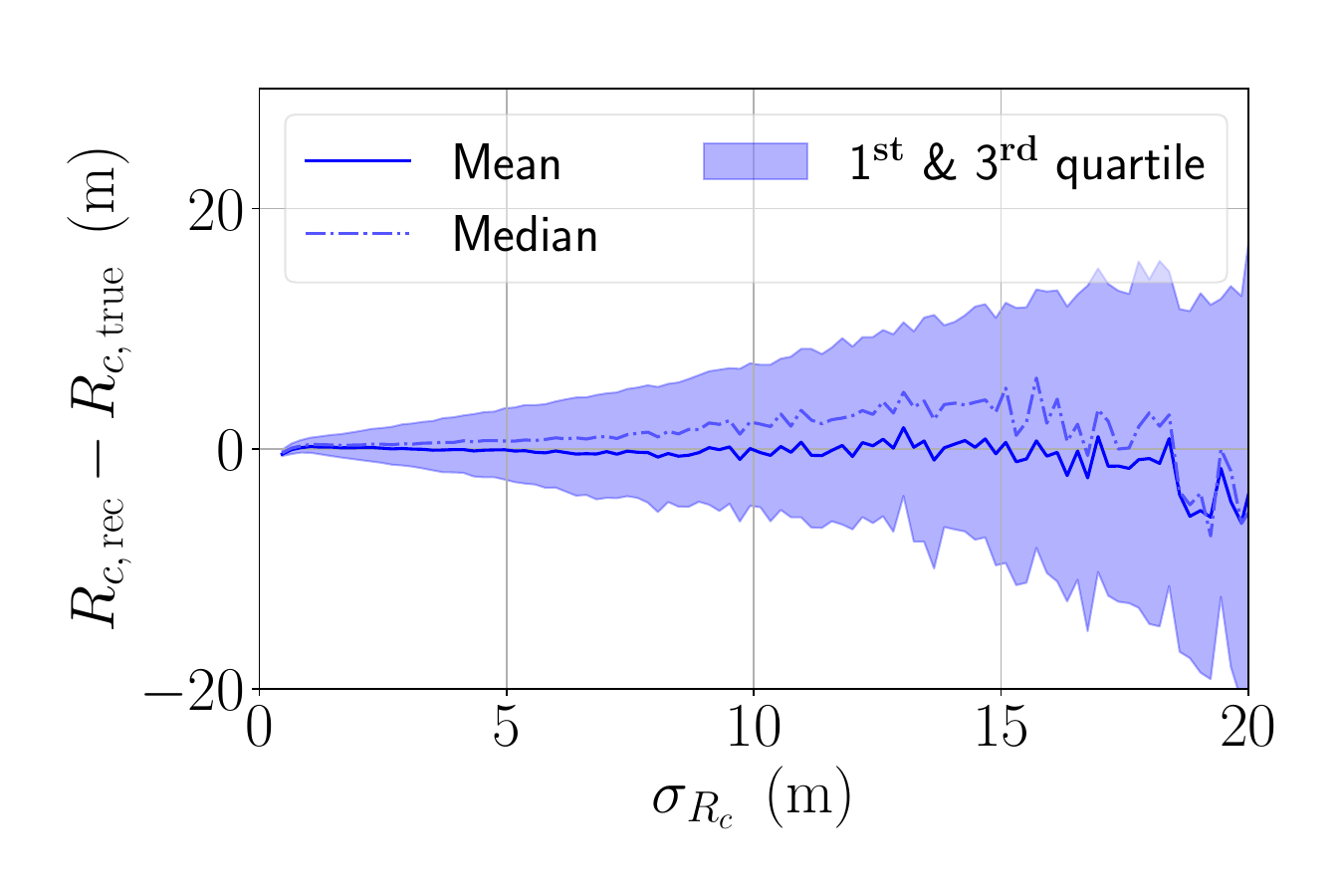}
	\includegraphics[width=.48\textwidth]{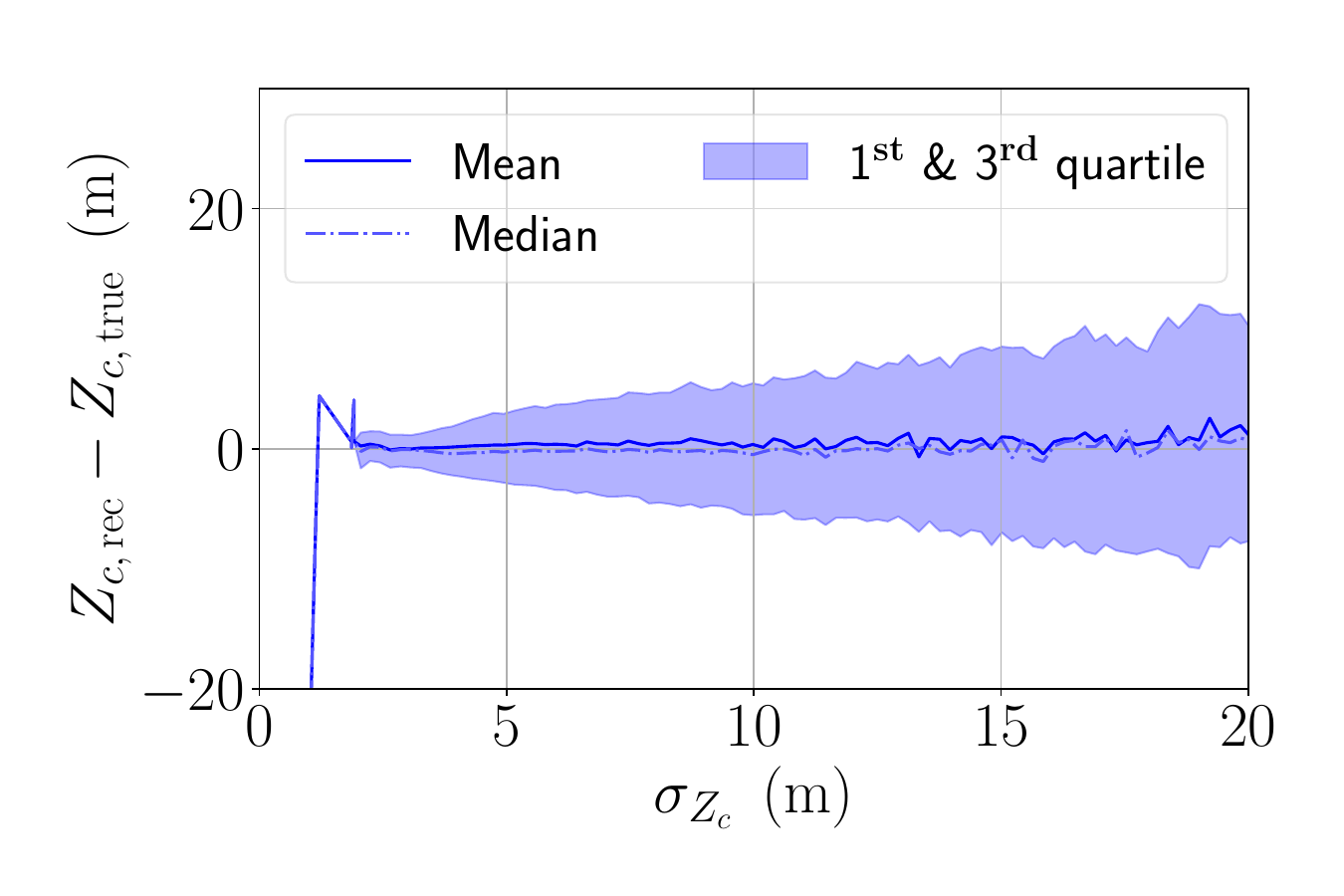}
	\caption{\label{fig:sigma_RZ}Error on $R_c$ (left) and $Z_c$ (right) as a function of the predicted uncertainty. The mean and median error stay close to zero with no significant bias. Second and third quartiles behave as expected for a Gaussian distribution.}
\end{figure}

\begin{figure}[htbp]
	\centering
	\includegraphics[width=.48\textwidth]{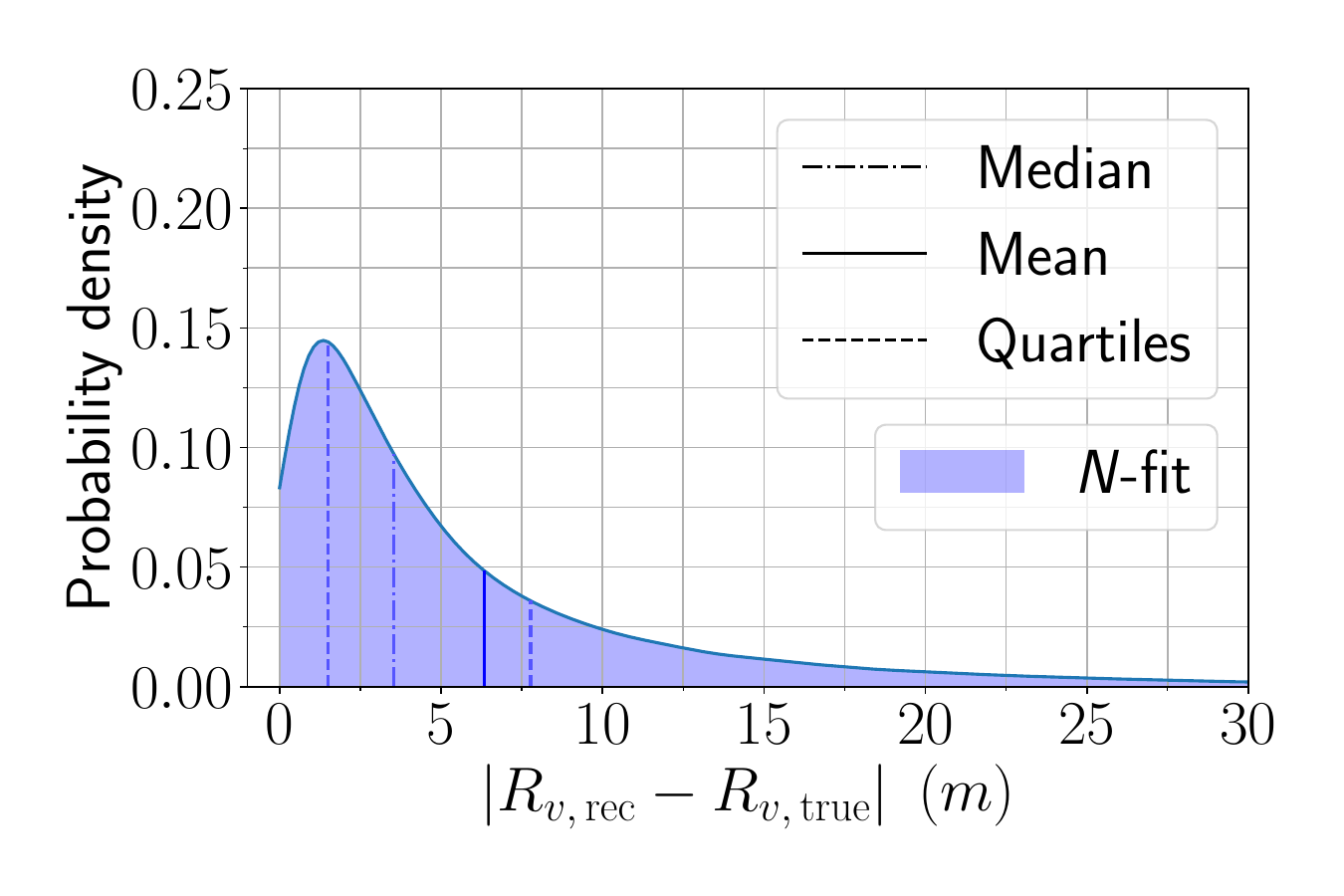}
	\includegraphics[width=.48\textwidth]{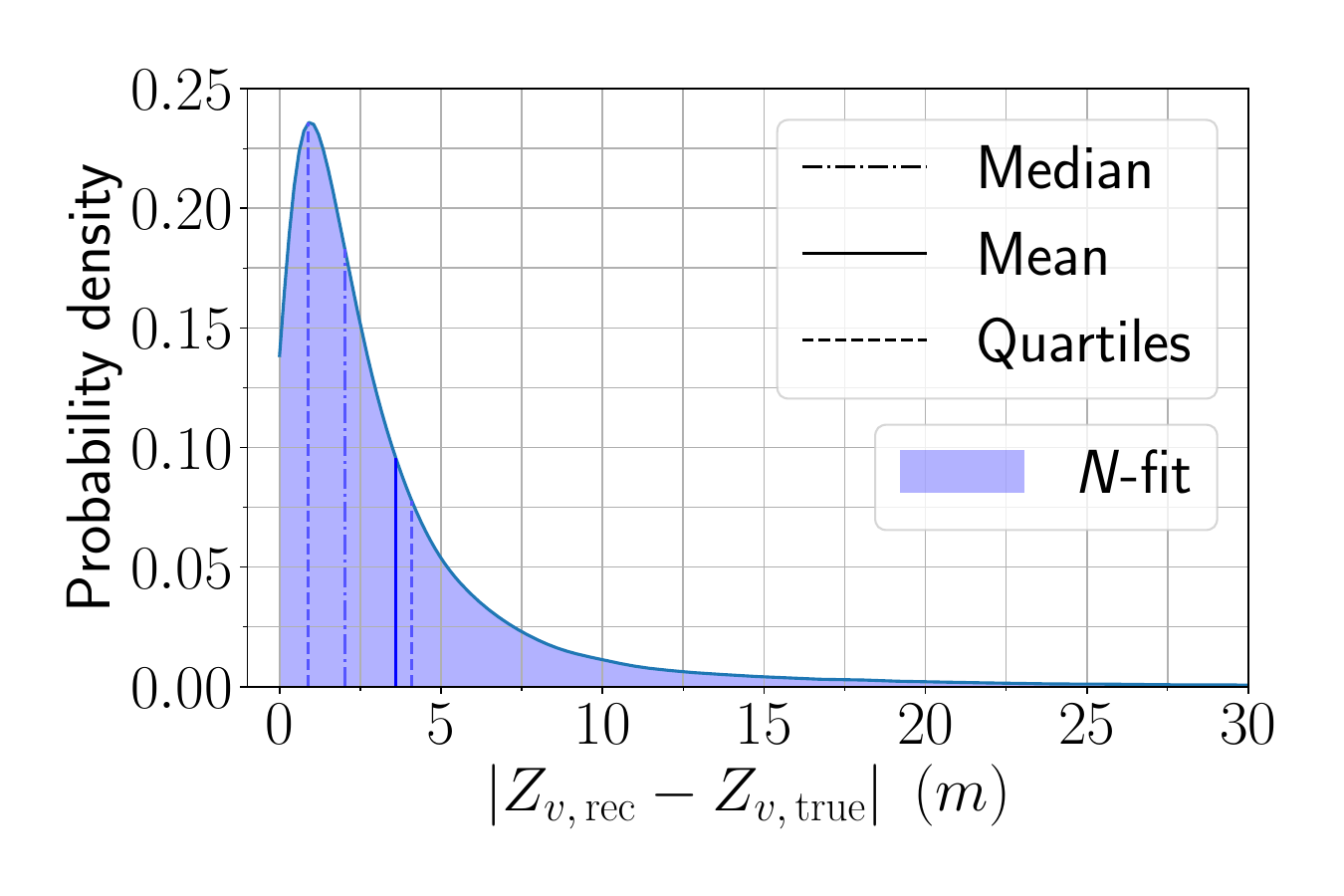}
	\caption{\label{fig:err_dist_RZ_sh}$R_v$ and $Z_v$ absolute error distributions.}
\end{figure}

\begin{figure}[htbp]
	\centering
	\includegraphics[width=.48\textwidth]{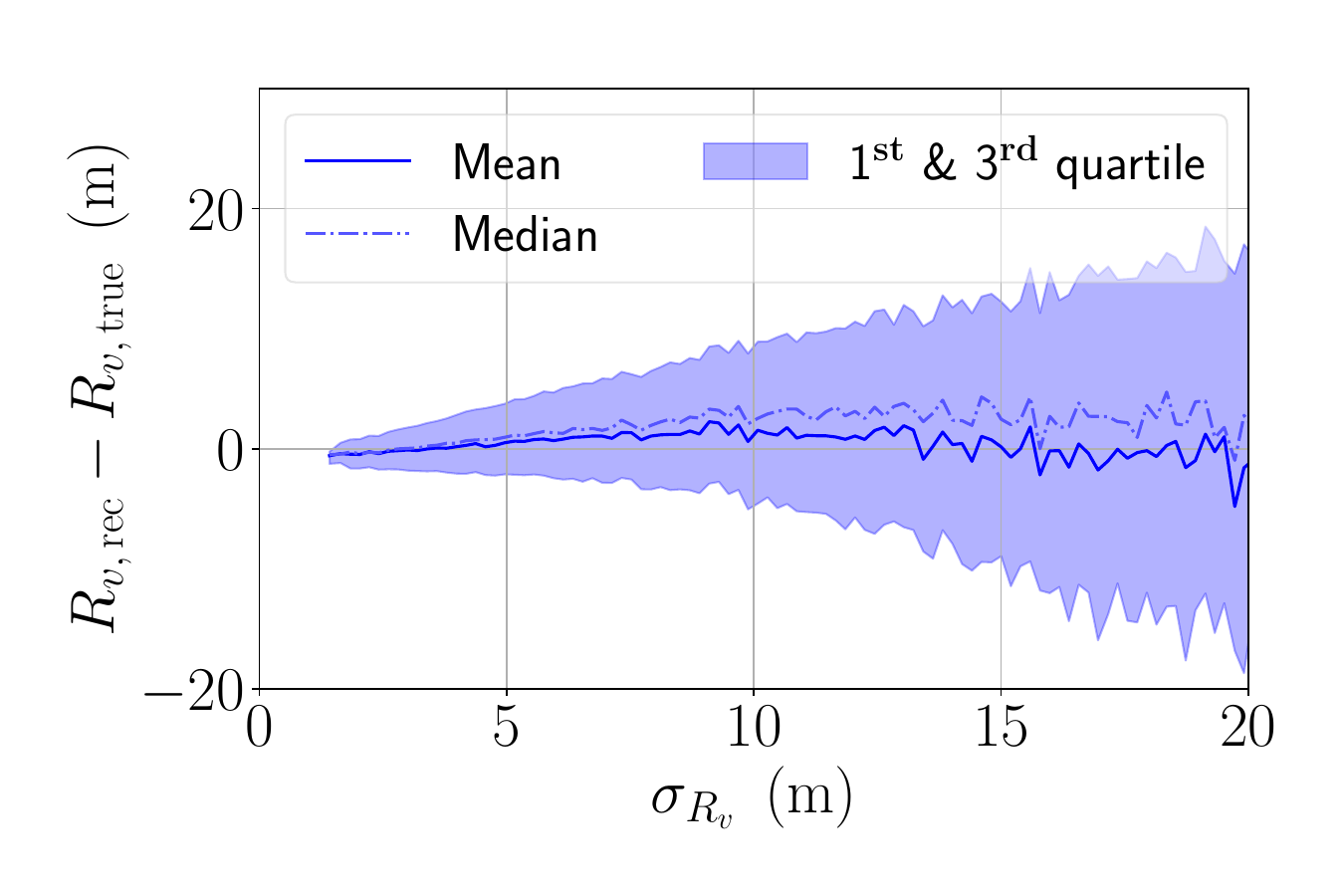}
	\includegraphics[width=.48\textwidth]{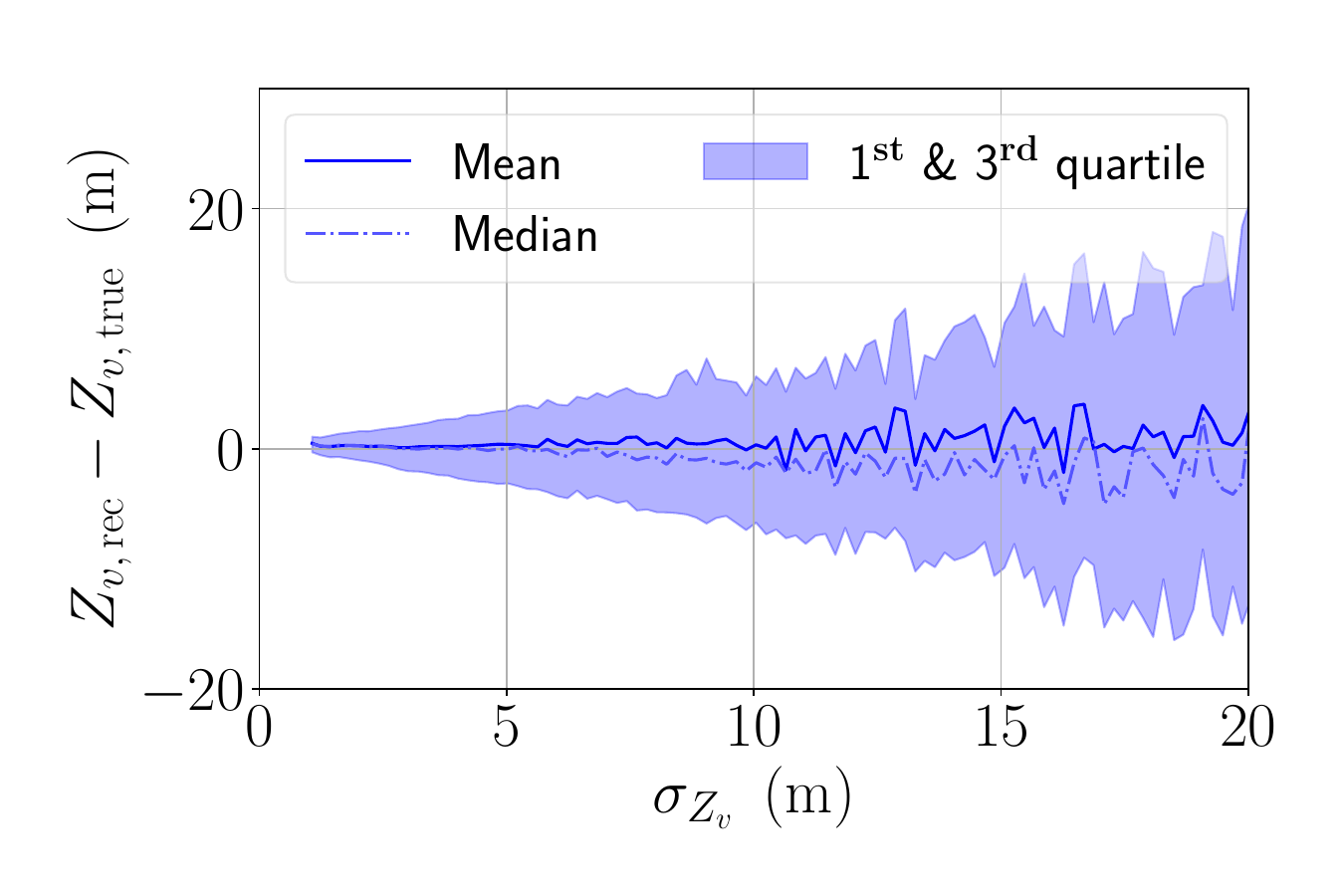}
	\caption{\label{fig:sigma_RZ_sh}Error on $R_v$ (left) and $Z_v$ (right) as a function of the predicted uncertainty. The mean and median error stay close to zero with no significant bias. Second and third quartiles behave as expected for a Gaussian distribution.}
\end{figure}

%
%
%
%
%
%
%
%

\clearpage

\section{Robustness tests}
\label{app:robust}

\subsection{K-fold cross validation}
\label{subsubsec:kfold}

The K-fold cross validation (KCV) technique \cite{KCV} is largely used in ML to check the robustness of the networks and their possible dependence with the dataset chosen for the training process. The technique consists in splitting the dataset in $K$ folds and use some of them for training while the others are used for validation and test. Then, the data is shifted circularly 1 fold and a second training is performed. The process is repeated until reaching $K$ circular shifts. This ensures that all the folds take all the different positions in training, validation and test.

We have randomly split the dataset in 5 folds with 20\% of the data in each one. Then, we have used 3 folds for training, 1 for validation and 1 for test. Consequently, we have performed the training process 5 times. Statistics of the results can be seen in table \ref{tab:5-fold}. The errors are stable and similar to those in table \ref{tab:results}, showing that the method has no dependence on a particular data split.

\begin{table}[htbp]
	\centering
	\renewcommand{\arraystretch}{1.2}
	\begin{tabular}{cccccc}
		\hline
		MAE & fold 5 & fold 4 & fold 3 & fold 2 & fold 1 \\ \hline \hline
		\multicolumn{1}{c}{$\theta$}  &  7.4$^\circ$ & 7.3$^\circ$ & 7.4$^\circ$ & 7.4$^\circ$ & 7.4$^\circ$ \\
		\multicolumn{1}{c}{$\phi$} & 41.5$^\circ$ & 41.3$^\circ$  & 41.6$^\circ$ & 41.4$^\circ$ & 41.3$^\circ$ \\
		\multicolumn{1}{c}{$\Omega$}   & 28.2$^\circ$ & 28.1$^\circ$ & 28.4$^\circ$ & 28.1$^\circ$ & 28.1$^\circ$ \\ \hline
	\end{tabular}
    \caption{\label{tab:5-fold}Mean absolute error of predicted neutrino angles $\theta$ and $\phi$ for the 5-fold cross-validation. The fold number denotes the fold used as test.}
\end{table}

Since the conditions of the ANTARES telescope have changed through the years, we need to ensure that the network model is able to fit different stages of the telescope properly. To that end, we performed another KCV, but this time with folds sorted by date. Thus, we split the data in 5 folds, each one containing events within the following dates:

\begin{itemize}
	\item Fold 1: from 06/02/2007 to 13/01/2010
	\item Fold 2: from 13/01/2010 to 17/05/2011
	\item Fold 3: from 22/05/2011 to 02/09/2012
	\item Fold 4: from 03/09/2012 to 24/04/2015
	\item Fold 5: from 24/04/2015 to 27/12/2017
\end{itemize}

Statistics of the results can be seen in \autoref{tab:5-fold_dates}. We acknowledge modest fluctuations in the results that emerge from the fact that not all lines and OMs were always operative in ANTARES at all times. The performance was slightly better if the first stages of the telescope are used as the test dataset. This is possibly due to the fact that these data had better quality, compared to the others, since the telescope was not affected by ageing effects. Thus, having better data in the test set gives raise to better performance.

\begin{table}[htbp]
	\centering
    \renewcommand{\arraystretch}{1.2}
	\begin{tabular}{cccccc}
		\hline
		MAE & fold 5 & fold 4 & fold 3 & fold 2 & fold 1 \\ \hline \hline
		\multicolumn{1}{c}{$\theta$}  &  8.1$^\circ$ & 7.6$^\circ$ & 7.4$^\circ$ & 7.5$^\circ$ & 6.9$^\circ$ \\ 
		\multicolumn{1}{c}{$\phi$} & 44.4$^\circ$ & 42.2$^\circ$ & 41.4$^\circ$s & 41.7$^\circ$ & 39.7$^\circ$ \\ 
		\multicolumn{1}{c}{$\Omega$}   & 30.0$^\circ$ & 28.7$^\circ$ & 28.4$^\circ$ & 29.1$^\circ$ & 26.3$^\circ$ \\ \hline
	\end{tabular}
    \caption{\label{tab:5-fold_dates}Mean absolute error of predicted neutrino angles $\theta$ and $\phi$ for the 5-fold dates-sorted cross-validation. The fold number is the one used as test.}
\end{table}

\subsection{Background noise}
\label{subsubsec:background}

\begin{figure}[htbp]
	\centering
	\includegraphics[width=.48\textwidth]{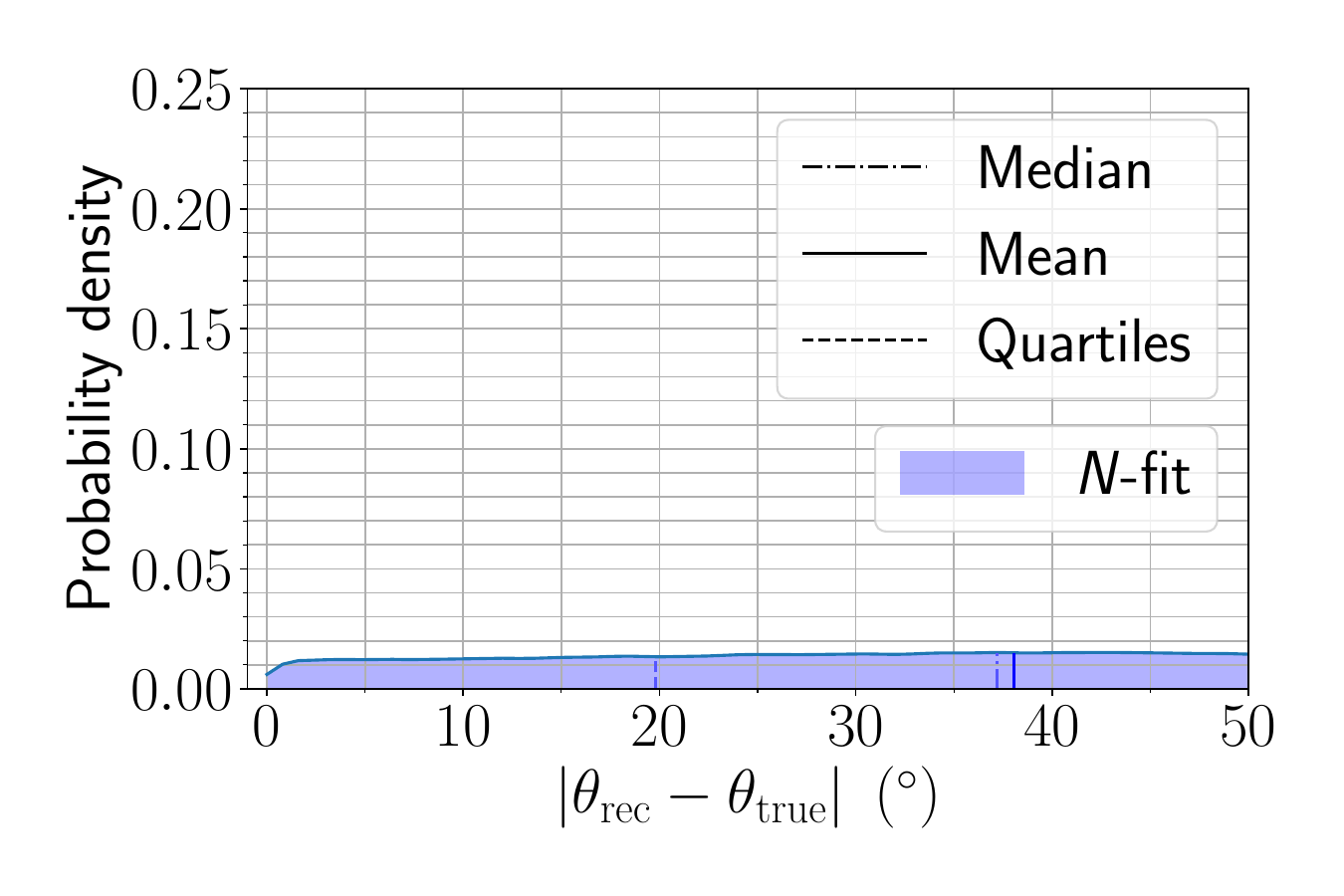}~
	\includegraphics[width=.48\textwidth]{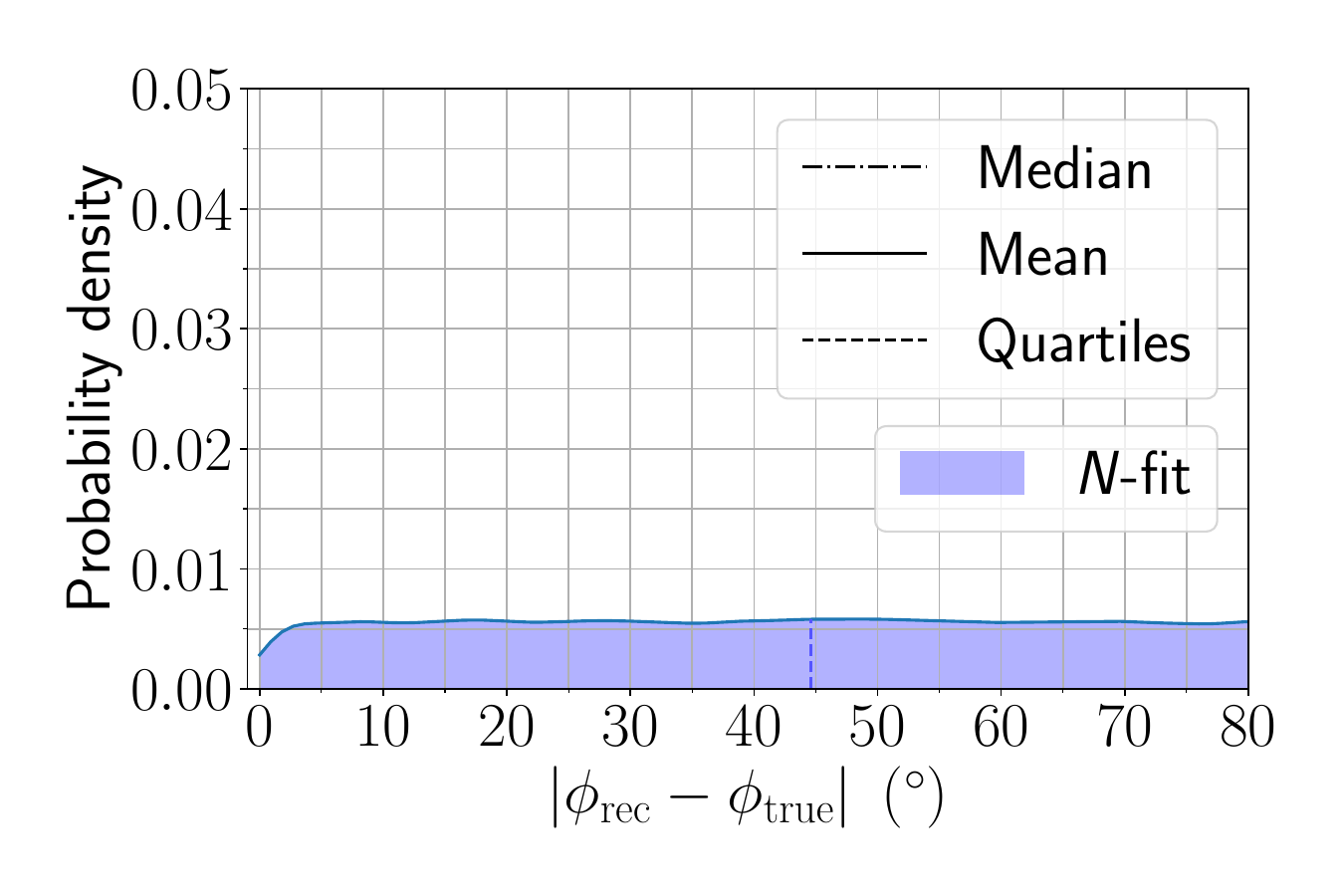}~
	
	\includegraphics[width=.35\textwidth]{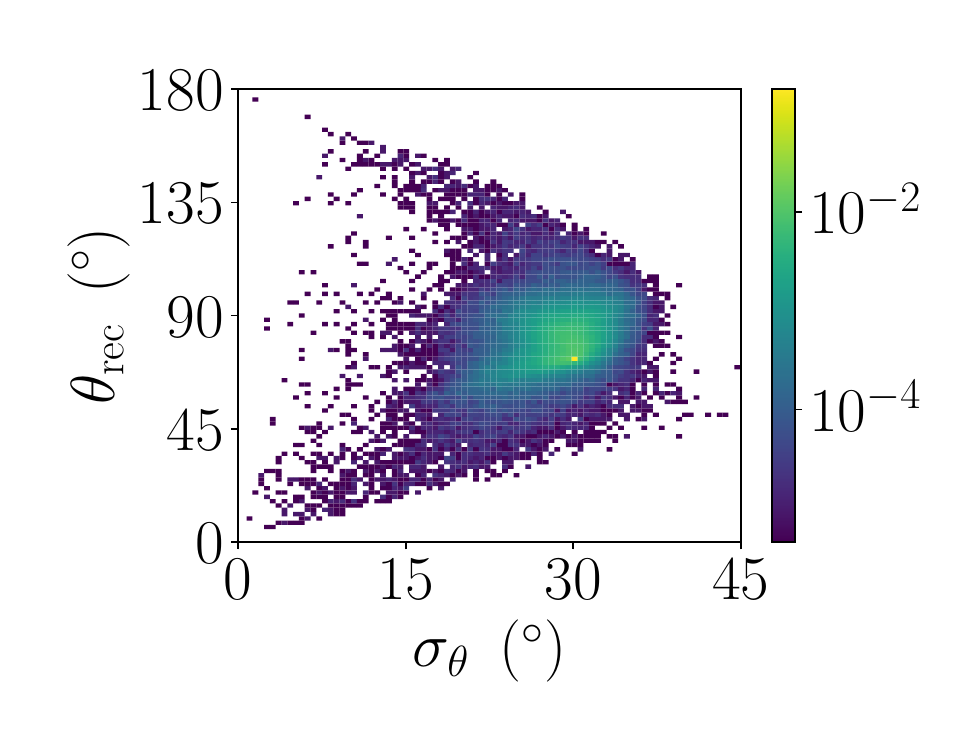}
	\caption{\label{fig:background}Error distributions if the input of the networks is background noise in the track branch.}
\end{figure}

As a control of the predictions generated by our model, we processed images of the background and fed them to the $N$-fit track direction networks. The expectation of using the background as the signal was to obtain direction reconstructions following a random uniform distribution, and that was indeed what we obtained as reported in \autoref{fig:background}.

Moreover, the predicted uncertainty was typically large, as seen in the bottom panel of Figure \autoref{fig:background} for the $\theta$ angle.

\subsection{Comparison with data}
\label{subsubsec:dataMC}

The last test performed is a comparison between data and MC simulations. Our model was fit entirely on MC simulations, so it could be biased for data in ways that could not be easily anticipated. Shall we want to use this approach in physics analyses, we must be sure that the results of the simulations can be applied to data. This is true for every reconstruction algorithm and it is the reason why the simulations are done \textit{run-by-run}. Furthermore, every event of a simulation is provided with a set of weights to directly compare simulations and data \cite{weights}. These weights take into account the interaction, propagation and detection physics of the events, as well as the expected flux of every type of event.

\begin{figure}[htbp]
	\centering{
	\includegraphics[width=.49\textwidth]{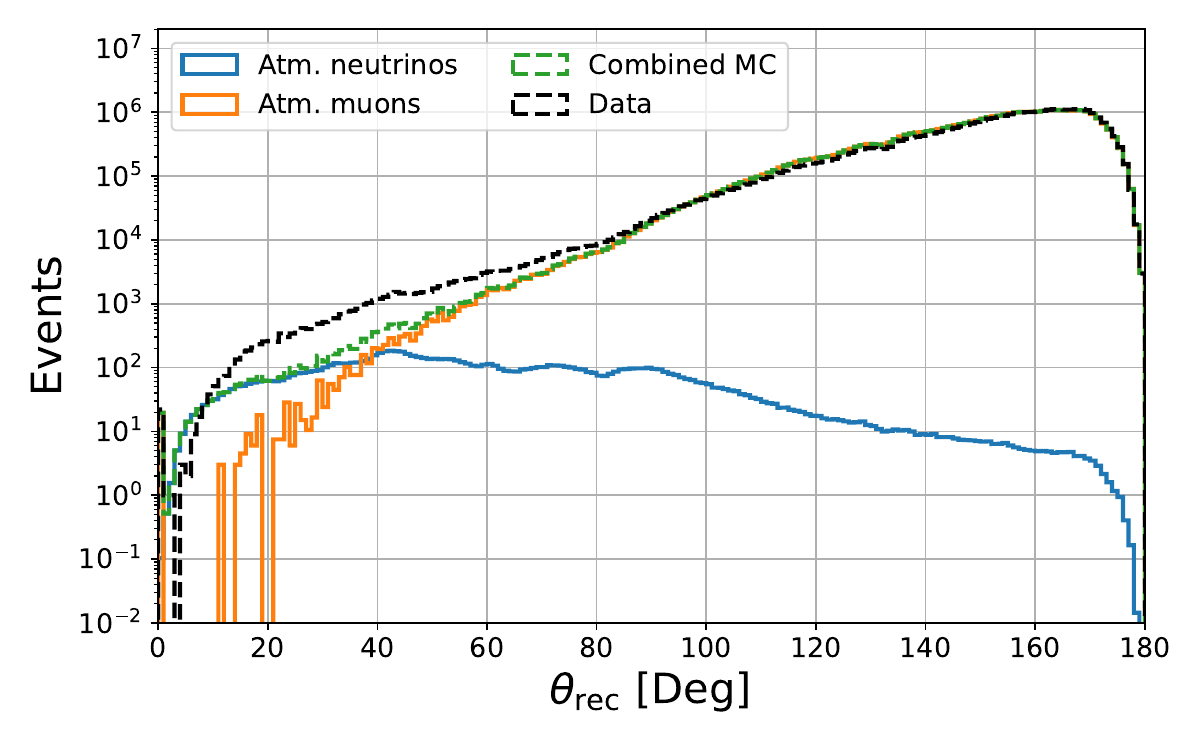}
    \hfill
	\includegraphics[width=.49\textwidth]{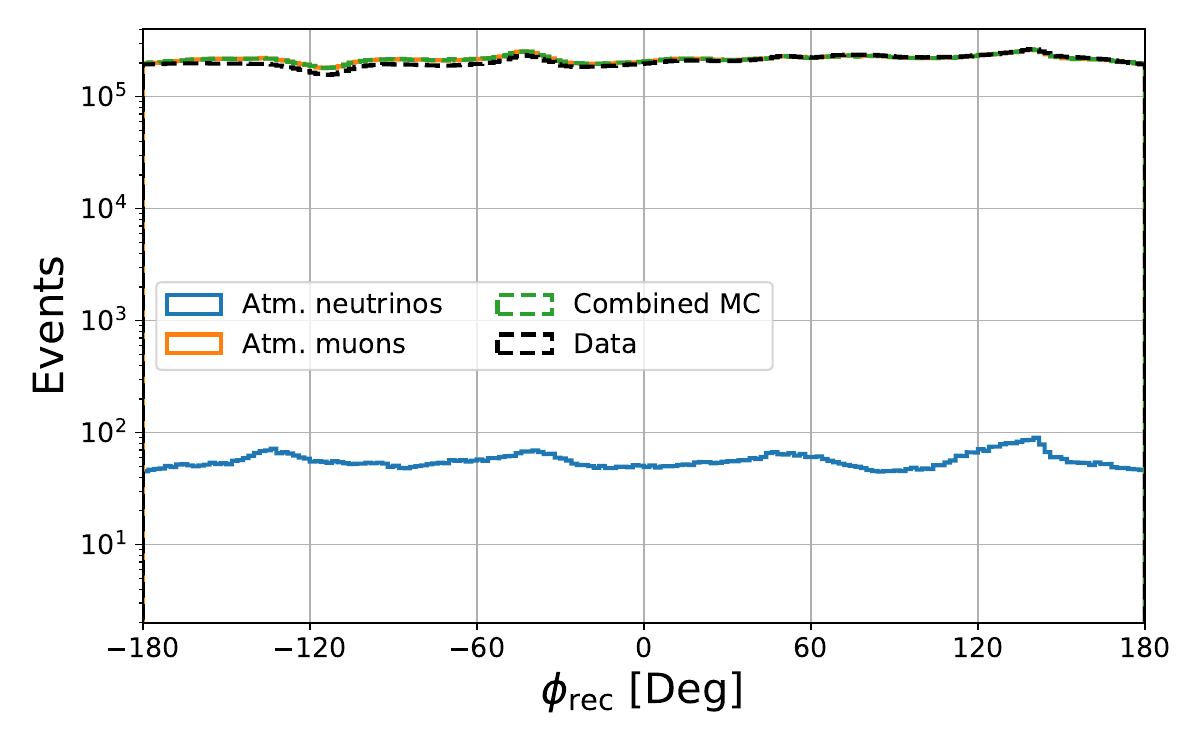}
	\caption{\label{fig:real}Comparison between data and MC simulations.}}
\end{figure}

\begin{figure}[htbp]
	\centering{
	\includegraphics[width=.49\textwidth]{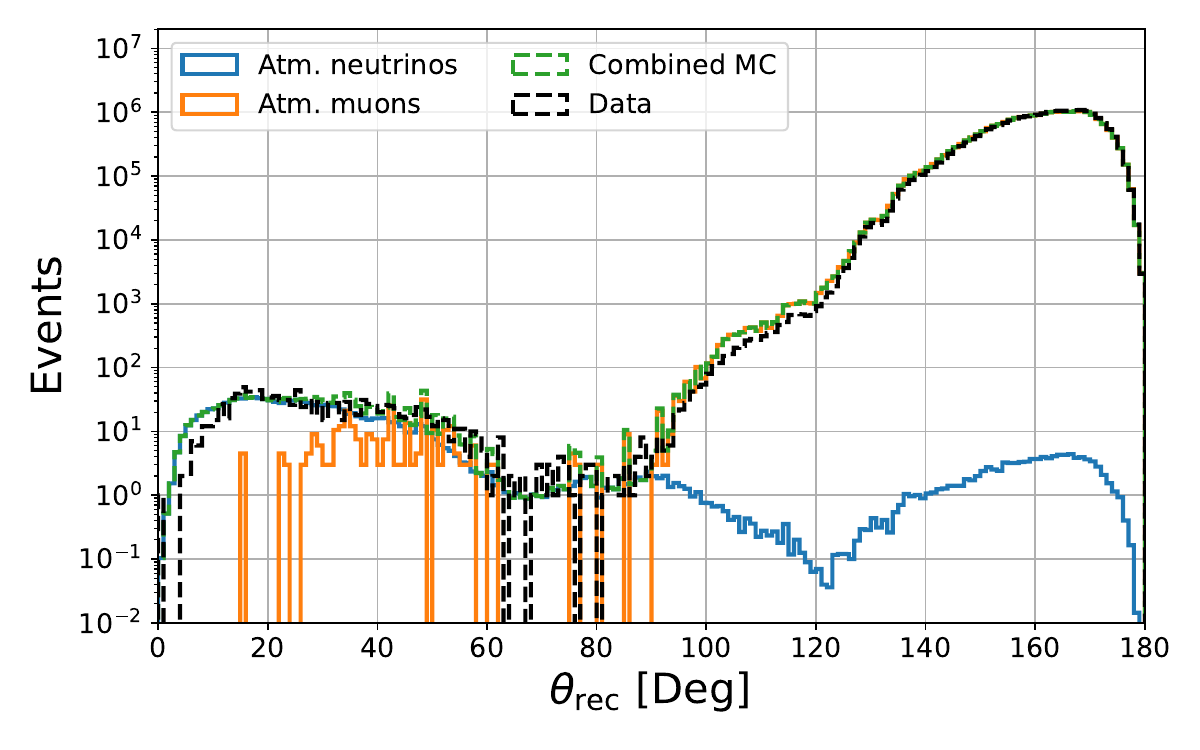}
    \hfill
	\includegraphics[width=.49\textwidth]{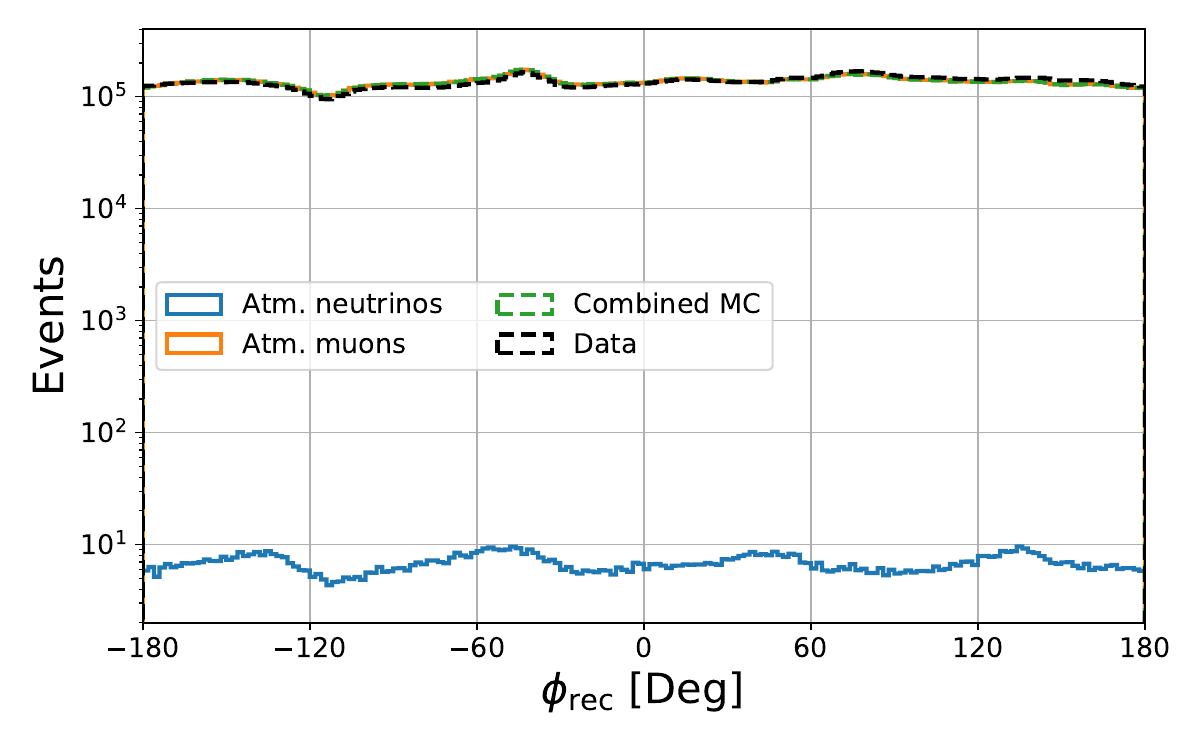}
	\caption{\label{fig:real_cut}Comparison between data and MC simulations using a cut of $\sigma_\Omega<21^\circ$.}}
\end{figure}

For the comparison, we selected new runs that were not used for training. We also made sure that, for every selected run, all the MC simulation files were available. \autoref{fig:real} and \autoref{fig:real_cut} illustrate the results from this comparison.

We observe that the distributions are similar, specially when a cutoff based on the quality parameter $\sigma_\Omega$ is applied (\autoref{fig:real_cut}), rejecting much of the background noise data and atmospheric muons. These results demonstrate that our method is ready to be used in physics analyses.





\end{document}